\documentclass[aps,pra,10pt,twocolumn,nolongbibliography]{revtex4-2}%
\pdfoutput=1%
\usepackage{amsmath, amssymb}%
\usepackage{mathphyscmd}%
\everymath{\displaystyle}%
\usepackage{hyperref}%
\hypersetup{%
	pdfinfo={%
		Author       = {David Gaspard},
		Title        = {Quantum master equations for a fast particle in a gas},
		Subject      = {Quantum transport theory},
		CreationDate = {D:20220720162402+0200},
	},
	pdfborderstyle={/S/U/W 1},
	breaklinks=true
}

\begin{document}%
\title{Quantum master equations for a fast particle in a gas}%

\author{David Gaspard}%
\affiliation{Nuclear Physics and Quantum Physics, CP229, Universit\'e libre de Bruxelles (ULB), B-1050 Brussels, Belgium}%
\date{\today}%

\begin{abstract}%
The propagation of a fast particle in a low-density gas at thermal equilibrium is studied in the context of quantum mechanics.
A quantum master equation in the Redfield form governing the reduced density matrix of the particle is derived explicitly from first principles.
Under some approximations, this equation reduces to the linear Boltzmann equation.
The issue of the positivity of the time evolution is also discussed by means of a Lindblad form.
The Born and Markov assumptions underlying these equations, as well as other approximations regarding the bath correlation function, are discussed in details.
Furthermore, all these master equations are shown to be equivalent with each other if the density matrix of the particle is diagonal in the momentum basis, or if the collision rate is independent of the particle momentum.
\end{abstract}%
\maketitle%

\section{Introduction}\label{sec:introduction}
How does a quantum particle propagate in a particle detector~?
This question is not obvious from the perspective of quantum mechanics because, in principle, it requires including the detector and all its constituents in the wave function of the system.
However, the many degrees of freedom of such a system makes any direct solution of the Schrödinger equation impossible in general.
A convenient way to address this issue is to resort to approximate evolution equations for the reduced density matrix of the particle, often referred to as quantum master equations~\cite{Gardiner2000, Breuer2002, Weiss2008}.
Famous examples include the Redfield equation and the Lindblad equation~\cite{Gardiner2000, Breuer2002, Weiss2008}.
The Redfield equation is a Markovian equation governing the evolution of systems weakly coupled to an environment.
It was historically developed in the context of nuclear magnetic resonance~\cite{Redfield1957, Redfield1965}.
However, the Redfield equation is known to violate the positivity of the reduced density matrix for certain initial conditions~\cite{Breuer2002, Weiss2008}.
This means that, in special circumstances, some eigenvalues of the density matrix, representing probabilities, could be negative or larger than one.
This issue is a very active topic in the literature~\cite{Pechukas1994, GaspardP1999a, Farina2019, Mozgunov2020, Davidovic2020, Hartmann2020}, and is typically addressed by resorting to a master equation in the Lindblad form~\cite{Gorini1976, Lindblad1976, Manzano2020}, at the cost of additional assumptions.
Indeed, as shown in the literature~\cite{Gorini1976, Lindblad1976}, the Lindblad equation is the most general Markovian master equation preserving the positivity, and even the complete positivity, of the reduced density matrix.
\par Regarding the propagation of a particle in a gaseous environment, a good candidate is the Boltzmann equation~\cite{Boltzmann1872, Weinberg1958, Balescu1975, Huang1987, Harris2004}.
This essential equation of non-equilibrium statistical mechanics is especially well suited to describe the kinetics of gases.
Attempts to generalize the nonlinear Boltzmann equation to quantum gases date back to the 1920s with the historical papers by Nordheim~\cite{Nordheim1928} and Uehling and Uhlenbeck~\cite{Uehling1933}.
These quantum Boltzmann equations include modifications to account for the Fermi-Dirac or the Bose-Einstein statistics of the quantum gas~\cite{Kadanoff1962}.
However, they are nonlinear in the distribution function, and actually even more nonlinear than the original Boltzmann equation.
They are beyond the scope of the present work, since the quantum statistics of gas molecules may generally be neglected in particle detectors.
\par In general, quantum master equations can be used to study decoherence in open systems.
Decoherence is defined as the decay over time of the off-diagonal elements of the reduced density matrix of the system due to the entanglement with a quantum environment~\cite{Gardiner2000, Breuer2002, Weiss2008, Schlosshauer2019}.
In the context of collisional decoherence~\cite{Hornberger2003a, Hornberger2003b, Hornberger2006a, Hornberger2008, Vacchini2009, Hornberger2009, Diosi1995, Diosi2009, Diosi2022, Halliwell2007, Kamleitner2010, Breuer2002}, entanglement is caused by the collisions between the particle of interest and the gas scatterers.
In practice, the resulting diagonalization of the density matrix leads to the decrease of the visibility of interference patterns~\cite{Hornberger2003a, Hornberger2009, Vacchini2009, Schlosshauer2019}.
Decoherence was originally proposed by Zeh and Joos as a key ingredient to understand quantum measurement~\cite{Zeh1970, Zeh1973, Joos1985, Zurek1991}.
Indeed, these processes share several common characteristics, especially the fact that they are both irreversible in most practical situations~\cite{Zurek1991, Kiefer1999, Allahverdyan2013, Schlosshauer2019, Breuer2002, Weiss2008}.
Collisional decoherence is now a well established theory which is successfully confirmed by experiments on matter-wave interferometry, even for very large molecules~\cite{Arndt1999, Arndt2009-arxiv, Hornberger2003a, Hackermuller2003, Hackermuller2004, Stibor2005, Sonnentag2007, Hasselbach2010, Nimmrichter2011b, Gerlich2011, Juffmann2013, Eibenberger2013, Fein2019, Brand2020, Schrinski2020}.
However, this theory has never been applied to the case of fast particles of a few MeVs which is considered in the present work.
\par The main purpose of this paper is to derive a quantum master equation for the reduced density matrix of a fast particle propagating in a gas at thermal equilibrium.
In particular, this equation is desired to be consistent in the Wigner representation with the linear Boltzmann equation.
To this end, the derivation will resort to the Redfield equation, and will assume that the interaction potential between the particle and the scatterers is invariant under spatial translations.
In the framework of the Lindblad equation, it is known that such an assumption provides constraints on the structure of the master equation~\cite{Holevo1993, Vacchini2001, Vacchini2005a}.
A similar structure is found in this paper but for the Redfield equation.
Furthermore, it turns out that the results of this paper are consistent with the master equations obtained in the literature on the quantum Brownian motion of a slow particle~\cite{Hornberger2003b, Hornberger2006a, Hornberger2008, Vacchini2009, Hornberger2009, Diosi1995, Diosi2009}.
This consistency suggests that the assumption of a fast particle does not play a significant role in the derivation. This is indeed the case, as shown in this paper.
The long-term goal of this work is to develop a fully quantum model for the propagation of ionizing fast particles, including alpha particles from typical radioactive sources, in a gaseous detector such as a cloud chamber or an ionization chamber~\cite{Segre1977, Sigmund2006, Sigmund2014}.
\par This paper is organized as follows.
Preliminary remarks and assumptions introducing the quantum system are presented in Sec.~\ref{sec:presentation}.
In particular, the momentum states are defined in Sec.~\ref{sec:momentum-states}, the full Hamiltonian of the system in Sec.~\ref{sec:hamiltonian}, and the thermal state of the gas in Sec.~\ref{sec:thermal-state}.
Section~\ref{sec:binary-collision} gives a reminder about binary collisions, especially the Fermi golden rule and the cross section.
Then, the derivation of the quantum master equations is presented in Sec.~\ref{sec:derivations}, and is based on the Redfield equation introduced in Sec.~\ref{sec:redfield}.
The expansion of the collision terms ultimately leads to the simplified Redfield equation in the end of Sec.~\ref{sec:collision-terms}, which is the main result of this paper.
The issue of the non-positivity of this equation is discussed in Sec.~\ref{sec:positivity} with respect to an approximate Lindblad form.
Finally, the simplified Redfield equation is shown to reduce to the linear Boltzmann equation in Sec.~\ref{sec:boltzmann}.
Conclusions are drawn in Sec.~\ref{sec:conclusions}.
\par Throughout this paper, SI units are used.
In particular, $h$ is the Planck constant, $\hbar=h/2\pi$ is the reduced Planck constant, and $\kbol$ is the Boltzmann constant.
All the calculations will be performed in arbitrary dimension $d\in\{1,2,3,\ldots\}$.
Quantum operators will be denoted as $\op{A}$ to distinguish them from the associated eigenvalue $A$.

\section{Presentation of the model}\label{sec:presentation}

\subsection{Momentum states}\label{sec:momentum-states}
First, one assumes that all the particles in the system are contained in the cubic region $\mathcal{V}$ of side $L$.
Consequently, the momentum eigenstates are defined for all $\vect{r}\in\mathcal{V}$ as
\begin{equation}\label{eq:def-momentum-state}
\braket{\vect{r}}{\vect{k}} = \frac{1}{\sqrt{V}}\E^{\I\vect{k}\cdot\vect{r}}  \:,
\end{equation}
where $V=L^d$ is the volume of the region $\mathcal{V}$.
If periodic boundary conditions are imposed on the wave function, then the momentum is quantized to the cubic Bravais lattice
\begin{equation}\label{eq:quantized-momentum}
\vect{k} = \frac{2\pi}{L}\tran{(n_1,n_2,\ldots,n_d)}  \:,
\end{equation}
where $(n_1,n_2,\ldots,n_d)\in\mathbb{Z}^d$.
Therefore, the orthogonality relation reads
\begin{equation}\label{eq:momentum-orthogonality}
\braket{\vect{k}}{\vect{k}'} = \int_{\mathcal{V}} \frac{1}{V} \E^{\I(\vect{k}'-\vect{k})\cdot\vect{r}} \D\vect{r} = \delta_{\vect{k}-\vect{k}'}  \:,
\end{equation}
where $\delta_{\vect{x}}$ denotes the single-argument Kronecker delta, which is equal to one if $\vect{x}=\vect{0}$, and zero otherwise.
Furthermore, according to Eq.~\eqref{eq:momentum-orthogonality}, the norm of the momentum states is just $\braket{\vect{k}}{\vect{k}}=1$.
The momentum states $\ket{\vect{k}}$ are thus dimensionless, as it should be for a properly normalized quantum state.
In the limit of infinite quantization volume ($V\rightarrow\infty$), the momentum spectrum becomes continuous and the orthogonality relation~\eqref{eq:momentum-orthogonality} can be approximately expressed in term of the Dirac delta
\begin{equation}\label{eq:momentum-orthogonality-continuous}
\braket{\vect{k}}{\vect{k}'} \simeq \frac{(2\pi)^d}{V} \delta(\vect{k}-\vect{k}')  \:.
\end{equation}
This is only approximate because the limit $V\rightarrow\infty$ cannot be rigorously taken at this stage.
Note that, in contrast to $\delta(\vect{x})$, the square of $\delta_{\vect{x}}$ is properly defined: $\delta_{\vect{x}}^2=\delta_{\vect{x}}$.
This is why the discrete basis~\eqref{eq:def-momentum-state} will be preferred for the calculations.
\par This way of normalizing the momentum eigenstates prevents possible infinities from occurring when evaluating integrals~\cite{Hornberger2003b, Hornberger2008}.
This is also physically motivated by the fact that plane waves are actually idealizations of wave packets with finite spatial extension, especially when the particles are confined in a sealed enclosure representing the gaseous detector.
The confinement is crucial to properly define the density of the gas $n=N/V$, but also to ensure that the collision rate with the incident particle is finite.
This is why the incident particle is also assumed to be contained in the region $\mathcal{V}$.
\par The resolution of identity resulting from the orthogonality relation~\eqref{eq:momentum-orthogonality} reads
\begin{equation}\label{eq:momentum-closure}
\op{1} = \sum_{\vect{k}} \ket{\vect{k}}\bra{\vect{k}} \simeq \frac{V}{(2\pi)^d} \int_{\mathbb{R}^d} \ket{\vect{k}}\bra{\vect{k}} \D\vect{k}  \:.
\end{equation}
The sum in Eq.~\eqref{eq:momentum-closure} implicitly runs over the wave vectors of the cubic Bravais lattice~\eqref{eq:quantized-momentum}.
Except otherwise mentioned, all the sums over the momentum in this paper will run over the cubic Bravais lattice.
Another corollary of Eq.~\eqref{eq:def-momentum-state} is the trace over the momentum states
\begin{equation}\label{eq:momentum-trace}
\Tr\op{A} = \sum_{\vect{k}} \bra{\vect{k}}\op{A}\ket{\vect{k}} \simeq \frac{V}{(2\pi)^d} \int_{\mathbb{R}^d} \bra{\vect{k}}\op{A}\ket{\vect{k}} \D\vect{k}  \:.
\end{equation}
More generally, any sum over the discrete momentum basis can be replaced in the continuum limit by an integral according to the rule:
\begin{equation}\label{eq:discrete-sum-to-continuum}
\sum_{\vect{k}} \rightarrow \int_{\mathbb{R}^d} \frac{V}{(2\pi)^d} \D\vect{k}  \:.
\end{equation}
However, each factor $V$ that will appear in this way will have to be compensated by a $V$ in the denominator in order to regularize the limit $V\rightarrow\infty$.

\subsection{Hamiltonian and assumptions}\label{sec:hamiltonian}
One considers a model for a spinless quantum particle of mass $m_\sys$ interacting with a gas composed of $N$ mobile scatterers of mass $m_\bath$.
In the nonrelativistic regime, the Hamiltonian of the whole system reads
\begin{equation}\label{eq:general-hamiltonian}
\op{H} = \underbrace{\frac{\op{\vect{p}}_\sys^2}{2m_\sys}}_{\op{H}_\sys} + \underbrace{\sum_{i=1}^N \frac{\op{\vect{p}}_i^2}{2m_\bath}}_{\op{H}_\bath} + \underbrace{\sum_{i=1}^N u(\op{\vect{r}} - \op{\vect{x}}_i)}_{\op{U}}  \:,
\end{equation}
where $(\op{\vect{r}},\op{\vect{p}}_\sys)$ are the position and the momentum of the particle, and $(\op{\vect{x}}_1,\op{\vect{x}}_2,\ldots,\op{\vect{x}}_N)$ and $(\op{\vect{p}}_1,\op{\vect{p}}_2,\ldots,\op{\vect{p}}_N)$ are the positions and the momenta of the scatterers, respectively.
The wave functions of both the particle and the scatterers are subject to the periodic boundary conditions discussed in Sec.~\ref{sec:momentum-states}.
This allows the use of the quantized momentum states $\ket{\vect{k}}$ defined in Eq.~\eqref{eq:def-momentum-state} and the subsequent formalism.
\par Furthermore, the potential $u(\vect{r})$ in Eq.~\eqref{eq:general-hamiltonian} is supposed to be a spherically symmetric bump function of range $R$ such that
\begin{equation}\label{eq:short-range-assumption}
R \ll \varsigma  \:,
\end{equation}
where $\varsigma$ (sigma) denotes the mean distance between nearest neighboring scatterers defined as
\begin{equation}\label{eq:mean-interscatterer-distance}
\varsigma = \left(\frac{V}{N}\right)^{\frac{1}{d}}  \:.
\end{equation}
Note that the short-range constraint of Eq.~\eqref{eq:short-range-assumption} excludes Coulomb interactions which nevertheless play an important role in the propagation of fast charged particles in matter~\cite{Segre1977, Sigmund2006, Sigmund2014}.
\par In addition, the mean free path $\lscat$ defined as~\cite{Akkermans2007, ShengP2006, Weiss2008, Segre1977, Sigmund2006, Sigmund2014, Harris2004, Weinberg1958, GaspardD2022a, GaspardD2022b}
\begin{equation}\label{eq:def-mean-free-path}
\lscat = \frac{1}{n\sigma}  \:,
\end{equation}
is assumed to be large enough compared to the particle wavelength:
\begin{equation}\label{eq:weak-scattering}
k_{\sys,0}\lscat \gg 1  \:.
\end{equation}
In Eq.~\eqref{eq:weak-scattering}, $k_{\sys,0}=2\pi/\lambda_{\sys,0}$ denotes the initial wavenumber of the particle.
The condition~\eqref{eq:weak-scattering} is generally referred to in the literature as the weak scattering regime~\cite{ShengP2006} or the weak disorder regime~\cite{Akkermans2007}.
This condition will be particularly important in Sec.~\ref{sec:collision-terms} and in Appendix~\ref{app:principal-values}. 
\par The density matrices of the entire system, of the gas, and of the particle will be denoted as $\op{\rho}$, $\op{\rho}_\bath$, and $\op{\rho}_\sys$, respectively.
In particular, the reduced density matrix of the incident particle is given by
\begin{equation}\label{eq:def-rho-system}
\op{\rho}_\sys(t) = \Tr_\bath \op{\rho}(t)  \:,
\end{equation}
where $\Tr_\bath$ denotes the partial trace over the states of the scatterers.
At every time $t$, the quantum state of the particle is completely described by the density matrix $\op{\rho}_\sys(t)$.
At the beginning ($t=0$), the particle is assumed to be in the pure momentum state $\op{\rho}_\sys(0)=\ket{\vect{k}_{\sys,0}}\bra{\vect{k}_{\sys,0}}$.
Under the effect of the collisions between the particle and the scatterers, the partial trace~\eqref{eq:def-rho-system} is expected to decrease the purity of the density matrix $\op{\rho}_\sys(t)$.
This process is known as the collisional decoherence~\cite{Hornberger2003a, Hornberger2003b, Hornberger2006a, Vacchini2009, Hornberger2009, Kamleitner2010}, and is the focus of this paper.
\par Furthermore, having in mind a gaseous particle detector, the particle of mass $m_\sys$ represents the incident ionizing radiation.
This particle is thus significantly more energetic than the scatterers at the beginning of the interaction.
In this way, the particle slows down under the effect of collisions until it reaches thermal equilibrium with the gas.
According to Eq.~\eqref{eq:general-hamiltonian}, the only way for the particle to loose its energy is through the recoil of the scatterers.
This recoil is supposed to approximate more realistic energy loss processes such as the excitation or the ionization of the gas molecules.
\par Finally, the Hamiltonian~\eqref{eq:general-hamiltonian} neglects the possible interactions between the scatterers of the gas themselves.
This assumption is reasonable for an ideal dilute gas.

\subsection{Thermal state of the gas}\label{sec:thermal-state}
Regarding the gas of scatterers, it is characterized by the thermal de Broglie wavelength $\lth$, and equivalently by the thermal wavenumber $\kth$.
They are respectively defined as~\cite{Huang1987, Cohen2020-vol3}
\begin{equation}\label{eq:thermal-wavenumber}
\lth = \frac{h}{\sqrt{2\pi m_\bath\kbol T}}  \quad\text{and}\quad
\kth = \frac{1}{\hbar}\sqrt{2\pi m_\bath\kbol T}  \:,
\end{equation}
where $T$ is the absolute temperature.
One assumes that the thermal wavelength is much smaller than the mean interscatterer distance
\begin{equation}\label{eq:classical-gas}
\lth \ll \varsigma  \:,
\end{equation}
so that the gas may be described at equilibrium by the classical Maxwell-Boltzmann statistics, instead of quantum statistics such as the Bose-Einstein or the Fermi-Dirac statistics.
Although quantum master equations may also be derived without this assumption~\cite{Kadanoff1962}, it is perfectly reasonable in the framework of gaseous particle detectors where condition~\eqref{eq:classical-gas} is generally fulfilled.
Therefore, one assumes that the gas is at thermal equilibrium and that its density matrix is given by
\begin{equation}\label{eq:bath-thermal-state}
\op{\rho}_\bath = \frac{1}{Z}\E^{-\beta\op{H}_\bath}  \:,
\end{equation}
where $Z=Z(\beta)$ is the partition function and $\beta=1/\kbol T$ is the inverse temperature.
Given the equilibrium assumption~\eqref{eq:bath-thermal-state}, the gas will often be referred to as the \emph{bath} in which the particle is immersed.
In fact, this assumption is not necessary for the development of the quantum master equations made in Sec.~\ref{sec:derivations}.
It mainly helps to interpret the partial trace~\eqref{eq:def-rho-system}, as one will see soon.
\par The partition function of Eq.~\eqref{eq:bath-thermal-state} is given by
\begin{equation}\label{eq:bath-partition-function}
Z = \Tr(\E^{-\beta\op{H}_\bath}) = \sum_{\vect{k}_1,\ldots,\vect{k}_N} \E^{-\frac{\hbar^2}{2m_\bath\kbol T}\sum_{i=1}^N \vect{k}_i^2}  \:.
\end{equation}
With the factor~\eqref{eq:bath-partition-function}, the density matrix~\eqref{eq:bath-thermal-state} is normalized according to $\Tr_\bath\op{\rho}_\bath=1$.
In principle, the overcount of indistinguishable quantum states under the exchange of particles should be corrected by the permutation factor $N!$.
However, this correction is here omitted because it is compensated nearly everywhere and thus has no consequence on the sought quantum master equations.
Using the continuum approximation~\eqref{eq:momentum-trace} of the sum over $\vect{k}_1,\ldots,\vect{k}_N$ and the thermal wavenumber~\eqref{eq:thermal-wavenumber}, the partition function~\eqref{eq:bath-partition-function} becomes
\begin{equation}\label{eq:bath-partition-function-result}
Z = \left(\frac{L}{2\pi} \int_{\mathbb{R}} \E^{-\pi k^2/\kth^2} \D k\right)^{Nd}
 = \left(\frac{L}{2\pi} \kth\right)^{Nd} = \left(\frac{V}{\lth^d}\right)^N  \:.
\end{equation}
Note that the ratio $V/\lth^d$ can be interpreted as the number of ways that a gas particle whose quantum state extends over the effective volume $\lth^d$ can occupy the volume $V$.
The power $N$ comes from the $N$ independent particles to be placed in the volume $V$, knowing that the particles are independent of each other.
As mentioned above, the permutation factor $N!$ has been omitted in Eq.~\eqref{eq:bath-partition-function-result}.
\par The density matrix~\eqref{eq:bath-thermal-state} has the particularity of being diagonal and factorizable in the momentum basis since $[\op{\rho}_\bath,\op{\vect{p}}_i]=0~\forall i$ but also $[\op{\vect{p}}_i,\op{\vect{p}}_j]=0~\forall i,j$.
Therefore, one can write
\begin{equation}\label{eq:bath-momentum-full-distrib}\begin{split}
\bra{\vect{k}_1,\ldots,\vect{k}_N}\op{\rho}_\bath\ket{\vect{k}_1,\ldots,\vect{k}_N} & = \rho_\bath(\vect{k}_1,\vect{k}_2,\ldots,\vect{k}_N)  \\
 & = \prod_{i=1}^N \rho_\bath(\vect{k}_i)  \:,
\end{split}\end{equation}
where $\rho_\bath(\vect{k}_\bath)$ denotes the Maxwell-Boltzmann distribution for a generic bath particle of momentum $\vect{k}_\bath$.
This distribution is given by
\begin{equation}\label{eq:bath-momentum-distrib}
\rho_\bath(\vect{k}_\bath) = \frac{\lth^d}{V} \E^{-\pi\vect{k}_\bath^2/\kth^2}  \:.
\end{equation}
The distribution~\eqref{eq:bath-momentum-distrib} is normalized according to
\begin{equation}\label{eq:bath-momentum-distrib-norm}
\sum_{\vect{k}_\bath} \rho_\bath(\vect{k}_\bath) = 1  \:.
\end{equation}
It is also useful to look at the position-basis representation of the density matrix~\eqref{eq:bath-thermal-state}.
In contrast to the momentum representation, the position representation is not diagonal, but can nevertheless be factorized as follows
\begin{equation}\label{eq:bath-position-full-matrix}
\bra{\vect{x}_1,\ldots,\vect{x}_N}\op{\rho}_\bath\ket{\dual{\vect{x}}_1,\ldots,\dual{\vect{x}}_N} = \prod_{i=1}^N \rho_\bath(\vect{x}_i,\dual{\vect{x}}_i)  \:.
\end{equation}
The single-particle density matrix, $\rho_\bath(\vect{x},\dual{\vect{x}})$, in Eq.~\eqref{eq:bath-position-full-matrix} is given by the Fourier transform
\begin{equation}\label{eq:bath-position-matrix}\begin{split}
\rho_\bath(\vect{x},\dual{\vect{x}}) & = \frac{1}{(2\pi)^d} \int_{\mathbb{R}^d} \rho_\bath(\vect{k}_\bath) \E^{\I\vect{k}_\bath\cdot(\vect{x}-\dual{\vect{x}})} \D\vect{k}_\bath  \\
 & = \frac{1}{V} \E^{-\frac{1}{4\pi}\kth^2(\vect{x}-\dual{\vect{x}})^2}  \:.
\end{split}\end{equation}
The density matrix~\eqref{eq:bath-position-matrix} is symmetric with respect to the matrix transpose $\vect{x}\leftrightarrow\dual{\vect{x}}$, and is equal to $1/V$ along the diagonal ($\vect{x}=\dual{\vect{x}}$).
It also quickly vanishes for large separation distance $\norm{\vect{x}-\dual{\vect{x}}}\gg\lth$.
The characteristic decay length is known as the \emph{coherence length} and can be defined as~\cite{Barnett2000, Born2019}
\begin{equation}\label{eq:def-coherence-length}
\lcoh(\dual{\vect{x}})^2 = \frac{\int_{\mathbb{R}^d} \norm{\vect{x}-\dual{\vect{x}}}^2 \abs{\rho_\bath(\vect{x},\dual{\vect{x}})}^2 \D\vect{x}}{\int_{\mathbb{R}^d} \abs{\rho_\bath(\vect{x},\dual{\vect{x}})}^2 \D\vect{x}}  \:.
\end{equation}
In the case of the thermal density matrix~\eqref{eq:bath-position-matrix}, one finds the constant value
\begin{equation}\label{eq:bath-coherence-length}
\lcoh^2 = \frac{d}{4\pi}\lth^2  \:.
\end{equation}
This shows that the coherence length of one of the gas particle is of the order of the thermal wavelength $\lth$.
Therefore, the quantum-wave nature of the gas particle is only meaningful for distances smaller than $\lth$.
This fact is also supported by the remark below Eq.~\eqref{eq:bath-partition-function-result} that the quantum state of the gas particle occupies the effective volume $\lth^d$ in the medium.
\par Last but not least, it is illuminating to calculate the partial trace of the potential $\textstyle\op{U}=\sum_{i=1}^N u(\op{\vect{r}}-\op{\vect{x}}_i)$ over the bath, in order to get a better understanding of the partial trace~\eqref{eq:def-rho-system}.
Using the facts that $\op{U}$ is diagonal in the position basis and that the density matrix can be factorized with Eq.\ \eqref{eq:bath-position-full-matrix}, one gets
\begin{equation}\label{eq:bath-average-potential-2}
\Tr_\bath\!\left(\op{\rho}_\bath \op{U}\right) = N \int_{\mathcal{V}} \rho_\bath(\vect{x},\vect{x}) u(\op{\vect{r}}-\vect{x}) \D\vect{x}  \:.
\end{equation}
In the case of a thermal state, the scatterer density is $\rho_\bath(\vect{x},\vect{x})=1/V$ according to Eq.~\eqref{eq:bath-position-matrix}.
Therefore, the average~\eqref{eq:bath-average-potential-2} is just
\begin{equation}\label{eq:bath-average-potential-3}
\Tr_\bath\!\left(\op{\rho}_\bath \op{U}\right) = \frac{N}{V} \int_{\mathcal{V}} u(\op{\vect{r}}-\vect{x}) \D\vect{x} = N\avg{u}  \:,
\end{equation}
which is practically independent of the position $\op{\vect{r}}$ of the particle for short-range potentials.
\par One notices that the integrals in Eqs.~\eqref{eq:bath-average-potential-2} and~\eqref{eq:bath-average-potential-3} can be interpreted as the average potential generated by the scatterers.
More generally, this means that the partial trace over the bath states essentially reduces to an average over the scatterer positions.
This is an important remark, because it shows that the density matrix of the particle of interest, which is given by the partial trace~\eqref{eq:def-rho-system}, is completely analogous to the average of the density matrix over the random configurations of the scatterers considered in the framework of the Lorentz gas model in Refs.~\cite{GaspardD2022a, GaspardD2022b}.
However, in contrast to those papers, the average is here physically motivated by the quantum uncertainty over the scatterer positions in the gas.
Indeed, from the physical point of view, the partial trace~\eqref{eq:def-rho-system} makes more sense than an abstract statistical average which does not necessarily represent the actual situation in a given random realization of the positions $(\vect{x}_1,\vect{x}_2,\ldots,\vect{x}_N)$.
\par It should be noted that this physical interpretation of the configurational average is only valid for gases in which the disorder is of dynamical origin.
This would not be the case for random impurities in solids at low temperature, for instance, because this kind of disorder is quenched due to the absence of free motion.
In this latter case, the configurational average is less physically motivated than in gases.

\subsection{Binary collision and cross section}\label{sec:binary-collision}
Before going to the derivation of the quantum master equations, let us take a closer look at the binary collision between the particle and a single scatterer.
To this end, the general Hamiltonian~\eqref{eq:general-hamiltonian} must be restricted to $N=1$
\begin{equation}\label{eq:binary-hamiltonian}
\op{H} = \frac{\op{\vect{p}}_\sys^2}{2m_\sys} + \frac{\op{\vect{p}}_\bath^2}{2m_\bath} + u(\op{\vect{r}} - \op{\vect{x}})  \:,
\end{equation}
where $(\op{\vect{r}}, \op{\vect{p}}_\sys)$ are the position and the momentum of the particle, and $(\op{\vect{x}}, \op{\vect{p}}_\bath)$ are the position and the momentum of the scatterer.
The effect of the potential term in Eq.~\eqref{eq:binary-hamiltonian} can be determined at the leading order of perturbation theory by treating $\op{H}_0=\op{H}_\sys+\op{H}_\bath$ as the unperturbed Hamiltonian and $\op{U}=u(\op{\vect{r}}-\op{\vect{x}})$ as the ideally small perturbation.
The result of this calculation is the well-known Fermi golden rule which yields the rate of the transition $\ket{\alpha}\rightarrow\ket{\beta}$ between eigenstates of the unperturbed Hamiltonian~\cite{Dirac1927, Fermi1950, Visser2009, Landau1967, Sakurai2020}
\begin{equation}\label{eq:fermi-golden-rule}
w(\beta\mid\alpha) = \frac{2\pi}{\hbar} \abs{\bra{\beta}\op{U}\ket{\alpha}}^2 \delta(E_\beta-E_\alpha)  \:.
\end{equation}
In Eq.~\eqref{eq:fermi-golden-rule}, $E_\alpha$ and $E_\beta$ are the energy eigenvalues associated with the eigenstates $\ket{\alpha}$ and $\ket{\beta}$, respectively.
One can write
\begin{equation}\label{eq:energy-eigenvalues}
\op{H}_0 \ket{\alpha} = E_\alpha \ket{\alpha}  \quad\text{and}\quad
\op{H}_0 \ket{\beta}  = E_\beta  \ket{\beta}   \:.
\end{equation}
At higher order of perturbation theory, the Fermi golden rule~\eqref{eq:fermi-golden-rule} still holds formally by replacing the potential $\op{U}$ by the transition operator $\op{T}(E_\alpha)$ defined by the Dyson series~\cite{Joachain1979, Newton1982, Taylor2006, ShengP2006, Akkermans2007}
\begin{equation}\label{eq:transition-dyson-series}
\op{T}(E) = \op{U} + \op{U}\op{G}_0(E)\op{U} + \op{U}\op{G}_0(E)\op{U}\op{G}_0(E)\op{U} + \cdots  \:,
\end{equation}
where $\op{G}_0(E)=(E-\op{H}_0)^{-1}$ is the Green operator associated with the unperturbed Hamiltonian.
If the states $\ket{\beta}$ constitute a quasi-continuum basis, such as the momentum basis, then the Dirac delta in Eq.~\eqref{eq:fermi-golden-rule} can be eliminated by integration over $\ket{\beta}$, as one will see soon.
\par When applied to the binary collision governed by the Hamiltonian~\eqref{eq:binary-hamiltonian}, the Fermi golden rule~\eqref{eq:fermi-golden-rule} leads to the differential cross section, which is a key ingredient of the master equations, especially of the Boltzmann equation.
Although it is a relatively standard result of scattering theory, this derivation is presented here because closely related calculations are invoked in Sec.~\ref{sec:derivations} for the many-scatterer Hamiltonian~\eqref{eq:general-hamiltonian}.
This follows from the fact that collisions involving different scatterers are independent.
According to Eq.~\eqref{eq:fermi-golden-rule}, the transition rate of the collision process $(\vect{k}_\sys,\vect{k}_\bath)\rightarrow(\vect{k}'_\sys,\vect{k}'_\bath)$ is given by
\begin{equation}\label{eq:binary-collision-rate-1}
w(\vect{k}'_\sys,\vect{k}'_\bath\mid\vect{k}_\sys,\vect{k}_\bath)
 = \frac{2\pi}{\hbar} \abs{\bra{\vect{k}'_\sys,\vect{k}'_\bath}\op{U}\ket{\vect{k}_\sys,\vect{k}_\bath}}^2 \delta(D)  \:,
\end{equation}
where $D$ is a compact notation for the energy difference
\begin{equation}\label{eq:def-energy-diff}
D = E_{\vect{k}'_\sys} + E_{\vect{k}'_\bath} - E_{\vect{k}_\sys} - E_{\vect{k}_\bath}  \:.
\end{equation}
The energies in Eq.~\eqref{eq:def-energy-diff} are related to the momenta by $E_{\vect{k}_\sys}=\tfrac{\hbar^2\vect{k}_\sys^2}{2m_\sys}$ and $E_{\vect{k}_\bath}=\tfrac{\hbar^2\vect{k}_\bath^2}{2m_\bath}$, and similarly for $E_{\vect{k}'_\sys}$ and $E_{\vect{k}'_\bath}$.
One considers separately the matrix element and the energy-conservation Dirac delta in Eq.~\eqref{eq:binary-collision-rate-1}.
First, to calculate the matrix element of the potential in Eq.~\eqref{eq:binary-collision-rate-1}, one uses the Fourier expansion of the potential
\begin{equation}\label{eq:single-potential-from-fourier}
\op{U} = u(\op{\vect{r}} - \op{\vect{x}}) = \frac{1}{V} \sum_{\vect{q}} \fourier{u}(\vect{q}) \E^{\I\vect{q}\cdot(\op{\vect{r}} - \op{\vect{x}})}  \:,
\end{equation}
where $\fourier{u}(\vect{q})$ is defined as
\begin{equation}\label{eq:fourier-single-potential}
\fourier{u}(\vect{q}) = \int_{\mathcal{V}} u(\vect{r}) \E^{-\I\vect{q}\cdot\vect{r}} \D\vect{r}  \:.
\end{equation}
Note that the Fourier decomposition~\eqref{eq:single-potential-from-fourier} is discrete due to the finite quantization volume $V$.
From Eq.~\eqref{eq:single-potential-from-fourier}, the matrix element of $\op{U}$ in the momentum basis can be evaluated using the fundamental property
\begin{equation}\label{eq:momentum-translation}
\E^{\I\vect{q}\cdot\op{\vect{r}}} \ket{\vect{k}_\sys} = \ket{\vect{k}_\sys + \vect{q}}  \:.
\end{equation}
This property is derived by projecting both sides of Eq.~\eqref{eq:momentum-translation} onto the position basis, and simply means that $\E^{\I\vect{q}\cdot\op{\vect{r}}}$ adds up the momentum $\vect{q}$ to the particle momentum.
Of course, a similar operator, $\E^{\I\vect{q}\cdot\op{\vect{x}}}$, also exists for the scatterer.
The momentum translation operator $\E^{\I\vect{q}\cdot\op{\vect{r}}}$ defined in Eq.~\eqref{eq:momentum-translation} will play a key role in Sec.~\ref{sec:derivations}.
Using Eqs.~\eqref{eq:single-potential-from-fourier} and~\eqref{eq:momentum-translation}, one gets
\begin{equation}\label{eq:potential-matrix-elem}
\bra{\vect{k}'_\sys,\vect{k}'_\bath}\op{U}\ket{\vect{k}_\sys,\vect{k}_\bath}
 = \frac{1}{V} \fourier{u}(\vect{k}'_\sys-\vect{k}_\sys) \delta_{\vect{k}'_\sys + \vect{k}'_\bath - \vect{k}_\sys - \vect{k}_\bath}  \:.
\end{equation}
In Eq.~\eqref{eq:potential-matrix-elem}, it is clear that the Kronecker delta expresses the conservation of the total momentum and is due to the translational invariance of the potential $\op{U}$.
Eliminating one of the momenta with the delta, the transition rate~\eqref{eq:binary-collision-rate-1} becomes
\begin{equation}\label{eq:binary-collision-rate-2}
w_{\vect{q}}(\vect{k}_\sys,\vect{k}_\bath) = \frac{2\pi}{\hbar} \frac{1}{V^2} \abs{\fourier{u}(\vect{q})}^2 \delta(D_{\vect{q}})  \:,
\end{equation}
with
\begin{equation}\label{eq:energy-diff-in-q-1}
D_{\vect{q}} = E_{\vect{k}_\sys + \vect{q}} + E_{\vect{k}_\bath - \vect{q}} - E_{\vect{k}_\sys} - E_{\vect{k}_\bath}  \:.
\end{equation}
In Eq.~\eqref{eq:binary-collision-rate-2}, the notation $\vect{q}$ stands for the momentum transferred to particle ``$\sys$'' by the collision with ``$\bath$''.
The energy difference can also be expressed in the center-of-mass frame as
\begin{equation}\label{eq:energy-diff-in-q-2}
D_{\vect{q}} = \frac{\hbar^2}{2m} \left[ (\vect{k} + \vect{q})^2 - \vect{k}^2 \right]  \:,
\end{equation}
where
\begin{equation}\label{eq:def-reduced-mass}
m = \frac{m_\sys m_\bath}{m_\sys + m_\bath} 
\end{equation}
is the reduced mass of the binary system, and
\begin{equation}\label{eq:def-relative-momentum}
\vect{k} = \frac{m_\bath\vect{k}_\sys - m_\sys\vect{k}_\bath}{m_\sys + m_\bath} 
\end{equation}
is the relative momentum between the colliding particles.
This momentum $\vect{k}$ also represents the momentum of particle ``$\sys$'' in the center-of-mass frame.
Since $D_{\vect{q}}=0$ due to energy conservation, Eq.~\eqref{eq:energy-diff-in-q-2} implies that the transferred momentum $\vect{q}$ is constrained to a sphere of center $-\vect{k}$ and of radius $k=\norm{\vect{k}}$.
Accordingly, the final relative momentum $\vect{k}'=\vect{k}+\vect{q}$ is constrained to a sphere centered at the origin and of radius $k$, and can thus be written as $\vect{k}'=k\vect{\Omega}$ with $\norm{\vect{\Omega}}=1$.
This means that, in the center-of-mass frame, the particle momentum changes only in direction but not in magnitude.
\par To get rid of the Dirac delta in Eq.~\eqref{eq:binary-collision-rate-2}, one considers that the containment volume $V$ is so large that the momentum spectrum in Eq.~\eqref{eq:quantized-momentum} is quasi-continuous.
The collision rate then becomes a differential element defined on the continuum as
\begin{equation}\label{eq:diff-collision-rate}
\D w(\vect{k}'\mid\vect{k}) = \frac{V}{(2\pi)^d} w(\vect{k}'\mid\vect{k}) \D\vect{k}'  \:,
\end{equation}
where the differential element $\D\vect{k}'$ represents the volume $(2\pi)^d/V$ occupied by the final momentum state, and $w(\vect{k}'\mid\vect{k})$ is given by
\begin{equation}\label{eq:binary-collision-rate-3}
w(\vect{k}'\mid\vect{k}) = \frac{2\pi}{\hbar} \frac{1}{V^2} \abs{\fourier{u}(\vect{k}'-\vect{k})}^2 \delta(E_{\vect{k}'} - E_{\vect{k}})  \:,
\end{equation}
with $E_{\vect{k}}=\tfrac{\hbar^2\vect{k}^2}{2m}$ and similarly for $E_{\vect{k}'}$.
In this way, the rate~\eqref{eq:binary-collision-rate-3} is equal to $w_{\vect{q}}(\vect{k}_\sys,\vect{k}_\bath)$ in Eq.~\eqref{eq:binary-collision-rate-2}.
Expressing the volume element in spherical coordinates with $\D\vect{k}'={k'}^{d-1}\D k'\D\Omega$, dividing each side by $\D\Omega$, and integrating over $k'$, one gets the angular collision rate
\begin{equation}\label{eq:angular-collision-rate}
\der{w}{\Omega}(\vect{\Omega}\mid\vect{k}) = \frac{V}{(2\pi)^d} \int_0^\infty w(k'\vect{\Omega}\mid\vect{k}) {k'}^{d-1}\D k'  \:.
\end{equation}
Furthermore, one introduces the differential cross section which is defined as the ratio between the differential collision rate and the magnitude of the relative flux $\vect{J}=\tfrac{1}{V}(\vect{v}_\sys-\vect{v}_\bath)$ of incident particles before the collision~\cite{Joachain1979, Newton1982}
\begin{equation}\label{eq:diff-cross-sec-from-diff-rate}
\D\sigma(\vect{\Omega}\mid\vect{k}) = \frac{\D w(\vect{\Omega}\mid\vect{k})}{\frac{1}{V} \norm{\vect{v}_\sys-\vect{v}_\bath}}  \:,
\end{equation}
where the velocities are related to the momenta by $\vect{v}_\sys=\tfrac{\hbar\vect{k}_\sys}{m_\sys}$ and $\vect{v}_\bath=\tfrac{\hbar\vect{k}_\bath}{m_\bath}$.
One can also use the fact that the relative velocity between the particle and the scatterer is related to the relative momentum by
\begin{equation}\label{eq:relative-velocity}
\vect{v}_\sys - \vect{v}_\bath = \frac{\hbar\vect{k}}{m}  \:,
\end{equation}
where $m$ is the reduced mass defined in Eq.~\eqref{eq:def-reduced-mass}. One gets from Eq.~\eqref{eq:angular-collision-rate}
\begin{equation}\label{eq:cross-section-from-rate}
\der{\sigma}{\Omega}(\vect{\Omega}\mid\vect{k}) = \frac{V^2}{(2\pi)^d} \frac{m}{\hbar k} \int_0^\infty w(k'\vect{\Omega}\mid\vect{k}) {k'}^{d-1}\D k' \:.
\end{equation}
Therefore, evaluating the integral~\eqref{eq:cross-section-from-rate} with the collision rate~\eqref{eq:binary-collision-rate-3} leads to the differential cross section
\begin{equation}\label{eq:cross-section-from-potential}
\der{\sigma}{\Omega}(\vect{\Omega}\mid\vect{k}) = \frac{\pi}{2} \frac{k^{d-3}}{(2\pi)^d} \abs{\frac{2m}{\hbar^2} \fourier{u}(k\vect{\Omega}-\vect{k})}^2  \:.
\end{equation}
In dimension three ($d=3$), Eq.~\eqref{eq:cross-section-from-potential} reduces to the known expression for the cross section at the leading order of perturbation theory~\cite{Joachain1979, Newton1982, Taylor2006}.
Finally, from Eq.~\eqref{eq:cross-section-from-potential}, one can determine the total cross section in the standard way~\cite{Joachain1979, Newton1982, Taylor2006}:
\begin{equation}\label{eq:def-total-cross-section}
\sigma(k) = \oint_{\mathcal{S}_d} \der{\sigma}{\Omega}(\vect{\Omega}\mid\vect{k}) \D\Omega  \:,
\end{equation}
where $\mathcal{S}_d$ represents the unit sphere in the space $\mathbb{R}^d$.

\section{Derivations of master equations}\label{sec:derivations}
In this section, the derivations of several quantum master equations for the density matrix of particle ``$\sys$'' are presented as well as the relations between them.
Assumptions focus on the case of an incident particle faster than the scatterers of the gas.
\par The derivation proceeds in four steps: first, one derives the Redfield equation, which is a general Markovian master equation obtained at next-to-leading order of perturbation theory~\cite{Redfield1957, Redfield1965, Breuer2002, Weiss2008, Hornberger2009}.
To this end, the procedure of Ref.~\cite{GaspardP1999a} is followed.
Second, one exploits the Fourier expansion~\eqref{eq:single-potential-from-fourier} to expand the collision terms in the formalism of quantum operators and to highlight the bath correlation function.
Under the assumption of weak scattering regime, this step leads to a simplified Redfield equation.
Third, a reduction of the Redfield equation to the Lindblad form is presented and discussed.
Finally, the Wigner transform is applied to restore the spatial dependence of the master equation and reveal a linear Boltzmann equation.
These steps turn out particularly useful to stress the underlying assumptions behind the master equations.

\subsection{Redfield equation}\label{sec:redfield}
The starting point is the quantum Liouville equation governing the time evolution of the density matrix of the full system in the Schrödinger picture
\begin{equation}\label{eq:general-liouville}
\pder{\op{\rho}}{t} = \supop{L}\op{\rho}(t)  \:,
\end{equation}
where $\supop{L}$ is the Liouvillian superoperator~\cite{Breuer2002, GaspardP1999a, Manzano2020} defined as
\begin{equation}\label{eq:def-liouville-supop}
\supop{L}\op{X} = \tfrac{1}{\I\hbar} [\op{H}, \op{X}]  \:,
\end{equation}
where $\op{H}$ is the general Hamiltonian~\eqref{eq:general-hamiltonian} and $\op{X}$ means any operator, but especially the density matrix $\op{\rho}(t)$.
Analogously, one defines the following Liouvillians
\begin{equation}\label{eq:all-liouville-supop}
\supop{L}_\sys \op{X} = \tfrac{1}{\I\hbar} [\op{H}_\sys,  \op{X}]  \;,\quad%
\supop{L}_\bath\op{X} = \tfrac{1}{\I\hbar} [\op{H}_\bath, \op{X}]  \:,\quad%
\supop{L}_0 = \supop{L}_\sys + \supop{L}_\bath  \:,
\end{equation}
and the potential superoperator
\begin{equation}\label{eq:potential-supop}
\supop{L}_{\rm U}\op{X} = \tfrac{1}{\I\hbar} [\op{U}, \op{X}]  \:.
\end{equation}
In Eq.~\eqref{eq:general-liouville}, the Liouvillian superoperator $\supop{L}$ can also be decomposed into a free part and an interaction part
\begin{equation}\label{eq:general-liouville-decomposition}
\pder{\op{\rho}}{t} = \left(\supop{L}_0 + \supop{L}_{\rm U}\right)\op{\rho}(t)  \:.
\end{equation}
Treating $\supop{L}_{\rm U}$ as a perturbation of $\supop{L}_0$, one introduces the interaction-picture density matrix $\op{\rho}_\ipic(t)$ as
\begin{equation}\label{eq:inter-picture-rho}
\op{\rho}(t) = \E^{\supop{L}_0t} \op{\rho}_\ipic(t)  \:.
\end{equation}
Inserting Eq.~\eqref{eq:inter-picture-rho} into the Liouville equation~\eqref{eq:general-liouville-decomposition} gives us
\begin{equation}\label{eq:liouville-inter-picture}
\pder{\op{\rho}_\ipic}{t} = \supop{L}_\ipic(t) \op{\rho}_\ipic(t)  \:,
\end{equation}
where the interaction Liouvillian is given by
\begin{equation}\label{eq:def-interaction-supop}
\supop{L}_\ipic(t) = \E^{-\supop{L}_0t} \supop{L}_{\rm U} \E^{\supop{L}_0t}  \:.
\end{equation}
Note that $\supop{L}_\ipic(t)$ explicitly depends on time.
The superoperator $\supop{L}_\ipic(t)$ in Eq.~\eqref{eq:def-interaction-supop} can be expressed more directly in terms of a single commutator.
Indeed, one has
\begin{equation}\label{eq:interaction-supop-from-u}
\supop{L}_\ipic(t)\op{X} = \tfrac{1}{\I\hbar}\left[\op{U}_\ipic(t), \op{X}\right]  \:,
\end{equation}
where $U_\ipic(t)$ denotes the interaction-picture potential defined as
\begin{equation}\label{eq:def-inter-pic-potential}
U_\ipic(t) = \E^{+\frac{\I}{\hbar}\op{H}_0t} \op{U} \E^{-\frac{\I}{\hbar}\op{H}_0t}  \:.
\end{equation}
The Liouville equation~\eqref{eq:liouville-inter-picture} can be integrated in time to get
\begin{equation}\label{eq:rho-inter-pic-integrated}
\op{\rho}_\ipic(t) = \op{\rho}(0) + \int_0^t\D t' \supop{L}_\ipic(t')\op{\rho}_\ipic(t')  \:,
\end{equation}
where $\op{\rho}(0)=\op{\rho}_\ipic(0)$ is the initial condition at $t=0$.
Substituting Eq.~\eqref{eq:rho-inter-pic-integrated} back into the right-hand side of the Liouville equation~\eqref{eq:liouville-inter-picture} leads to
\begin{equation}\label{eq:liouville-inter-pic-integrated}
\pder{\op{\rho}_\ipic}{t} = \supop{L}_\ipic(t) \op{\rho}(0) + \int_0^t\D t' \supop{L}_\ipic(t)\supop{L}_\ipic(t')\op{\rho}_\ipic(t')  \:.
\end{equation}
Note that Eq.~\eqref{eq:liouville-inter-pic-integrated} is still exact as it does not rely on a perturbative approximation.
\par If the particle is supposed to be independent of the environment at the beginning, then the initial state factorizes as
\begin{equation}\label{eq:redfield-initial-state}
\op{\rho}(0) = \op{\rho}_\sys(0)\otimes\op{\rho}_\bath  \:,
\end{equation}
where the initial bath state $\op{\rho}_\bath$ is taken to be the thermal equilibrium state~\eqref{eq:bath-thermal-state}.
The outer product symbol ``$\otimes$'' will be omitted in the following calculations.
In principle, the factorization property~\eqref{eq:redfield-initial-state} cannot be preserved at all time for $\op{\rho}(t)$ because of the quick entanglement with the scatterers due to the collisions.
At later times ($t>0$), the density matrix of the particle should be given by the partial trace over the bath
\begin{equation}\label{eq:def-rho-system-inter-pic}
\op{\rho}_{\sys,\ipic}(t) = \Tr_\bath \op{\rho}_\ipic(t)  \:.
\end{equation}
Note that this definition applies to the interaction picture, but is also consistent with definition~\eqref{eq:def-rho-system} in the standard picture.
Under the partial trace, Eq.~\eqref{eq:liouville-inter-pic-integrated} becomes
\begin{equation}\label{eq:liouville-series-trace}
\pder{\op{\rho}_{\sys,\ipic}}{t} = \Tr_\bath\!\big(\supop{L}_\ipic(t)\op{\rho}(0)\big) 
 + \int_0^t \D t' \Tr_\bath\!\big(\supop{L}_\ipic(t)\supop{L}_\ipic(t')\op{\rho}_\ipic(t')\big)  \:.
\end{equation}
Let us consider the first term in the right-hand side of Eq.~\eqref{eq:liouville-series-trace}.
According to Eq.~\eqref{eq:def-interaction-supop}, this term reads
\begin{equation}\label{eq:first-order-term-1}
\Tr_\bath\!\big(\supop{L}_\ipic(t)\op{\rho}(0)\big) = \Tr_\bath\!\left(\E^{-\supop{L}_0t} \supop{L}_{\rm U} \E^{\supop{L}_0t} \op{\rho}_\sys(0)\op{\rho}_\bath\right)  \:.
\end{equation}
Expression~\eqref{eq:first-order-term-1} can be simplified using several properties.
The first one is
\begin{equation}\label{eq:supop-property-1}
\E^{\supop{L}_0t} = \E^{\supop{L}_\sys t} \E^{\supop{L}_\bath t}  \:,
\end{equation}
and comes from the commutation relation $[\op{H}_\sys,\op{H}_\bath]=0$.
Note that the quantities related to particle ``$\sys$'', such as $\E^{\supop{L}_\sys t}$ and $\op{\rho}_\sys$, can get out of the partial trace $\Tr_\bath$.
The second one is the time invariance of the thermal equilibrium state
\begin{equation}\label{eq:supop-property-2}
\E^{\supop{L}_\bath t}\op{\rho}_\bath = \op{\rho}_\bath  \:.
\end{equation}
More generally, property~\eqref{eq:supop-property-2} also applies to any stationary state of the form $\op{\rho}_\bath=f(\op{H}_\bath)$, which is not necessary an equilibrium state.
The third one is due to the cyclic property of the trace, and the unitarity of the mapping $\E^{\supop{L}_\bath t}\op{X}=\E^{-\frac{\I}{\hbar}\op{H}_\bath t}\op{X}\E^{+\frac{\I}{\hbar}\op{H}_\bath t}$. Whatever the operator $\op{X}$, it reads
\begin{equation}\label{eq:supop-property-3}
\Tr_\bath\!\left(\E^{\supop{L}_\bath t}\op{X}\right) = \Tr_\bath\op{X}  \:.
\end{equation}
Note that the cyclic property of $\Tr_\bath$ only concerns the operators associated with the bath.
Using Eqs.~\eqref{eq:supop-property-1}--\eqref{eq:supop-property-3}, the first-order term~\eqref{eq:first-order-term-1} reduces to
\begin{equation}\label{eq:first-order-term-2}
\Tr_\bath\!\big(\supop{L}_\ipic(t)\op{\rho}(0)\big) = \E^{-\supop{L}_\sys t} \Tr_\bath\!\big(\supop{L}_{\rm U} \op{\rho}_\bath\big) \E^{\supop{L}_\sys t}\op{\rho}_\sys(0)  \:.
\end{equation}
Expression~\eqref{eq:first-order-term-2} can be rewritten more explicitly using Eq.~\eqref{eq:potential-supop}. One gets
\begin{equation}\label{eq:first-order-term-3}
\Tr_\bath\!\big(\supop{L}_\ipic(t)\op{\rho}(0)\big) = \E^{-\supop{L}_\sys t} \frac{1}{\I\hbar} \left[\Tr_\bath\!\left(\op{\rho}_\bath\op{U}\right), \E^{\supop{L}_\sys t}\op{\rho}_\sys(0)\right]  \:.
\end{equation}
According to Eq.~\eqref{eq:bath-average-potential-3}, the term $\Tr_\bath(\op{\rho}_\bath\op{U})$ is equal to the average potential $N\tavg{u}$.
Since the medium is uniform and subject to periodic boundary conditions, the average potential $\tavg{u}$ is a constant independent from the position.
Therefore, the commutator in Eq.~\eqref{eq:first-order-term-3} identically vanishes:
\begin{equation}\label{eq:first-order-term-final}
\Tr_\bath\!\big(\supop{L}_\ipic(t)\op{\rho}(0)\big) = 0  \:.
\end{equation}
In other words, the first-order term in Eq.~\eqref{eq:liouville-series-trace} does not contribute to the time evolution of the density matrix.
It only changes the zero energy reference, but without affecting the equation of motion.
From Eq.~\eqref{eq:liouville-series-trace}, the relevant equation is thus
\begin{equation}\label{eq:second-order-term-1}
\pder{\op{\rho}_{\sys,\ipic}}{t} = \int_0^t \D t' \Tr_\bath\!\big(\supop{L}_\ipic(t)\supop{L}_\ipic(t')\op{\rho}_\ipic(t')\big)  \:.
\end{equation}
Using Eqs.~\eqref{eq:def-interaction-supop}, \eqref{eq:supop-property-1} and~\eqref{eq:supop-property-3}, the integral term in Eq.~\eqref{eq:second-order-term-1} reads
\begin{equation}\label{eq:second-order-term-2}
\pder{\op{\rho}_{\sys,\ipic}}{t} = \E^{-\supop{L}_\sys t} \int_0^t\D t' \Tr_\bath\!\left(\supop{L}_{\rm U} \E^{\supop{L}_0(t-t')} \supop{L}_{\rm U} \E^{\supop{L}_0t'} \op{\rho}_\ipic(t')\right)  \:.
\end{equation}
Now, one considers a first approximation to close Eq.~\eqref{eq:second-order-term-2} for $\op{\rho}_\sys(t)$.
The density matrix in the right-hand side of Eq.~\eqref{eq:second-order-term-2} is assumed to be reasonably approached by
\begin{equation}\label{eq:born-approx-rho}
\op{\rho}_\ipic(t') = \op{\rho}_\ipic(t) + \bigo(\op{U})  \:,
\end{equation}
for all time $t'\in[0,t]$.
Note that the approximation~\eqref{eq:born-approx-rho} can be understood as a perturbative approximation at zeroth order of $\op{U}$.
The fact that $\op{\rho}_\ipic(t')$ is replaced by $\op{\rho}_\ipic(t)$ in this approximation, instead of $\op{\rho}_\ipic(0)$ for instance, is motivated by the fast expected decay in $t'$ of the integral~\eqref{eq:second-order-term-2} around the current time $t$.
This expectation is closely related to the Markov assumption which is further discussed in Sec.~\ref{sec:collision-terms} for a fast incident particle.
Using the fact that $\op{\rho}_\ipic(t)=\E^{-\supop{L}_0t}\op{\rho}(t)$, according to Eq.~\eqref{eq:inter-picture-rho}, and considering the change of integration variable $\tau=t-t'$~\cite{GaspardP1999a}, one can write from Eq.~\eqref{eq:second-order-term-2}
\begin{equation}\label{eq:second-order-term-3}\begin{split}
\pder{\op{\rho}_{\sys,\ipic}}{t} & = \E^{-\supop{L}_\sys t} \int_0^t\D\tau \Tr_\bath\!\left(\supop{L}_{\rm U} \E^{\supop{L}_0\tau} \supop{L}_{\rm U} \E^{-\supop{L}_0\tau} \op{\rho}(t)\right)  \\
 & + \bigo(\op{U}^3)  \:.
\end{split}\end{equation}
The density matrix $\op{\rho}(t)$ in the integral of Eq.~\eqref{eq:second-order-term-3} can be further approximated at zeroth order of $\op{U}$ with
\begin{equation}\label{eq:born-approx-rho-init}
\op{\rho}(t) = \op{\rho}_\sys(t)\op{\rho}_\bath + \bigo(\op{U})  \:,
\end{equation}
where $\op{\rho}_\bath$ is the thermal equilibrium state~\eqref{eq:bath-thermal-state}.
This expression derives from the initial condition~\eqref{eq:redfield-initial-state} and Eq.~\eqref{eq:born-approx-rho}, but it does not mean that the particle and the bath can be factorized at any time because the correction term $\bigo(\op{U})$ always couples the two subsystems.
Inserting Eq.~\eqref{eq:born-approx-rho-init} into Eq.~\eqref{eq:second-order-term-3} yields
\begin{equation}\label{eq:second-order-term-4}
\pder{\op{\rho}_{\sys,\ipic}}{t} = \E^{-\supop{L}_\sys t} \int_0^t\D\tau \supop{K}(\tau) \op{\rho}_\sys(t) + \bigo(\op{U}^3)  \:,
\end{equation}
where $\supop{K}(\tau)$ is the correlation superoperator, which acts only on $\op{\rho}_\sys(t)$ and is defined as~\cite{GaspardP1999a}
\begin{equation}\label{eq:redfield-kernel}
\supop{K}(\tau) = \Tr_\bath\!\left(\supop{L}_{\rm U} \E^{\supop{L}_0\tau} \supop{L}_{\rm U} \E^{-\supop{L}_0\tau} \op{\rho}_\bath\right)  \:.
\end{equation}
\par Finally, the derivative in the left-hand side of Eq.~\eqref{eq:second-order-term-3} can also be related to the Schrödinger-picture state at time $t$ according to
\begin{equation}\label{eq:rho-system-from-inter-pic}
\op{\rho}_\sys(t) = \Tr_\bath\op{\rho}(t) = \Tr_\bath\!\left(\E^{\supop{L}_0t} \op{\rho}_\ipic(t)\right) = \E^{\supop{L}_\sys t} \op{\rho}_{\sys,\ipic}(t)  \:.
\end{equation}
This series of equalities comes from Eqs.~\eqref{eq:def-rho-system}, \eqref{eq:inter-picture-rho}, \eqref{eq:supop-property-1}, \eqref{eq:supop-property-3} and~\eqref{eq:def-rho-system-inter-pic} in that order.
According to Eq.~\eqref{eq:rho-system-from-inter-pic}, one has
\begin{equation}\label{eq:time-derivative-inter-pic}
\pder{\op{\rho}_{\sys,\ipic}}{t} = \E^{-\supop{L}_\sys t} \left(\pder{\op{\rho}_\sys}{t} - \supop{L}_\sys\op{\rho}_\sys(t)\right)  \:.
\end{equation}
Substituting Eq.~\eqref{eq:time-derivative-inter-pic} into Eq.~\eqref{eq:second-order-term-4}, one finds~\cite{Redfield1957, Redfield1965, GaspardP1999a, Hornberger2009, Breuer2002, Weiss2008}
\begin{equation}\label{eq:redfield-equation-varying}
\pder{\op{\rho}_\sys}{t} = \supop{L}_\sys\op{\rho}_\sys(t) + \int_0^t \D\tau \supop{K}(\tau) \op{\rho}_\sys(t) + \bigo(\op{U}^3)  \:.
\end{equation}
Note that Eq.~\eqref{eq:redfield-equation-varying} is based solely on the perturbative approximation, but not yet on the Markovian approximation.
Furthermore, if the particle dynamics is considered on a much longer time than the characteristic decay time of $\supop{K}(\tau)$, known as the bath correlation time, then it is justified to take the limit $\textstyle\int_0^t\rightarrow\int_0^\infty$ in Eq.~\eqref{eq:redfield-equation-varying} \cite{Redfield1957, Redfield1965, Breuer2002, Weiss2008, GaspardP1999a, Manzano2020, Mozgunov2020, Davidovic2020, Hartmann2020}.
Therefore, the particle dynamics will be resolved only on a time scale much longer than the bath correlation time.
In this way, one obtains the \emph{Redfield equation}~\cite{Redfield1957, Redfield1965, Breuer2002, Weiss2008}
\begin{equation}\label{eq:redfield-equation-final}
\pder{\op{\rho}_\sys}{t} = \supop{L}_\sys\op{\rho}_\sys(t) + \int_0^\infty \D\tau \supop{K}(\tau) \op{\rho}_\sys(t) + \bigo(\op{U}^3)  \:.
\end{equation}
This equation is Markovian, in contrast to Eq.~\eqref{eq:redfield-equation-varying}.
\par Regarding the other fundamental properties of Eq.\ \eqref{eq:redfield-equation-final}, one can check that it preserves the trace of the density matrix through $\partial_t\Tr_\sys\op{\rho}_\sys=0$, as a consequence of $\Tr_\sys\supop{L}_\sys\op{\rho}_\sys=0$ and $\Tr\supop{L}_{\rm U}\op{X}=0$.
Therefore, the Redfield equation conserves the total probability
\begin{equation}\label{eq:redfield-total-proba}
\Tr_\sys\op{\rho}_\sys(t) = 1 \qquad\forall t>0  \:.
\end{equation}
Moreover, it also preserves the Hermiticity of the density matrix: $\herm{\op{\rho}}_\sys(t)=\op{\rho}_\sys(t)$.
However, it is not guaranteed to preserve the positivity of $\op{\rho}_\sys(t)$~\cite{Breuer2002, Weiss2008}.

\subsection{Collision terms}\label{sec:collision-terms}
In this subsection, one expands the collision terms of the Redfield equation~\eqref{eq:redfield-equation-final}.
According to Eqs.~\eqref{eq:potential-supop} and~\eqref{eq:def-interaction-supop}, this expression is actually a double commutator with the potential
\begin{equation}\label{eq:redfield-expansion-1}
\pder{\op{\rho}_\sys}{t} = \supop{L}_\sys\op{\rho}_\sys + \frac{1}{(\I\hbar)^2} \int_0^\infty \D\tau \Tr_\bath\left[ \op{U}, \left[ \op{U}_\ipic(-\tau), \op{\rho}_\sys \op{\rho}_\bath \right] \right]  \:,
\end{equation}
where $\op{U}_\ipic(t)$ is the interaction-picture potential defined in Eq.~\eqref{eq:def-inter-pic-potential}.
Even more explicitly, the double commutator in Eq.~\eqref{eq:redfield-expansion-1} contains four terms which can be compactly written as
\begin{equation}\label{eq:redfield-expansion-2}
\pder{\op{\rho}_\sys}{t} = \supop{L}_\sys\op{\rho}_\sys + \op{C}_{\rm G} + \herm{\op{C}}_{\rm G} - \op{C}_{\rm L} - \herm{\op{C}}_{\rm L}  \:,
\end{equation}
with the terms
\begin{equation}\label{eq:def-gain-term}
\op{C}_{\rm G} = \frac{1}{\hbar^2} \int_0^\infty \D\tau \Tr_\bath\!\left( \op{U}_\ipic(-\tau) \op{\rho}_\sys \op{\rho}_\bath \op{U} \right)  \:,
\end{equation}
and
\begin{equation}\label{eq:def-loss-term}
\op{C}_{\rm L} = \frac{1}{\hbar^2} \int_0^\infty \D\tau \Tr_\bath\!\left( \op{U} \op{U}_\ipic(-\tau) \op{\rho}_\sys \op{\rho}_\bath \right)  \:.
\end{equation}
These terms can be physically interpreted based on the sign of their contribution: plus sign means a gain term, and minus sign means a loss term, hence the notation.
As discussed later in Sec.~\ref{sec:boltzmann}, the two terms of each type combine to give the gain or loss term in the classical Boltzmann equation.

\subsubsection{Gain term}
First, let us take a closer look at the gain term~\eqref{eq:def-gain-term}.
In order to expand this term, one approach is to use the Fourier decomposition of the full particle-scatterer potential
\begin{equation}\label{eq:full-potential-from-fourier}
\op{U} = \sum_{i=1}^N u(\op{\vect{r}} - \op{\vect{x}}_i) = \frac{1}{V} \sum_{i=1}^N \sum_{\vect{q}} \fourier{u}(\vect{q}) \E^{\I\vect{q}\cdot(\op{\vect{r}} - \op{\vect{x}}_i)}  \:,
\end{equation}
where $\fourier{u}(\vect{q})$ is defined according to Eq.~\eqref{eq:fourier-single-potential}.
Expression~\eqref{eq:full-potential-from-fourier} if very handy because the imaginary exponential $\E^{\I\vect{q}\cdot(\op{\vect{r}}-\op{\vect{x}}_i)}$ can be factored into the system and the bath operators.
This factorization is allowed by the commutation $[\op{\vect{r}}, \op{\vect{x}}_i]=0$.
Substituting Eq.~\eqref{eq:full-potential-from-fourier} into Eq.~\eqref{eq:def-gain-term}, one gets
\begin{equation}\label{eq:gain-term-1}\begin{split}
\op{C}_{\rm G} & = \int_0^\infty \frac{\D\tau}{\hbar^2V^2} \sum_{i,j}^N \sum_{\vect{q},\dual{\vect{q}}} \fourier{u}(\vect{q}) \cc{\fourier{u}}(\dual{\vect{q}})  \\
\times & \Tr_\bath\!\left( \E^{-\frac{\I}{\hbar}\op{H}_0\tau} \E^{\I\vect{q}\cdot(\op{\vect{r}} - \op{\vect{x}}_i)} \E^{\frac{\I}{\hbar}\op{H}_0\tau} \op{\rho}_\sys \op{\rho}_\bath \E^{-\I\dual{\vect{q}}\cdot(\op{\vect{r}} - \op{\vect{x}}_j)} \right)  \:.
\end{split}\end{equation}
It is possible to simplify Eq.~\eqref{eq:gain-term-1} without projecting everything onto the eigenbasis of the free Hamiltonian $\op{H}_0$.
One option is to commute the first two exponentials in the trace using the momentum translation property
\begin{equation}\label{eq:translation-property}
\E^{-\I\vect{q}\cdot(\op{\vect{r}}-\op{\vect{x}}_i)} f(\op{\vect{k}}_\sys, \op{\vect{k}}_i) \E^{\I\vect{q}\cdot(\op{\vect{r}}-\op{\vect{x}}_i)} = f(\op{\vect{k}}_\sys + \vect{q}, \op{\vect{k}}_i - \vect{q})  \:,
\end{equation}
for any function $f(\vect{x},\vect{y})$. Property~\eqref{eq:translation-property} solely derives from Eq.~\eqref{eq:momentum-translation}.
According to Eq.~\eqref{eq:translation-property}, one can write
\begin{equation}
\E^{-\frac{\I}{\hbar}\op{H}_0\tau} \E^{\I\vect{q}\cdot(\op{\vect{r}} - \op{\vect{x}}_i)} = \E^{\I\vect{q}\cdot(\op{\vect{r}} - \op{\vect{x}}_i)} \E^{-\frac{\I}{\hbar}\op{H}'_0\tau}  \:,
\end{equation}
where $\op{H}'_0$ is the modified Hamiltonian of the form
\begin{equation}\label{eq:modified-hamiltonian}
\op{H}'_0 = E_{\op{\vect{k}}_\sys + \vect{q}} + E_{\op{\vect{k}}_i - \vect{q}} + \sum_{j(\neq i)}^N E_{\op{\vect{k}}_j}  \:.
\end{equation}
Equation~\eqref{eq:modified-hamiltonian} represents the system energy after the collision process $(\vect{k}_\sys,\vect{k}_i)\rightarrow(\vect{k}_\sys+\vect{q},\vect{k}_i-\vect{q})$ with the $i$-th scatterer.
Furthermore, it is convenient to define the Hamiltonian difference
\begin{equation}\label{eq:energy-diff-operator-1}
\op{D}_{\vect{q},i} = \op{H}'_0-\op{H}_0 = E_{\op{\vect{k}}_\sys + \vect{q}} + E_{\op{\vect{k}}_i - \vect{q}} - E_{\op{\vect{k}}_\sys} - E_{\op{\vect{k}}_i}  \:,
\end{equation}
in the same way as Eq.~\eqref{eq:energy-diff-in-q-1}. Note, however, that $\op{D}_{\vect{q},i}$ is a quantum operator.
With this notation, Eq.~\eqref{eq:gain-term-1} reads
\begin{equation}\label{eq:gain-term-2}\begin{split}
\op{C}_{\rm G} & = \int_0^\infty \frac{\D\tau}{\hbar^2V^2} \sum_{i,j}^N \sum_{\vect{q},\dual{\vect{q}}} \fourier{u}(\vect{q}) \cc{\fourier{u}}(\dual{\vect{q}})  \\
\times & \E^{\I\vect{q}\cdot\op{\vect{r}}} \Tr_\bath\!\left( \E^{\I\dual{\vect{q}}\cdot\op{\vect{x}}_j} \E^{-\I\vect{q}\cdot\op{\vect{x}}_i} \E^{-\frac{\I}{\hbar}\op{D}_{\vect{q},i}\tau} \op{\rho}_\bath \right) \op{\rho}_\sys \E^{-\I\dual{\vect{q}}\cdot\op{\vect{r}}}  \:,
\end{split}\end{equation}
where one has used the cyclicity of the bath operators within the trace $\Tr_\bath$.
Expression~\eqref{eq:gain-term-2} can be simplified further by means of a useful additional property.
Letting $\textstyle\op{A}=\sum_{\vect{k}} A_{\vect{k}} \ket{\vect{k}}\bra{\vect{k}}$ be an operator diagonal in the momentum basis, the trace of the translated operator $\E^{\I\vect{q}\cdot\op{\vect{r}}} \op{A}$ will be zero, except for $\vect{q}=\vect{0}$. In other words, one has the property
\begin{equation}\label{eq:trace-property-1}
\Tr\!\left( \E^{\I\vect{q}\cdot\op{\vect{r}}} \op{A} \right) = \delta_{\vect{q}} \Tr\op{A}  \:.
\end{equation}
Applied to the trace of Eq.~\eqref{eq:gain-term-2} with $\op{A}=\E^{-\frac{\I}{\hbar}\op{D}_{\vect{q},i}\tau}\op{\rho}_\bath$ playing the role of the diagonal operator in the momentum basis, Eq.~\eqref{eq:trace-property-1} becomes
\begin{equation}\label{eq:trace-property-2}
\Tr_\bath\!\left( \E^{\I\dual{\vect{q}}\cdot\op{\vect{x}}_j} \E^{-\I\vect{q}\cdot\op{\vect{x}}_i} \op{A} \right) = \left[\delta_{\vect{q}-\dual{\vect{q}}} \delta_{ij} + \delta_{\vect{q}}\delta_{\dual{\vect{q}}} (1-\delta_{ij})\right] \Tr_\bath\op{A}  \:.
\end{equation}
Expression~\eqref{eq:trace-property-2} translates the following statement: if $i=j$, then the momenta $\vect{q}$ and $\dual{\vect{q}}$ must be equal to each other so as to eliminate the exponentials, otherwise if $i\neq j$, then $\vect{q}$ and $\dual{\vect{q}}$ must both be equal to zero.
Obviously, the second case corresponds to a trivial collision with no actual change of the system state.
Although these terms $i\neq j$ are not zero, they can be omitted from the calculation, because they will be eliminated anyway by the corresponding opposite contributions from the loss terms in Eq.~\eqref{eq:redfield-expansion-2}.
After simplifying Eq.~\eqref{eq:gain-term-2} with Eq.~\eqref{eq:trace-property-2}, one last step comes from the observation that each term of given $i$ is identical.
Therefore, one can replace the sum over $i$ by a factor $N$, and rename $\vect{k}_i$ to $\vect{k}_\bath$ for convenience since it now corresponds to a generic bath particle.
Finally, one obtains
\begin{equation}\label{eq:gain-term-3}
\op{C}_{\rm G} = \sum_{\vect{q}} \E^{\I\vect{q}\cdot\op{\vect{r}}} \left( \int_0^\infty \D\tau \op{K}_{\vect{q}}(\tau) \right) \op{\rho}_\sys \E^{-\I\vect{q}\cdot\op{\vect{r}}}  \:,
\end{equation}
where $\op{K}_{\vect{q}}(\tau)$ is the system-bath interaction operator defined as
\begin{equation}\label{eq:def-bath-correlation}
\op{K}_{\vect{q}}(\tau) = \frac{N}{\hbar^2V^2} \abs{\fourier{u}(\vect{q})}^2 \Tr_\bath\!\left( \E^{-\frac{\I}{\hbar}\op{D}_{\vect{q}}\tau} \op{\rho}_\bath \right)  \:,
\end{equation}
in order to gather all the dependencies on the time $\tau$.
The Hamiltonian difference operator in Eq.~\eqref{eq:def-bath-correlation} reads
\begin{equation}\label{eq:energy-diff-operator-2}
\op{D}_{\vect{q}} = E_{\op{\vect{k}}_\sys + \vect{q}} + E_{\op{\vect{k}}_\bath - \vect{q}} - E_{\op{\vect{k}}_\sys} - E_{\op{\vect{k}}_\bath}  \:.
\end{equation}
It should be noted that $\op{K}_{\vect{q}}(\tau)$ in Eq.~\eqref{eq:def-bath-correlation} is actually a non-Hermitian operator which does not commute with $\op{\rho}_\sys$ or $\E^{\I\vect{q}\cdot\op{\vect{r}}}$ in general.

\subsubsection{System-bath interaction operator}
In this subsection, one shows that the operator~\eqref{eq:def-bath-correlation} decreases fast enough in $\tau$ for the time integral in Eq.~\eqref{eq:gain-term-3} to converge.
This operator can be written as
\begin{equation}\label{eq:bath-correlation-1}
\op{K}_{\vect{q}}(\tau) = \frac{N}{\hbar^2V^2} \abs{\fourier{u}(\vect{q})}^2 \E^{-\frac{\I}{\hbar}(E_{\op{\vect{k}}_\sys+\vect{q}} - E_{\op{\vect{k}}_\sys})\tau} \kappa_{\vect{q}}(\tau)  \:,
\end{equation}
where $\kappa_{\vect{q}}(\tau)$ is the bath correlation function given by the trace over the bath in Eq.~\eqref{eq:def-bath-correlation}
\begin{equation}\label{eq:def-bath-subfunction}
\kappa_{\vect{q}}(\tau) = \sum_{\vect{k}_\bath} \E^{-\frac{\I}{\hbar}(E_{\vect{k}_\bath-\vect{q}} - E_{\vect{k}_\bath})\tau} \rho_\bath(\vect{k}_\bath)  \:,
\end{equation}
and the single-particle bath distribution $\rho_\bath(\vect{k}_\bath)$ introduced in Eq.~\eqref{eq:bath-momentum-distrib}.
In the continuum limit ($V\rightarrow\infty$), one gets in terms of the velocity $\vect{v}_\bath=\tfrac{\hbar\vect{k}_\bath}{m_\bath}$
\begin{equation}\label{eq:bath-subfunction-1}
\kappa_{\vect{q}}(\tau) = \E^{-\I\frac{\hbar\vect{q}^2}{2m_\bath}\tau} \int_{\mathbb{R}^d} \E^{\I\vect{v}_\bath\cdot\vect{q}\tau} f_\bath(\vect{v}_\bath) \D\vect{v}_\bath  \:,
\end{equation}
where $f_\bath(\vect{v}_\bath)$ is the usual Maxwell-Boltzmann velocity distribution normalized to unity.
The result of the integral in Eq.~\eqref{eq:bath-subfunction-1} is
\begin{equation}\label{eq:bath-subfunction-result}
\kappa_{\vect{q}}(\tau) = \E^{-\I\frac{\hbar\vect{q}^2}{2m_\bath}\tau} \E^{-\frac{\vect{q}^2\tau^2}{2\beta m_\bath}}  \:.
\end{equation}
This shows that Eq.~\eqref{eq:bath-correlation-1} decays with $\tau$ and that the time integral in Eq.~\eqref{eq:gain-term-3} is meaningful, as it should be.
The characteristic time of this decay is known as the bath correlation time and is defined in this paper for the given momentum transfer $\vect{q}$ as
\begin{equation}\label{eq:def-bath-correlation-time}
\tau_\bath = \sqrt{\frac{\beta m_\bath}{\vect{q}^2}} = \frac{\sqrt{d}}{v_\bath q}  \:,
\end{equation}
where $v_\bath^2=\tavg{\vect{v}_\bath^2}=d/(\beta m_\bath)$ is the mean square velocity of the scatterers.
On the one hand, one notices that the time $\tau_\bath$ has no upper bound because the transferred momentum $q$ can be arbitrarily small.
On the other hand, $\tau_\bath$ possesses a rough lower bound given by
\begin{equation}\label{eq:bath-time-lower-bound}
\tau_\bath \gtrsim \frac{R}{v_\bath}  \:,
\end{equation}
where $R$ is the range of the potential $u(\vect{r})$.
Indeed, the correlation function Eq.~\eqref{eq:bath-correlation-1} is weighted by the Fourier transform $\fourier{u}(\vect{q})$, which, for well behaved potentials, is expected to decay at momenta larger than $1/R$.
Furthermore, the lower bound $R/v_\bath$ can be compared to the duration of a single collision between the incident particle and a scatterer: $\tau_\sys\propto R/v_\sys$.
If the incident particle is fast ($v_\sys\geq v_\bath$), then Eq.~\eqref{eq:bath-time-lower-bound} implies that $\tau_\bath\gtrsim\tau_\sys$.
If, in addition, there is no longer time scale relevant to the collision than $\tau_\sys$, then one can consider $\tau_\bath$ as arbitrarily large and safely take the limit $\tau_\bath\rightarrow\infty$ in the calculations.
\par This result contrasts with the assumption of infinitely small $\tau_\bath$, which would be needed to consider the bath as delta-correlated and to motivate the reduction to a Lindblad equation~\cite{GaspardP1999a, Strunz2004, Hornberger2009, Hartmann2020}.
According to Eq.~\eqref{eq:def-bath-correlation-time}, the assumption $\tau_\bath\rightarrow 0$ would hold only if the bath velocities are very large compared to the incident particle ($v_\bath\gg v_\sys$).
However, assuming a delta-correlated bath in the present calculation would lead to a flawed equation continuously increasing the particle energy with no friction and no thermalization.
Therefore, this approach is not followed here.
\par Now, one has to account for the time integral in the Redfield equation~\eqref{eq:gain-term-3}.
The integral over $\tau$ will be given by~\cite{Vladimirov1971}
\begin{equation}\label{eq:exp-imag-integral}
\int_0^\infty \E^{\pm\frac{\I}{\hbar}\op{D}_{\vect{q}}\tau} \D\tau = \lim_{\varepsilon\rightarrow 0^+} \frac{\pm\I\hbar}{\op{D}_{\vect{q}}\pm\I\varepsilon} = \pi\hbar\delta(\op{D}_{\vect{q}}) \pm\I\hbar\Pv\frac{1}{\op{D}_{\vect{q}}}  \:,
\end{equation}
where $\Pv$ denotes the Cauchy principal value.
It should be noted that Eq.~\eqref{eq:exp-imag-integral} is very general and can be used to integrate Eq.~\eqref{eq:gain-term-3} regardless of the particle velocity.
Therefore, the rest of the derivation is not restricted to a fast particle.
According to Eq.~\eqref{eq:exp-imag-integral}, the time integral in Eq.~\eqref{eq:gain-term-3} splits into two terms:
\begin{equation}\label{eq:correlation-split}
\int_0^\infty \D\tau \op{K}_{\vect{q}}(\tau) = \frac{1}{2} \op{W}_{\vect{q}} - \I\op{Y}_{\vect{q}}  \:.
\end{equation}
The operator $\op{W}_{\vect{q}}$ contains the Dirac delta of Eq.~\eqref{eq:exp-imag-integral}, and $\op{Y}_{\vect{q}}$ contains the principal value.
These operators respectively read
\begin{equation}\label{eq:def-rate-operator}
\op{W}_{\vect{q}} = W_{\vect{q}}(\op{\vect{k}}_\sys) = \frac{2\pi}{\hbar} \frac{n}{V} \abs{\fourier{u}(\vect{q})}^2 \Tr_\bath\!\left( \delta(\op{D}_{\vect{q}}) \op{\rho}_\bath \right)  \:,
\end{equation}
and
\begin{equation}\label{eq:def-pval-operator}
\op{Y}_{\vect{q}} = Y_{\vect{q}}(\op{\vect{k}}_\sys) = \frac{n}{\hbar V} \abs{\fourier{u}(\vect{q})}^2 \Tr_\bath\!\left( \Pv\frac{1}{\op{D}_{\vect{q}}} \op{\rho}_\bath \right)  \:,
\end{equation}
where $n=N/V$ is the number of scatterers per unit volume.
Note that $\op{W}_{\vect{q}}$ and $\op{Y}_{\vect{q}}$ are Hermitian and have the units of an inverse time.
Moreover, $\op{W}_{\vect{q}}$ can be interpreted as a rate operator for the collision $\vect{k}_\sys\rightarrow\vect{k}_\sys+\vect{q}$ and is defined this way to be consistent with the binary collision rate~\eqref{eq:binary-collision-rate-2} up to a factor $N$.
In particular, $\op{W}_{\vect{q}}$ can be expressed directly in terms of Eq.~\eqref{eq:binary-collision-rate-2} as follows
\begin{equation}\label{eq:rate-operator-from-binary}
\op{W}_{\vect{q}} = N \Tr_\bath\!\left( w_{\vect{q}}(\op{\vect{k}}_\sys, \op{\vect{k}}_\bath) \op{\rho}_\bath \right)  \:.
\end{equation}
With the notations~\eqref{eq:def-rate-operator} and~\eqref{eq:def-pval-operator}, the gain term~\eqref{eq:gain-term-3} becomes
\begin{equation}\label{eq:gain-term-4}
\op{C}_{\rm G} = \sum_{\vect{q}} \E^{\I\vect{q}\cdot\op{\vect{r}}} \left( \frac{1}{2} \op{W}_{\vect{q}} - \I\op{Y}_{\vect{q}} \right) \op{\rho}_\sys \E^{-\I\vect{q}\cdot\op{\vect{r}}}  \:.
\end{equation}
Finally, one can also split the gain term notation $\op{C}_{\rm G}$ into the rate and principal value parts for easier manipulation
\begin{equation}\label{eq:gain-term-split}
\op{C}_{\rm G} = \op{R}_{\rm G} - \I\op{P}_{\rm G}  \:,
\end{equation}
with the notations
\begin{equation}\label{eq:def-gain-rate}
\op{R}_{\rm G} = \frac{1}{2} \sum_{\vect{q}} \E^{\I\vect{q}\cdot\op{\vect{r}}} \op{W}_{\vect{q}} \op{\rho}_\sys \E^{-\I\vect{q}\cdot\op{\vect{r}}}  \:,
\end{equation}
and
\begin{equation}\label{eq:def-gain-pval}
\op{P}_{\rm G} = \sum_{\vect{q}} \E^{\I\vect{q}\cdot\op{\vect{r}}} \op{Y}_{\vect{q}} \op{\rho}_\sys \E^{-\I\vect{q}\cdot\op{\vect{r}}}  \:.
\end{equation}

\subsubsection{Simplified Redfield equation}
The calculation of the loss term~\eqref{eq:def-loss-term} follows a very similar reasoning to that of $\op{C}_{\rm G}$. One finds the result
\begin{equation}\label{eq:loss-term-result}
\op{C}_{\rm L} = \sum_{\vect{q}} \left( \frac{1}{2} \op{W}_{\vect{q}} - \I\op{Y}_{\vect{q}} \right) \op{\rho}_\sys  \:,
\end{equation}
where the operators are given by Eqs.~\eqref{eq:def-rate-operator} and~\eqref{eq:def-pval-operator}.
As the gain term, one can split $\op{C}_{\rm L}$ into the rate and principal value parts
\begin{equation}\label{eq:loss-term-split}
\op{C}_{\rm L} = \op{R}_{\rm L} - \I\op{P}_{\rm L}  \:,
\end{equation}
with
\begin{equation}\label{eq:def-loss-rate}
\op{R}_{\rm L} = \frac{1}{2} \sum_{\vect{q}} \op{W}_{\vect{q}} \op{\rho}_\sys  \:,
\end{equation}
and
\begin{equation}\label{eq:def-loss-pval}
\op{P}_{\rm L} = \sum_{\vect{q}} \op{Y}_{\vect{q}} \op{\rho}_\sys  \:.
\end{equation}
Now, all the collision terms can be grouped into Eq.~\eqref{eq:redfield-expansion-2} to get
\begin{equation}\label{eq:redfield-expansion-3}\begin{split}
\pder{\op{\rho}_\sys}{t} = \supop{L}_\sys\op{\rho}_\sys & + \op{R}_{\rm G} + \herm{\op{R}}_{\rm G} - \op{R}_{\rm L} - \herm{\op{R}}_{\rm L}   \\
 & - \I \left( \op{P}_{\rm G} - \herm{\op{P}}_{\rm G} - \op{P}_{\rm L} + \herm{\op{P}}_{\rm L} \right)  \:.
\end{split}\end{equation}
The principal value terms in the second line of Eq.~\eqref{eq:redfield-expansion-3} can be interpreted as coherent quantum contributions.
It is shown in Appendix~\ref{app:principal-values} that these principal value terms are negligible in the weak scattering regime ($k_\sys\lscat\gg 1$).
Therefore, it is reasonable to omit them at this point, and only retain the rate terms in the first line of Eq.~\eqref{eq:redfield-expansion-3}.
\par Finally, from Eqs.~\eqref{eq:def-gain-rate}, \eqref{eq:def-loss-rate} and~\eqref{eq:redfield-expansion-3}, one obtains the sought quantum master equation
\begin{equation}\label{eq:simplified-redfield}
\pder{\op{\rho}_\sys}{t} = \supop{L}_\sys\op{\rho}_\sys + \frac{1}{2} \sum_{\vect{q}} \left( \E^{\I\vect{q}\cdot\op{\vect{r}}} \{\op{W}_{\vect{q}}, \op{\rho}_\sys\} \E^{-\I\vect{q}\cdot\op{\vect{r}}} - \{\op{W}_{\vect{q}}, \op{\rho}_\sys\} \right)  \:,
\end{equation}
where $\{\op{A},\op{B}\}=\op{A}\op{B}+\op{B}\op{A}$ denotes the anticommutator.
Equation~\eqref{eq:simplified-redfield} will also be referred to as the simplified Redfield equation because it neglects the principal value terms.
Due to this approximation, it is not equivalent to the original Redfield equation~\eqref{eq:redfield-equation-final}.
In the limit of infinite quantization volume ($V\rightarrow\infty$), Eq.~\eqref{eq:simplified-redfield} can be expressed on the continuum spectrum of momenta using the replacement rule~\eqref{eq:discrete-sum-to-continuum}.
However, this step is deferred to Sec.~\ref{sec:boltzmann}.

\subsection{Comment on positivity}\label{sec:positivity}
The Redfield equation~\eqref{eq:simplified-redfield} is not guaranteed to preserve the positivity of the particle density matrix $\op{\rho}_\sys$, since it is not of the Lindblad form~\cite{Lindblad1976, Gorini1976, Manzano2020, Breuer2002, Weiss2008}.
The reason is that the rate operator $\op{W}_{\vect{q}}$ in the gain term of Eq.~\eqref{eq:simplified-redfield} acts on one side of $\op{\rho}_\sys$ or the other, but not on both sides at the same time as in the Lindblad equation.
As a consequence, some of the eigenvalues of $\op{\rho}_\sys$ can possibly reach negative values.
This can be considered as a problem or not depending on the physical context~\cite{Pechukas1994, GaspardP1999a, Farina2019, Mozgunov2020, Davidovic2020, Hartmann2020, Breuer2002, Weiss2008}.
In this section, it is shown by means of an approximate Lindblad form that this issue does not compromise the validity of Eq.~\eqref{eq:simplified-redfield} in the framework of fast particles.
Indeed, it is possible to obtain an approximate Lindblad equation from Eq.~\eqref{eq:simplified-redfield} by factoring the rate operator as
\begin{equation}\label{eq:rate-operator-factor}
\op{W}_{\vect{q}} = \herm{\op{A}}_{\vect{q}} \op{A}_{\vect{q}} = \op{A}_{\vect{q}} \herm{\op{A}}_{\vect{q}}  \:,
\end{equation}
for some non-Hermitian operator $\op{A}_{\vect{q}}$. In general, this operator can be expressed as
\begin{equation}\label{eq:amplitude-operator-from-rate}
\op{A}_{\vect{q}} = \sqrt{W_{\vect{q}}(\op{\vect{k}}_\sys)} \E^{\I\phi_{\vect{q}}(\op{\vect{k}}_\sys)}  \:,
\end{equation}
where $\phi_{\vect{q}}$ is a real function which possibly depends on $\vect{k}_\sys$.
It should be noted that $\op{A}_{\vect{q}}$ is not unique because the complex phase $\phi_{\vect{q}}$ cannot be fixed in this way.
More generally, instead of Eq.~\eqref{eq:rate-operator-factor}, one could consider the following factorization of the rate operator:
\begin{equation}\label{eq:rate-operator-factor-general}
\op{W}_{\vect{q}} = \Tr_\bath\!\left( \herm{A_{\vect{q}}(\op{\vect{k}}_\sys, \op{\vect{k}}_\bath)} A_{\vect{q}}(\op{\vect{k}}_\sys, \op{\vect{k}}_\bath) \op{\rho}_\bath \right)  \:.
\end{equation}
Doing so, a different scattering amplitude can be attributed to each collision process $(\vect{k}_\sys,\vect{k}_\bath)\rightarrow(\vect{k}_\sys+\vect{q},\vect{k}_\bath-\vect{q})$.
In particular, the factorization~\eqref{eq:rate-operator-factor-general} would be needed to relate Eq.~\eqref{eq:simplified-redfield} to the quantum Boltzmann equation in Lindblad form of Ref.~\cite{Hornberger2006a}.
The downside of Eq.~\eqref{eq:rate-operator-factor-general} is that it requires to factor the Dirac delta of energy conservation in Eq.~\eqref{eq:def-rate-operator}.
However, the square root of a delta cannot be properly defined.
This problem is closely related to the delta squaring issue encountered in the collisional decoherence literature~\cite{Joos1985, Hornberger2003a, Hornberger2003b, Hornberger2006a, Adler2006, Hornberger2009, Kamleitner2010}.
This cannot be addressed by the approach presented in this paper. This is why Eq.~\eqref{eq:rate-operator-factor-general} will not be used here.
On the other hand, this issue can be avoided in general with the use of Eq.~\eqref{eq:rate-operator-factor}, because it amounts to evaluate the square root of a smooth distribution of $\vect{q}$ averaged over the thermal bath $\op{\rho}_\bath$.
This does not pose the mathematical problem encountered with the factorization~\eqref{eq:rate-operator-factor-general}.
However, as mentioned before, the complex phase of $\op{A}_{\vect{q}}$ in Eq.~\eqref{eq:amplitude-operator-from-rate} is arbitrary and hence the resulting Lindblad equation will not be uniquely determined.
\par Using Eq.~\eqref{eq:rate-operator-factor}, the first anticommutator in Eq.~\eqref{eq:simplified-redfield} can be written as a completely positive map plus some correction terms:
\begin{equation}\label{eq:lindblad-form-trick}
\frac{\{\op{W}_{\vect{q}}, \op{\rho}_\sys\}}{2} = \op{A}_{\vect{q}} \op{\rho}_\sys \herm{\op{A}}_{\vect{q}} + \frac{\op{A}_{\vect{q}}[\herm{\op{A}}_{\vect{q}}, \op{\rho}_\sys] - [\op{A}_{\vect{q}}, \op{\rho}_\sys]\herm{\op{A}}_{\vect{q}}}{2}  \:.
\end{equation}
Therefore, if one defines the quantum jump operator
\begin{equation}\label{eq:lindblad-operator}
\op{L}_{\vect{q}} = \E^{\I\vect{q}\cdot\op{\vect{r}}} \op{A}_{\vect{q}}  \:,
\end{equation}
then the Redfield equation~\eqref{eq:simplified-redfield} can be rewritten exactly as
\begin{equation}\label{eq:lindblad-equation}
\pder{\op{\rho}_\sys}{t} = \supop{L}_\sys\op{\rho}_\sys + \sum_{\vect{q}} \left( \op{L}_{\vect{q}} \op{\rho}_\sys \herm{\op{L}}_{\vect{q}} - \frac{1}{2}\{\herm{\op{L}}_{\vect{q}} \op{L}_{\vect{q}}, \op{\rho}_\sys\} \right) + \op{I} \:,
\end{equation}
where $\op{I}$ contains the correction coming from Eq.~\eqref{eq:lindblad-form-trick}, that is
\begin{equation}\label{eq:lindblad-correction-1}
\op{I} = \sum_{\vect{q}} \E^{\I\vect{q}\cdot\op{\vect{r}}} \frac{\op{A}_{\vect{q}}[\herm{\op{A}}_{\vect{q}}, \op{\rho}_\sys] - [\op{A}_{\vect{q}}, \op{\rho}_\sys]\herm{\op{A}}_{\vect{q}}}{2} \E^{-\I\vect{q}\cdot\op{\vect{r}}}  \:.
\end{equation}
If the correction $\op{I}$ is neglected, then the master equation~\eqref{eq:lindblad-equation} is of the Lindblad form, ensuring the completely positive evolution of $\op{\rho}_\sys$.
In order to interpret the nature of the correction $\op{I}$ in the context of a fast incident particle, it is useful to assume that the deviation of the particle momentum around some central momentum $\vect{k}_{\sys,0}$ is small:
\begin{equation}\label{eq:small-straggling-approx-1}
\op{\vect{k}}_\sys = \vect{k}_{\sys,0} + \Delta\op{\vect{k}}_\sys  \quad\text{with}\quad  \norm{\Delta\op{\vect{k}}_\sys} \ll \norm{\vect{k}_{\sys,0}}  \:.
\end{equation}
Therefore, the amplitude operator $A_{\vect{q}}(\op{\vect{k}}_\sys)$ can be expanded at the first order of $\Delta\op{\vect{k}}_\sys$. One has
\begin{equation}\label{eq:small-straggling-approx-2}
A_{\vect{q}}(\op{\vect{k}}_\sys) = A_{\vect{q}}(\vect{k}_{\sys,0}) + \Delta\op{\vect{k}}_\sys\cdot\grad_{\vect{k}_\sys}A_{\vect{q}}(\vect{k}_{\sys,0}) + \bigo(\Delta\op{\vect{k}}_\sys^2) \:.
\end{equation}
Substituting Eq.~\eqref{eq:small-straggling-approx-2} into Eq.~\eqref{eq:lindblad-correction-1} leads to the result
\begin{equation}\label{eq:lindblad-correction-2}\begin{split}
\op{I} = \sum_{\vect{q}} \E^{\I\vect{q}\cdot\op{\vect{r}}} & \left( -\I W_{\vect{q}}(\vect{k}_{\sys,0}) \pder{\phi_{\vect{q}}}{k_{\sys,i}}(\vect{k}_{\sys,0}) [\Delta\op{k}_{\sys,i}, \op{\rho}_\sys] \right.  \\
 & + \frac{S_{ij}}{2} \left[ \Delta\op{k}_{\sys,i}, [\Delta\op{k}_{\sys,j}, \op{\rho}_\sys] \right]  \\
 & + \left. \I\frac{A_{ij}}{2} \left\{\Delta\op{k}_{\sys,i}, [\Delta\op{k}_{\sys,j}, \op{\rho}_\sys] \right\} \right) \E^{-\I\vect{q}\cdot\op{\vect{r}}}  \:,
\end{split}\end{equation}
where the summations over the repeated indices $i$ and $j$ have been implied.
The quantities $S_{ij}$ and $A_{ij}$ in Eq.\ \eqref{eq:lindblad-correction-2} are respectively the symmetric and antisymmetric tensors defined as
\begin{equation}\label{eq:lindblad-correction-tensors}\begin{cases}
S_{ij} = \frac{1}{2} \left( \pder{A_{\vect{q}}}{k_{\sys,i}} \pder{\cc{A}_{\vect{q}}}{k_{\sys,j}} + \pder{\cc{A}_{\vect{q}}}{k_{\sys,i}} \pder{A_{\vect{q}}}{k_{\sys,j}} \right)_{\vect{k}_{\sys,0}}  \:,\\
A_{ij} = \frac{1}{2\I} \left( \pder{A_{\vect{q}}}{k_{\sys,i}} \pder{\cc{A}_{\vect{q}}}{k_{\sys,j}} - \pder{\cc{A}_{\vect{q}}}{k_{\sys,i}} \pder{A_{\vect{q}}}{k_{\sys,j}} \right)_{\vect{k}_{\sys,0}}  \:.
\end{cases}\end{equation}
The three terms of Eq.~\eqref{eq:lindblad-correction-2} can be interpreted respectively as a positional drift term, a positional diffusion term, and a momentum-dependent drift term.
In particular, a term of the form $[\Delta\op{k}_{\sys,i}, [\Delta\op{k}_{\sys,j}, \op{\rho}_\sys]]$, responsible for the particle diffusion in position space, is known in the literature to restore the complete positivity of the Caldeira-Leggett master equation~\cite{Breuer2002, Vacchini2009, Diosi1995, Diosi2009, Kamleitner2010, Hornberger2009}.
\par All the terms in Eq.~\eqref{eq:lindblad-correction-2} rely on gradients of $A_{\vect{q}}(\vect{k}_\sys)$ with respect to $\vect{k}_\sys$, and involve commutators $[\Delta\op{\vect{k}}_\sys, \op{\rho}_\sys]$.
Therefore, the correction $\op{I}$ can be neglected in two circumstances: either $A_{\vect{q}}(\vect{k}_\sys)$ slowly varies with $\vect{k}_\sys$, or $\op{\rho}_\sys$ is nearly diagonal in the momentum basis.
On the one hand, the former cannot be guaranteed in general because the complex phase of $A_{\vect{q}}$ is not known.
On the other hand, the latter is reasonable if the incident wave packet is much larger than its own wavelength.
As this condition can be fulfilled for ionizing fast particles, given their small wavelength of subatomic scale, this implies that the positivity of $\op{\rho}_\sys$ is approximately preserved by the Redfield equation~\eqref{eq:simplified-redfield}.
\par Finally, if the incident wave is a plane wave, then one has $[\Delta\op{\vect{k}}_\sys, \op{\rho}_\sys]=\vect{0}$ and equivalently $[\op{\vect{k}}_\sys, \op{\rho}_\sys]=\vect{0}$.
Furthermore, since the medium is uniform, $\op{\rho}_\sys$ remains diagonal in the momentum basis.
In that special case, the evolution prescribed by the Redfield equation~\eqref{eq:simplified-redfield} is guaranteed to be completely positive.

\subsection{Linear Boltzmann equation}\label{sec:boltzmann}
In this section, it is shown that the Redfield equation~\eqref{eq:simplified-redfield} reduces to a linear Boltzmann equation, and thus describes the transport of particle ``$\sys$'' within the gas.
For this purpose, the space dependence is restored through the Wigner transform which is defined as the Fourier transform of the off-diagonal part of the density matrix~\cite{Wigner1932, Moyal1949a, Basdevant2002, Cohen2020-vol3}.
When the density matrix is represented in momentum basis, the Wigner transform reads
\begin{equation}\label{eq:def-wigner-momentum}
f(\vect{r},\vect{k}) = \wigner{\op{\rho}} = \int_{\mathbb{R}^d} \bra{\vect{k} + \tfrac{\vect{s}}{2}} \op{\rho} \ket{\vect{k} - \tfrac{\vect{s}}{2}} \frac{\E^{\I\vect{s}\cdot\vect{r}}}{(2\pi)^d} \D\vect{s} \:,
\end{equation}
where $f(\vect{r},\vect{k})$ is known as the Wigner function. This is a real function of the position $\vect{r}$ and the momentum $\vect{k}$.
This function is also referred to as a quasi-probability distribution because of its similarity with the classical phase-space distribution.
However, in contrast to a usual probability distribution, $f(\vect{r},\vect{k})$ may be negative, typically in the presence of quantum interferences.
\par The Wigner transform can be directly applied to the Redfield equation~\eqref{eq:simplified-redfield}.
In particular, the free propagation term becomes
\begin{equation}\label{eq:wigner-free-motion}
\wigner{\supop{L}_\sys\op{\rho}_\sys} = -\vect{v}_\sys\cdot\grad_{\vect{r}} f_\sys(\vect{r},\vect{k}_\sys)  \:,
\end{equation}
where $\vect{v}_\sys=\tfrac{\hbar\vect{k}_\sys}{m_\sys}$ is the particle velocity.
Therefore, the Wigner transform of Eq.~\eqref{eq:simplified-redfield} reads
\begin{equation}\label{eq:quantum-boltzmann-1}
\pder{f_\sys}{t}(\vect{r},\vect{k}_\sys) + \vect{v}_\sys\cdot\grad_{\vect{r}} f_\sys(\vect{r},\vect{k}_\sys) = \wigner{\op{R}}  \:,
\end{equation}
where $\op{R}$ gathers all the collision terms in Eq.~\eqref{eq:simplified-redfield} which have to be transformed.
First, let us consider the loss term for a given value of $\vect{q}$:
\begin{equation}\label{eq:wigner-loss-term-1}\begin{split}
 & \wigner{\tfrac{\{\op{W}_{\vect{q}}, \op{\rho}_\sys\}}{2}}  \\
 & = \int_{\mathbb{R}^d} \bra{\vect{k}_\sys + \tfrac{\vect{s}}{2}} \tfrac{\{W_{\vect{q}}(\op{\vect{k}}_\sys), \op{\rho}_\sys\}}{2} \ket{\vect{k}_\sys - \tfrac{\vect{s}}{2}} \frac{\E^{\I\vect{s}\cdot\vect{r}}}{(2\pi)^d} \D\vect{s}  \:.
\end{split}\end{equation}
More explicitly, using the notation $\rho_\sys(\vect{x},\vect{y})=\bra{\vect{x}}\op{\rho}_\sys\ket{\vect{y}}$ for the density matrix, Eq.~\eqref{eq:wigner-loss-term-1} reads
\begin{equation}\label{eq:wigner-loss-term-2}\begin{split}
\wigner{\tfrac{\{\op{W}_{\vect{q}}, \op{\rho}_\sys\}}{2}} & = \int_{\mathbb{R}^d} \frac{W_{\vect{q}}(\vect{k}_\sys+\tfrac{\vect{s}}{2}) + W_{\vect{q}}(\vect{k}_\sys-\tfrac{\vect{s}}{2})}{2}  \\
 & \times \rho_\sys(\vect{k}_\sys + \tfrac{\vect{s}}{2}, \vect{k}_\sys - \tfrac{\vect{s}}{2}) \frac{\E^{\I\vect{s}\cdot\vect{r}}}{(2\pi)^d} \D\vect{s}  \:.
\end{split}\end{equation}
In general, the integral over $\vect{s}$ in Eq.~\eqref{eq:wigner-loss-term-2} cannot be evaluated and expressed in terms of $f_\sys(\vect{r},\vect{k}_\sys)$, as in the classical Boltzmann equation.
For this purpose, one has to make the additional assumption that the density matrix $\rho_\sys$ is close to being diagonal in momentum basis.
If the medium is uniform and if the envelope of the wave function does not vary too quickly in space, then this assumption is justified.
An important consequence of this assumption is that the relevant values of $\norm{\vect{s}}$ are $\norm{\vect{s}}\ll\norm{\vect{k}_\sys}$.
A complementary assumption is that the collision rate $W_{\vect{q}}(\vect{k}_\sys)$ little depends on the particle momentum $\vect{k}_\sys$.
This assumption is reasonable if there is no scattering resonance.
Therefore, the rate factor in Eq.~\eqref{eq:wigner-loss-term-2} can be expanded in series of $\vect{s}$ around $\vect{s}=\vect{0}$ as follows:
\begin{equation}\label{eq:rate-small-s-expansion}\begin{split}
 & \frac{W_{\vect{q}}(\vect{k}_\sys+\tfrac{\vect{s}}{2}) + W_{\vect{q}}(\vect{k}_\sys-\tfrac{\vect{s}}{2})}{2}  \\
 & = W_{\vect{q}}(\vect{k}_\sys) + \sum_{i,j}^d \frac{s_i s_j}{8} \frac{\partial^2W_{\vect{q}}(\vect{k}_\sys)}{\partial k_{\sys,i}\partial k_{\sys,j}} + \cdots  \:.
\end{split}\end{equation}
The zeroth order term in Eq.~\eqref{eq:rate-small-s-expansion} does no longer depend on $\vect{s}$, and thus one gets the Wigner function $f_\sys(\vect{r},\vect{k}_\sys)$.
The second order term quadratically depends on $\vect{s}$, leading to a Hessian matrix with respect to the position
\begin{equation}\label{eq:wigner-spatial-diffusion-term}
\int_{\mathbb{R}^d} s_i s_j \rho_\sys(\vect{k}_\sys+\tfrac{\vect{s}}{2}, \vect{k}_\sys-\tfrac{\vect{s}}{2}) \frac{\E^{\I\vect{s}\cdot\vect{r}}}{(2\pi)^d}  \D\vect{s} = -\frac{\partial^2}{\partial r_i\partial r_j} f_\sys(\vect{r},\vect{k}_\sys)  \:.
\end{equation}
Combining Eqs.~\eqref{eq:rate-small-s-expansion} and \eqref{eq:wigner-spatial-diffusion-term} into Eq.~\eqref{eq:wigner-loss-term-2} yields
\begin{equation}\label{eq:wigner-loss-term-3}\begin{split}
 & \wigner{\tfrac{\{\op{W}_{\vect{q}}, \op{\rho}_\sys\}}{2}}
   = W_{\vect{q}}(\vect{k}_\sys) f_\sys(\vect{r},\vect{k}_\sys)  \\
 & -\frac{1}{8} \sum_{i,j}^d \frac{\partial^2W_{\vect{q}}(\vect{k}_\sys)}{\partial k_{\sys,i} \partial k_{\sys,j}} \frac{\partial^2}{\partial r_i \partial r_j}f_\sys(\vect{r},\vect{k}_\sys) + \cdots \:.
\end{split}\end{equation}
Following the same approach, the Wigner transform of the gain term in Eq.~\eqref{eq:simplified-redfield} reads
\begin{equation}\label{eq:wigner-gain-term}\begin{split}
 & \wigner{\E^{\I\vect{q}\cdot\op{\vect{r}}}\tfrac{\{\op{W}_{\vect{q}}, \op{\rho}_\sys\}}{2}\E^{-\I\vect{q}\cdot\op{\vect{r}}}} 
   = W_{\vect{q}}(\vect{k}_\sys-\vect{q}) f_\sys(\vect{r},\vect{k}_\sys-\vect{q})  \\
 & -\frac{1}{8} \sum_{i,j}^d \frac{\partial^2W_{\vect{q}}(\vect{k}_\sys-\vect{q})}{\partial k_{\sys,i} \partial k_{\sys,j}} \frac{\partial^2}{\partial r_i \partial r_j}f_\sys(\vect{r},\vect{k}_\sys-\vect{q}) + \cdots \:.
\end{split}\end{equation}
This is the same expression as Eq.~\eqref{eq:wigner-loss-term-3} but replacing $\vect{k}_\sys$ by $\vect{k}_\sys-\vect{q}$ according to the unitary transformation~\eqref{eq:translation-property}.
In the following calculations, the second lines of Eqs.~\eqref{eq:wigner-loss-term-3} and~\eqref{eq:wigner-gain-term} will be neglected, because it can be made arbitrarily small for a sufficiently large wave packet, and is exactly zero for an incident plane wave.
In addition, these lines are also negligible if the total collision rate $W_{\vect{q}}(\vect{k}_\sys)$ does not depend on $\vect{k}_\sys$.
Remarkably, the same assumptions have already been exploited in Sec.~\ref{sec:positivity} to derive the Lindblad form~\eqref{eq:lindblad-equation}.
Therefore, under either of these assumptions, the linear Boltzmann equation obtained here also approximately describes a completely positive evolution for the density matrix.
\par Using Eqs.~\eqref{eq:wigner-loss-term-3} and~\eqref{eq:wigner-gain-term}, the Wigner transform of the collision terms reads
\begin{equation}\label{eq:wigner-collision-1}\begin{split}
\wigner{\op{R}} & = \sum_{\vect{q}} W_{\vect{q}}(\vect{k}_\sys-\vect{q}) f_\sys(\vect{r},\vect{k}_\sys-\vect{q})  \\
 & - \sum_{\vect{q}} W_{\vect{q}}(\vect{k}_\sys) f_\sys(\vect{r},\vect{k}_\sys)  \:.
\end{split}\end{equation}
The total collision rate $W_{\vect{q}}(\vect{k}_\sys)$ for the collision $\vect{k}_\sys\rightarrow\vect{k}_\sys+\vect{q}$ can be related to the binary collision rate $w_{\vect{q}}(\vect{k}_\sys,\vect{k}_\bath)$ by Eq.~\eqref{eq:rate-operator-from-binary} which is reformulated here
\begin{equation}\label{eq:boltzmann-rate-from-binary}
W_{\vect{q}}(\vect{k}_\sys) = N \sum_{\vect{k}_\bath} w_{\vect{q}}(\vect{k}_\sys,\vect{k}_\bath) \rho_\bath(\vect{k}_\bath)  \:,
\end{equation}
As a reminder, $w_{\vect{q}}(\vect{k}_\sys,\vect{k}_\bath)$ represents the rate of the collision $(\vect{k}_\sys,\vect{k}_\bath)\rightarrow(\vect{k}_\sys+\vect{q},\vect{k}_\bath-\vect{q})$.
In Eq.~\eqref{eq:boltzmann-rate-from-binary}, $\rho_\bath(\vect{k}_\bath)$ is normalized to unity as $\textstyle\sum_{\vect{k}_\bath}\rho_\bath(\vect{k}_\bath)=1$.
Inserting Eq.~\eqref{eq:boltzmann-rate-from-binary} into Eq.~\eqref{eq:wigner-collision-1} leads to
\begin{equation}\label{eq:wigner-collision-2}\begin{split}
 & \wigner{\op{R}}  \\
 & = \sum_{\vect{k}_\bath,\vect{q}} N w_{\vect{q}}(\vect{k}_\sys-\vect{q},\vect{k}_\bath) f_\sys(\vect{r},\vect{k}_\sys-\vect{q}) \rho_\bath(\vect{k}_\bath)  \\
 & - \sum_{\vect{k}_\bath,\vect{q}} N w_{\vect{q}}(\vect{k}_\sys,\vect{k}_\bath) f_\sys(\vect{r},\vect{k}_\sys) \rho_\bath(\vect{k}_\bath)  \:.
\end{split}\end{equation}
These two sums can be combined by tweaking the first one a bit.
In this regard, one successively performs the two substitutions $\vect{q}\rightarrow-\vect{q}$ and $\vect{k}_\bath\rightarrow\vect{k}_\bath-\vect{q}$ in the first sum of Eq.~\eqref{eq:wigner-collision-2}.
The result is
\begin{equation}\label{eq:wigner-collision-3}\begin{split}
 & \wigner{\op{R}}  \\
 & = \sum_{\vect{k}_\bath,\vect{q}} N w_{-\vect{q}}(\vect{k}_\sys+\vect{q},\vect{k}_\bath-\vect{q}) f_\sys(\vect{r},\vect{k}_\sys+\vect{q}) \rho_\bath(\vect{k}_\bath-\vect{q})  \\
 & - \sum_{\vect{k}_\bath,\vect{q}} N w_{\vect{q}}(\vect{k}_\sys,\vect{k}_\bath) f_\sys(\vect{r},\vect{k}_\sys) \rho_\bath(\vect{k}_\bath)  \:.
\end{split}\end{equation}
The interest of these substitutions is that the following property of the binary collision rate can be used:
\begin{equation}\label{eq:binary-rate-time-reversal}
w_{-\vect{q}}(\vect{k}_\sys+\vect{q}, \vect{k}_\bath-\vect{q}) = w_{\vect{q}}(\vect{k}_\sys, \vect{k}_\bath)  \:.
\end{equation}
Property~\eqref{eq:binary-rate-time-reversal} can be derived from definition~\eqref{eq:binary-collision-rate-2}, and is a consequence of the time-reversal symmetry of the microscopic collision.
Then, the two collision terms in Eq.~\eqref{eq:wigner-collision-3} can be gathered as follows
\begin{equation}\label{eq:wigner-collision-4}\begin{split}
\wigner{\op{R}} & = \sum_{\vect{k}_\bath,\vect{q}} N w_{\vect{q}}(\vect{k}_\sys,\vect{k}_\bath)  \\
 & \times \left[ f_\sys(\vect{r},\vect{k}'_\sys) \rho_\bath(\vect{k}'_\bath) - f_\sys(\vect{r},\vect{k}_\sys) \rho_\bath(\vect{k}_\bath) \right]  \:,
\end{split}\end{equation}
where the notations are $\vect{k}'_\sys=\vect{k}_\sys+\vect{q}$ and $\vect{k}'_\bath=\vect{k}_\bath-\vect{q}$, as in Sec.~\ref{sec:binary-collision}.
Now, one takes the limit of infinite quantization volume ($V\rightarrow\infty$) so that the sums over the momenta turn into integrals.
Using the differential collision rate~\eqref{eq:diff-collision-rate} and integrating over $\vect{k}'=\vect{k}+\vect{q}$ instead of $\vect{q}$, one gets
\begin{equation}\label{eq:wigner-collision-5}\begin{split}
\wigner{\op{R}} & = \int_{\mathbb{R}^d} \D\vect{k}_\bath \int_{\mathbb{R}^d} \D\vect{k}'\: N\der{w}{\vect{k}'}(\vect{k}'\mid\vect{k})  \\
 & \times\left[ f_\sys(\vect{r},\vect{k}'_\sys) f_\bath(\vect{k}'_\bath) - f_\sys(\vect{r},\vect{k}_\sys) f_\bath(\vect{k}_\bath) \right]  \:,
\end{split}\end{equation}
where the notations implicitly became $\vect{k}'_\sys=\vect{k}_\sys+(\vect{k}'-\vect{k})$ and $\vect{k}'_\bath=\vect{k}_\bath-(\vect{k}'-\vect{k})$.
Note that, from Eq.~\eqref{eq:wigner-collision-4} to Eq.~\eqref{eq:wigner-collision-5}, $\rho_\bath(\vect{k}_\bath)$ has been replaced by $f_\bath(\vect{k}_\bath)$, which is normalized according to $\textstyle\int_{\mathbb{R}^d}f_\bath(\vect{k}_\bath)\D\vect{k}_\bath=1$.
Then, splitting the radial and angular parts of the integral~\eqref{eq:wigner-collision-5} over $\vect{k}'$ with $\vect{k}'=k'\vect{\Omega}$, integrating over $k'$, and using Eq.~\eqref{eq:diff-cross-sec-from-diff-rate} to make appear the center-of-mass differential cross section, one finds
\begin{equation}\label{eq:wigner-collision-6}\begin{split}
\wigner{\op{R}} & = \int_{\mathbb{R}^d} \D\vect{k}_\bath \oint_{\mathcal{S}_d} \D\Omega\: nv\der{\sigma}{\Omega}(\vect{\Omega}\mid\vect{k})  \\
 & \times\left[ f_\sys(\vect{r},\vect{k}'_\sys) f_\bath(\vect{k}'_\bath) - f_\sys(\vect{r},\vect{k}_\sys) f_\bath(\vect{k}_\bath) \right]  \:,
\end{split}\end{equation}
where $v=\norm{\vect{v}_\sys-\vect{v}_\bath}$ is the relative velocity of the colliding particles, which is directly proportional to the relative momentum $\vect{k}$ according to Eq.~\eqref{eq:relative-velocity}.
The notations are now
\begin{equation}\label{eq:final-momenta}\begin{cases}
\vect{k}'_\sys  = \vect{k}_\sys  + (k\vect{\Omega}-\vect{k}) = \tfrac{m_\sys}{M}\vect{K}  + k\vect{\Omega}  \:,\\
\vect{k}'_\bath = \vect{k}_\bath - (k\vect{\Omega}-\vect{k}) = \tfrac{m_\bath}{M}\vect{K} - k\vect{\Omega}  \:,
\end{cases}\end{equation}
where $\vect{K}=\vect{k}_\sys+\vect{k}_\bath$ is the total momentum of the colliding particles, and $M=m_\sys+m_\bath$ is their total mass.
Finally, substituting Eq.~\eqref{eq:wigner-collision-6} into Eq.~\eqref{eq:quantum-boltzmann-1} leads to the traditional form of the Boltzmann equation in the absence of external forces~\cite{Boltzmann1872, Weinberg1958, Balescu1975, Huang1987, Harris2004}
\begin{equation}\label{eq:quantum-boltzmann-final}\begin{split}
\pder{f_\sys}{t} + \vect{v}_\sys\cdot\grad_{\vect{r}} f_\sys & = \int \D\vect{k}_\bath \D\Omega\: nv\der{\sigma}{\Omega}(\vect{\Omega}\mid\vect{k})  \\
 & \times\left[ f_\sys(\vect{r},\vect{k}'_\sys) f_\bath(\vect{k}'_\bath) - f_\sys(\vect{r},\vect{k}_\sys) f_\bath(\vect{k}_\bath) \right]  \:.
\end{split}\end{equation}
It may look surprising that Eq.~\eqref{eq:quantum-boltzmann-final} has the same form as the classical linear Boltzmann equation.
However, it should be noted that $f_\sys(\vect{r},\vect{k}_\sys)$ is still a Wigner function able to describe distributions of quantum nature.
In particular, nothing prevents $f_\sys(\vect{r},\vect{k}_\sys)$ from being locally negative due to quantum interferences.
In some way, one could say that the Boltzmann equation may also be thought of as a quantum master equation.

\section{Conclusions}\label{sec:conclusions}%
In this paper, several quantum master equations have been derived explicitly to describe the propagation of a fast quantum particle in a gas at thermal equilibrium, namely the simplified Redfield equation~\eqref{eq:simplified-redfield}, the Lindblad form~\eqref{eq:lindblad-equation}, and the Boltzmann equation~\eqref{eq:quantum-boltzmann-final}.
The starting point of the derivation was the quantum Liouville equation~\eqref{eq:general-liouville} of the full multiparticle problem.
The Hamiltonian of the system, given in Eq.~\eqref{eq:general-hamiltonian}, neglects the interaction between individual scatterers.
\par First, the Redfield equation~\eqref{eq:redfield-equation-final} was derived using perturbation theory on the interaction potential at next-to-leading order and the Markov assumption.
Then, the collision terms of the Redfield equation were expanded using the Fourier decomposition~\eqref{eq:full-potential-from-fourier} of the potential.
In the process, the system-bath interaction operator $\op{K}_{\vect{q}}(\tau)$ defined in Eq.~\eqref{eq:def-bath-correlation} was shown to decay to zero in time, hence ensuring the convergence of its time integral.
In the case of a fast particle, the time scale of this decay turns out to be much longer than the collision time.
This shows that the assumption of a delta-correlated bath, which could possibly be made for a very slow particle ($v_\sys\ll v_\bath$) and which would lead to a Lindblad equation, is not relevant for a fast particle.
Despite this, the time integral of $\op{K}_{\vect{q}}(\tau)$ can be evaluated regardless of the particle velocity, leading to energy conservation Dirac deltas and principal values.
In Appendix~\ref{app:principal-values}, the principal values were shown to be negligible in the weak scattering regime ($k_{\sys,0}\lscat\gg 1$).
This approximation led to the simplified Redfield equation~\eqref{eq:simplified-redfield} which is the central result of this paper.
The best feature of Eq.~\eqref{eq:simplified-redfield} is its four-term structure, made of two adjoint gain terms and two adjoint loss terms, which directly comes from Eq.~\eqref{eq:liouville-inter-pic-integrated}, the exact equation of the full problem.
Therefore, due to this similarity, one should expect Eq.~\eqref{eq:simplified-redfield} to reliably approach the populations and coherences of the full problem governed by the quantum Liouville equation~\eqref{eq:general-liouville}.
\par On the other hand, the four-term structure of Eq.~\eqref{eq:simplified-redfield} prevents it from being of the Lindblad form and thus from guaranteeing the completely positive evolution of the density matrix in all circumstances.
It was shown in Sec.~\ref{sec:positivity} that Eq.~\eqref{eq:simplified-redfield} can be cast in the Lindblad form~\eqref{eq:lindblad-equation} by factorization of the rate operator $W_{\vect{q}}(\op{\vect{k}}_\sys)$.
The reduction to a Lindblad equation is exact in the sense $\op{I}=0$ only when $\op{\rho}_\sys$ is diagonal in the momentum basis, or when $W_{\vect{q}}(\vect{k}_\sys)$ does not depend on $\vect{k}_\sys$.
In particular, the first condition seems reasonable in the framework of ionizing fast particles as their wavelength is typically much smaller than the spatial extent of the wave packet.
This supports the idea that Eq.~\eqref{eq:simplified-redfield} approximately preserves the complete positivity of the density matrix of fast particles.
\par Last but not least, the linear Boltzmann equation~\eqref{eq:quantum-boltzmann-final} was derived in Sec.~\ref{sec:boltzmann} from the simplified Redfield equation.
This derivation highlights the consistency between the simplified Redfield equation~\eqref{eq:simplified-redfield} and the Boltzmann equation regarding the transport of the particle.
In addition, this derivation is based on the same assumptions as for the Lindblad form~\eqref{eq:lindblad-equation}, namely either $\op{\rho}_\sys$ is diagonal in the momentum basis, or $W_{\vect{q}}(\vect{k}_\sys)$ is independent of $\vect{k}_\sys$.
This concordance shows that the evolution predicted by the linear Boltzmann equation is also completely positive in first approximation.
Furthermore, Eq.~\eqref{eq:quantum-boltzmann-final} has the same form as the classical Boltzmann equation, but it governs the evolution of the Wigner function of the particle, which is a quantum distribution.
Therefore, the Boltzmann equation can also be considered as a different kind of quantum master equation for the propagation of a particle in a gas, beside the Redfield and Lindblad equations.
\par In the future, it would be useful to study the differences of predictions between the master equations derived in this paper and the quantum Liouville equation of the full multiparticle problem.
One important issue concerns the spatial diffusion induced by the non-commutation of the rate operator $\op{W}_{\vect{q}}$ and the density matrix $\op{\rho}_\sys$.
This effect is expected to be significant for wave packets of small spatial extent compared to their central wavelength.
In this paper, terms contributing to spatial diffusion have been highlighted in the Redfield equation~\eqref{eq:simplified-redfield} and the Lindblad equation~\eqref{eq:lindblad-equation} with $\op{I}=0$, but seem absent from the Boltzmann equation~\eqref{eq:quantum-boltzmann-final}.
Such terms have long been conjectured in the collisional decoherence literature~\cite{Breuer2002, Weiss2008, Vacchini2009, Diosi1995, Diosi2009, Kamleitner2010, Hornberger2009, Diosi2022} to ensure the completely positive time evolution of the density matrix.
However, they have never been the subject of a precise comparison with the predictions of the quantum Liouville equation of the full problem, so that their physical significance is still an open question today.
\par Finally, in a later paper, one plans to study the properties of the Redfield equation~\eqref{eq:simplified-redfield} in more details, especially the friction, the deflection, and the decoherence of a fast particle in a gas.

\begin{acknowledgments}%
The present results were obtained within the framework of the author's doctoral thesis under the supervision of Prof.\ Jean-Marc Sparenberg.
The author is grateful to Profs.\ Pierre Gaspard and Jean-Marc Sparenberg for useful discussions and for reviewing this manuscript.
This work was funded by the Belgian National Fund for Scientific Research (F.R.S.-FNRS) as part of the ``Research Fellow'' (ASP - Aspirant) fellowship program.
\end{acknowledgments}%

\appendix%
\section{Principal value terms}\label{app:principal-values}
In this appendix, one estimates the total contribution of the four principal value terms which emerged in Sec.~\ref{sec:collision-terms} from the time integral $\textstyle\int_0^\infty\D\tau$ in the Redfield equation.
In particular, it is proved that this contribution is negligible in the weak scattering regime, that is, when the mean free path is much larger than the wavelength ($k\lscat\gg 1$).
The total contribution is given by the last bracket of Eq.~\eqref{eq:redfield-expansion-3}, that is
\begin{equation}\label{eq:total-pval-1}
\op{P} = \left( \op{P}_{\rm G} - \herm{\op{P}}_{\rm G} \right) - \left( \op{P}_{\rm L} - \herm{\op{P}}_{\rm L} \right)  \:.
\end{equation}
According to Eqs.~\eqref{eq:def-gain-pval} and~\eqref{eq:def-loss-pval}, one can write
\begin{equation}\label{eq:total-pval-2}
\op{P} = \sum_{\vect{q}} \E^{\I\vect{q}\cdot\op{\vect{r}}} [\op{Y}_{\vect{q}}, \op{\rho}_\sys] \E^{-\I\vect{q}\cdot\op{\vect{r}}} - [\op{Y}_{\vect{q}}, \op{\rho}_\sys]  \:.
\end{equation}
In addition, the unitary operators $\E^{\pm\I\vect{q}\cdot\op{\vect{r}}}$ can be applied directly in the commutator:
\begin{equation}\label{eq:total-pval-3}
\op{P} = \sum_{\vect{q}} \left[ \E^{\I\vect{q}\cdot\op{\vect{r}}}\op{Y}_{\vect{q}}\E^{-\I\vect{q}\cdot\op{\vect{r}}},  \E^{\I\vect{q}\cdot\op{\vect{r}}}\op{\rho}_\sys\E^{-\I\vect{q}\cdot\op{\vect{r}}} \right] - [\op{Y}_{\vect{q}}, \op{\rho}_\sys]  \:.
\end{equation}
In order to get an estimate of Eq.~\eqref{eq:total-pval-3}, one considers the following rough approximations for the density matrices:
\begin{equation}\label{eq:small-q-approx}\begin{aligned}
\E^{\I\vect{q}\cdot\op{\vect{r}}}\op{\rho}_\sys\E^{-\I\vect{q}\cdot\op{\vect{r}}}  & \approx \op{\rho}_\sys   \:,\\
\E^{\I\vect{q}\cdot\op{\vect{r}}}\op{\rho}_\bath\E^{-\I\vect{q}\cdot\op{\vect{r}}} & \approx \op{\rho}_\bath  \:.
\end{aligned}\end{equation}
In principle, these approximations require that $q$ is much smaller than the inverse coherence length $\lcoh^{-1}$ defined in Eq.~\eqref{eq:def-coherence-length} for both the particle and the scatterers.
Under the first approximation of Eq.~\eqref{eq:small-q-approx}, Eq.~\eqref{eq:total-pval-3} can be written as the commutator
\begin{equation}\label{eq:total-pval-4}
\op{P} = [\op{\Pi}, \op{\rho}_\sys]  \:,
\end{equation}
where the operator $\op{\Pi}$ is defined as
\begin{equation}\label{eq:def-pval-pi-operator}
\op{\Pi} = \sum_{\vect{q}} \E^{\I\vect{q}\cdot\op{\vect{r}}}\op{Y}_{\vect{q}}\E^{-\I\vect{q}\cdot\op{\vect{r}}} - \op{Y}_{\vect{q}}  \:.
\end{equation}
Furthermore, under the second approximation of Eq.~\eqref{eq:small-q-approx}, one finds the nontrivial approximate property
\begin{equation}\label{eq:translated-pval-property}
\E^{\I\vect{q}\cdot\op{\vect{r}}} \op{Y}_{\vect{q}} \E^{-\I\vect{q}\cdot\op{\vect{r}}} \approx -\op{Y}_{-\vect{q}}  \:,
\end{equation}
which comes from definition~\eqref{eq:def-pval-operator} and the facts that
\begin{equation}\label{eq:translated-diff-property}\begin{split}
\E^{\I\vect{q}\cdot(\op{\vect{r}}-\op{\vect{x}}_\bath)} & \frac{1}{E_{\op{\vect{k}}_\sys + \vect{q}} + E_{\op{\vect{k}}_\bath - \vect{q}} - E_{\op{\vect{k}}_\sys} - E_{\op{\vect{k}}_\bath}} \E^{-\I\vect{q}\cdot(\op{\vect{r}}-\op{\vect{x}}_\bath)} \\
 = & \frac{1}{E_{\op{\vect{k}}_\sys} + E_{\op{\vect{k}}_\bath} - E_{\op{\vect{k}}_\sys - \vect{q}} - E_{\op{\vect{k}}_\bath + \vect{q}}}  \:,
\end{split}\end{equation}
and that $\abs{\fourier{u}(\vect{q})}^2=\abs{\fourier{u}(-\vect{q})}^2$.
If one uses the change of variable $\vect{q}\rightarrow-\vect{q}$ under the summation symbol, then Eq.~\eqref{eq:def-pval-pi-operator} reduces to
\begin{equation}\label{eq:pval-pi-operator-1}
\op{\Pi} = -2 \sum_{\vect{q}} \op{Y}_{\vect{q}}  \:.
\end{equation}
Using Eq.~\eqref{eq:def-pval-operator} and the notation $\tavg{\op{X}}_\bath=\Tr_\bath(\op{\rho}_\bath\op{X})$ for the average over the bath states, one gets
\begin{equation}\label{eq:pval-pi-operator-2}
\op{\Pi} = -\frac{2n}{\hbar V} \sum_{\vect{q}} \abs{\fourier{u}(\vect{q})}^2 \avg{\Pv\frac{1}{\op{D}_{\vect{q}}}}_\bath  \:.
\end{equation}
The sum over $\vect{q}$ can be replaced by a sum over the final momentum by means of $\vect{k}'=\vect{k}+\vect{q}$. One writes
\begin{equation}\label{eq:pval-pi-operator-3}
\op{\Pi} = -\frac{2n}{\hbar V} \frac{2m}{\hbar^2} \sum_{\vect{k}'} \avg{\abs{\fourier{u}(\vect{k}'-\op{\vect{k}})}^2 \Pv\frac{1}{{\vect{k}'}^2 - \op{\vect{k}}^2}}_\bath  \:,
\end{equation}
where $\op{\vect{k}}=(m_\bath\op{\vect{k}}_\sys - m_\sys\op{\vect{k}}_\bath)/(m_\sys+m_\bath)$ is the relative momentum operator according to Eq.~\eqref{eq:def-relative-momentum}.
In the continuum limit ($V\rightarrow\infty$) and splitting the integral into the radial and angular parts, Eq.~\eqref{eq:pval-pi-operator-3} reads
\begin{equation}\label{eq:pval-pi-operator-4}\begin{split}
\op{\Pi} = & -\frac{2n}{\hbar (2\pi)^d} \frac{2m}{\hbar^2} \int_0^\infty \D k'\: {k'}^{d-1} \oint_{\mathcal{S}_d} \D\Omega  \\
 & \times \avg{\abs{\fourier{u}(k'\vect{\Omega} - \op{\vect{k}})}^2 \Pv\frac{1}{{k'}^2 - \op{k}^2}}_\bath  \:.
\end{split}\end{equation}
To evaluate these integrals, it is convenient to generalize the differential cross section initially defined in Eq.~\eqref{eq:cross-section-from-potential} to collisions off the energy shell:
\begin{equation}\label{eq:def-off-shell-diff-csec}
\der{\sigma}{\Omega}(k'\vect{\Omega}\mid\vect{k}) = \frac{\pi}{2} \frac{{k'}^{d-3}}{(2\pi)^d} \abs{\frac{2m}{\hbar^2} \fourier{u}(k'\vect{\Omega}-\vect{\vect{k}})}^2  \:.
\end{equation}
In this way, the angular part of the integral in Eq.~\eqref{eq:pval-pi-operator-4} reduces to the off-shell total cross section:
\begin{equation}\label{eq:def-off-shell-total-csec}
\sigma(k'\mid\vect{k}) = \oint_{\mathcal{S}_d} \der{\sigma}{\Omega}(k'\vect{\Omega}\mid\vect{k}) \D\Omega  \:.
\end{equation}
Therefore, Eq.~\eqref{eq:pval-pi-operator-4} simplifies into
\begin{equation}\label{eq:pval-pi-operator-5}
\op{\Pi} = -\frac{2n}{\hbar} \frac{2}{\pi} \frac{\hbar^2}{2m} \Pv \int_0^\infty \D k' \avg{\frac{{k'}^2}{{k'}^2 - \op{k}^2} \sigma(k'\mid\op{\vect{k}})}_\bath \:.
\end{equation}
The remaining integral in Eq.~\eqref{eq:pval-pi-operator-5} cannot be found in closed form in the general case, because of the dependence on an unknown cross section $\sigma(k'\mid\op{\vect{k}})$.
Since one is seeking for an order of magnitude for $\op{\Pi}$, one supposes that the integral in Eq.~\eqref{eq:pval-pi-operator-5} is of the order of $\op{k}\sigma(\op{k})$, where $\sigma(k)$ is the on-shell total cross section from Eq.~\eqref{eq:def-total-cross-section}.
Indeed, one expects $\sigma(k'\mid\op{\vect{k}})$ in Eq.~\eqref{eq:pval-pi-operator-5} to have a peak around $k'=\op{k}$, and to quickly vanish when $k'$ strongly deviates from $\op{k}$.
Therefore, one finds the approximation
\begin{equation}\label{eq:pval-pi-operator-6}
\op{\Pi} = -\frac{2n}{\hbar} \frac{\hbar^2}{2m} C \avg{\op{k}\sigma(\op{k})}_\bath  \:,
\end{equation}
where $C$ is a dimensionless prefactor.
As this paper focuses on the case of fast particles, one assumes that the incident particle travels much faster than the scatterers ($v_\sys\gg v_\bath$).
Therefore, the relative velocity, $\op{v}=\norm{\op{\vect{v}}_\sys-\op{\vect{v}}_\bath}=\hbar\op{k}/m$, can be approximated by $\op{v}_\sys=\hbar\op{k}_\sys/m_\sys$, and one can write from Eq.~\eqref{eq:pval-pi-operator-6}
\begin{equation}\label{eq:pval-pi-operator-7}
\op{\Pi} = -\frac{2n}{\hbar} \frac{\hbar^2}{2m_\sys} C\sigma_0 \op{k}_\sys  \:,
\end{equation}
where $\sigma_0=\tavg{\sigma(\op{k})}_\bath$.
Note that, strictly speaking, the total cross section $\sigma_0$ still depends on $\op{k}_\sys$.
However, this dependency is neglected so that $\sigma_0$ is treated as a constant.
One last approximation is that the deviation of the particle momentum around some central momentum is small:
\begin{equation}\label{eq:small-straggling-approx-alt-1}
\op{\vect{k}}_\sys = k_{\sys,0}\vect{\Omega}_0 + \Delta\op{\vect{k}}_\sys  \quad\text{with}\quad
\norm{\Delta\op{\vect{k}}_\sys} \ll k_{\sys,0}  \:.
\end{equation}
In Eq.~\eqref{eq:small-straggling-approx-alt-1}, $k_{\sys,0}\vect{\Omega}_0$ denotes the average momentum of the incident particle.
This approximation is consistent with the high-velocity assumption for the particle.
According to Eq.~\eqref{eq:small-straggling-approx-alt-1}, any power $\gamma\in\mathbb{R}$ of the momentum $\op{k}_\sys$ can be approximated as follows
\begin{equation}\label{eq:small-straggling-approx-alt-2}
\op{k}_\sys^\gamma = k_{\sys,0}^\gamma + \gamma k_{\sys,0}^{\gamma-1}\vect{\Omega}_0\cdot\Delta\op{\vect{k}}_\sys + \bigo(\Delta\op{\vect{k}}_\sys^2)  \:.
\end{equation}
In particular, expansion~\eqref{eq:small-straggling-approx-alt-2} can be used for $\gamma=1$ in Eq.~\eqref{eq:pval-pi-operator-7}.
The commutator in Eq.~\eqref{eq:total-pval-4} then reads
\begin{equation}\label{eq:total-pval-5}
\op{P} = [\op{\Pi}, \op{\rho}_\sys] = -\frac{2n}{\hbar} \frac{\hbar^2}{2m_\sys} C \sigma_0 \vect{\Omega}_0\cdot[\Delta\op{\vect{k}}_\sys, \op{\rho}_\sys]  \:.
\end{equation}
Expression~\eqref{eq:total-pval-5} turns out to be closely similar to the free propagation term $[\op{H}_\sys, \op{\rho}_\sys]$ in the quantum Liouville equation.
This similarity becomes even more apparent if one uses the approximation~\eqref{eq:small-straggling-approx-alt-2} for $\gamma=2$ to approach the Hamiltonian $\op{H}_\sys=\tfrac{\hbar^2\op{k}_\sys^2}{2m_\sys}$.
The result is
\begin{equation}\label{eq:free-hamiltonian-comparison}
[\op{H}_\sys, \op{\rho}_\sys] = \frac{\hbar^2}{2m_\sys} 2k_{\sys,0}\vect{\Omega}_0\cdot[\Delta\op{\vect{k}}_\sys, \op{\rho}_\sys]  \:.
\end{equation}
In fact, Eq.~\eqref{eq:free-hamiltonian-comparison} is proportional to Eq.~\eqref{eq:total-pval-5} according to
\begin{equation}\label{eq:total-pval-final}
\op{P} = [\op{\Pi}, \op{\rho}_\sys] = -\frac{1}{\hbar} C \frac{n\sigma_0}{k_{\sys,0}} [\op{H}_\sys, \op{\rho}_\sys]  \:.
\end{equation}
Therefore, the contribution of the principal value terms to the Redfield equation~\eqref{eq:redfield-expansion-3} can be approached by
\begin{equation}\label{eq:pval-contrib-final}
\pder{\op{\rho}_\sys}{t} = \left(1 - C \frac{n\sigma_0}{k_{\sys,0}}\right) \supop{L}_\sys\op{\rho}_\sys + \op{R}  \:.
\end{equation}
This result shows that the principal value terms affect the propagation velocity of the particle by a correction of the order of the dimensionless factor $Cn\sigma_0/k_{\sys,0}$.
This correction is small under the condition
\begin{equation}\label{eq:weak-scattering-alt}
\frac{n\sigma_0}{k_{\sys,0}} \ll 1  \:,
\end{equation}
which is equivalent to the weak scattering condition~\eqref{eq:weak-scattering}.
In this regime, the principal value terms are negligible compared to the free motion of the particle.
Since condition~\eqref{eq:weak-scattering-alt} is typically fulfilled for particles of a few MeVs in a gas, the approximation made near Eqs.~\eqref{eq:redfield-expansion-3} and~\eqref{eq:simplified-redfield} is well justified.


\begin{thebibliography}{81}%
\makeatletter
\providecommand \@ifxundefined [1]{%
 \@ifx{#1\undefined}
}%
\providecommand \@ifnum [1]{%
 \ifnum #1\expandafter \@firstoftwo
 \else \expandafter \@secondoftwo
 \fi
}%
\providecommand \@ifx [1]{%
 \ifx #1\expandafter \@firstoftwo
 \else \expandafter \@secondoftwo
 \fi
}%
\providecommand \natexlab [1]{#1}%
\providecommand \enquote  [1]{``#1''}%
\providecommand \bibnamefont  [1]{#1}%
\providecommand \bibfnamefont [1]{#1}%
\providecommand \citenamefont [1]{#1}%
\providecommand \href@noop [0]{\@secondoftwo}%
\providecommand \href [0]{\begingroup \@sanitize@url \@href}%
\providecommand \@href[1]{\@@startlink{#1}\@@href}%
\providecommand \@@href[1]{\endgroup#1\@@endlink}%
\providecommand \@sanitize@url [0]{\catcode `\\12\catcode `\$12\catcode
  `\&12\catcode `\#12\catcode `\^12\catcode `\_12\catcode `\%12\relax}%
\providecommand \@@startlink[1]{}%
\providecommand \@@endlink[0]{}%
\providecommand \url  [0]{\begingroup\@sanitize@url \@url }%
\providecommand \@url [1]{\endgroup\@href {#1}{\urlprefix }}%
\providecommand \urlprefix  [0]{URL }%
\providecommand \Eprint [0]{\href }%
\providecommand \doibase [0]{https://doi.org/}%
\providecommand \selectlanguage [0]{\@gobble}%
\providecommand \bibinfo  [0]{\@secondoftwo}%
\providecommand \bibfield  [0]{\@secondoftwo}%
\providecommand \translation [1]{[#1]}%
\providecommand \BibitemOpen [0]{}%
\providecommand \bibitemStop [0]{}%
\providecommand \bibitemNoStop [0]{.\EOS\space}%
\providecommand \EOS [0]{\spacefactor3000\relax}%
\providecommand \BibitemShut  [1]{\csname bibitem#1\endcsname}%
\let\auto@bib@innerbib\@empty
\bibitem [{\citenamefont {Gardiner}\ and\ \citenamefont
  {Zoller}(2000)}]{Gardiner2000}%
  \BibitemOpen
  \bibfield  {author} {\bibinfo {author} {\bibfnamefont {C.~W.}\ \bibnamefont
  {Gardiner}}\ and\ \bibinfo {author} {\bibfnamefont {P.}~\bibnamefont
  {Zoller}},\ }\href {https://link.springer.com/book/9783540223016} {\emph
  {\bibinfo {title} {Quantum Noise: A Handbook of Markovian and Non-Markovian
  Quantum Stochastic Methods with Applications to Quantum Optics}}},\ \bibinfo
  {edition} {2nd}\ ed.,\ \bibinfo {series} {Springer Series in Synergetics},
  Vol.~\bibinfo {volume} {56}\ (\bibinfo  {publisher} {Springer},\ \bibinfo
  {address} {Berlin},\ \bibinfo {year} {2000})\BibitemShut {NoStop}%
\bibitem [{\citenamefont {Breuer}\ and\ \citenamefont
  {Petruccione}(2002)}]{Breuer2002}%
  \BibitemOpen
  \bibfield  {author} {\bibinfo {author} {\bibfnamefont {H.-P.}\ \bibnamefont
  {Breuer}}\ and\ \bibinfo {author} {\bibfnamefont {F.}~\bibnamefont
  {Petruccione}},\ }\href
  {https://doi.org/10.1093/acprof:oso/9780199213900.001.0001} {\emph {\bibinfo
  {title} {The Theory of Open Quantum Systems}}}\ (\bibinfo  {publisher}
  {Oxford University Press},\ \bibinfo {year} {2002})\BibitemShut {NoStop}%
\bibitem [{\citenamefont {Weiss}(2008)}]{Weiss2008}%
  \BibitemOpen
  \bibfield  {author} {\bibinfo {author} {\bibfnamefont {U.}~\bibnamefont
  {Weiss}},\ }\href {https://doi.org/10.1142/6738} {\emph {\bibinfo {title}
  {Quantum Dissipative Systems}}},\ \bibinfo {edition} {3rd}\ ed.,\ \bibinfo
  {series} {Series in Modern Condensed Matter Physics}, Vol.~\bibinfo {volume}
  {13}\ (\bibinfo  {publisher} {World Scientific},\ \bibinfo {address}
  {Singapore},\ \bibinfo {year} {2008})\BibitemShut {NoStop}%
\bibitem [{\citenamefont {Redfield}(1957)}]{Redfield1957}%
  \BibitemOpen
  \bibfield  {author} {\bibinfo {author} {\bibfnamefont {A.~G.}\ \bibnamefont
  {Redfield}},\ }\href {https://doi.org/10.1147/rd.11.0019} {\bibfield
  {journal} {\bibinfo  {journal} {IBM J. Res. Dev.}\ }\textbf {\bibinfo
  {volume} {1}},\ \bibinfo {pages} {19} (\bibinfo {year} {1957})}\BibitemShut
  {NoStop}%
\bibitem [{\citenamefont {Redfield}(1965)}]{Redfield1965}%
  \BibitemOpen
  \bibfield  {author} {\bibinfo {author} {\bibfnamefont {A.~G.}\ \bibnamefont
  {Redfield}},\ }\href {https://doi.org/10.1016/B978-1-4832-3114-3.50007-6}
  {\bibfield  {journal} {\bibinfo  {journal} {Adv. Magn. Opt. Reson.}\ }\textbf
  {\bibinfo {volume} {1}},\ \bibinfo {pages} {1} (\bibinfo {year}
  {1965})}\BibitemShut {NoStop}%
\bibitem [{\citenamefont {Pechukas}(1994)}]{Pechukas1994}%
  \BibitemOpen
  \bibfield  {author} {\bibinfo {author} {\bibfnamefont {P.}~\bibnamefont
  {Pechukas}},\ }\href {https://doi.org/10.1103/PhysRevLett.73.1060} {\bibfield
   {journal} {\bibinfo  {journal} {Phys. Rev. Lett.}\ }\textbf {\bibinfo
  {volume} {73}},\ \bibinfo {pages} {1060} (\bibinfo {year}
  {1994})}\BibitemShut {NoStop}%
\bibitem [{\citenamefont {Gaspard}\ and\ \citenamefont
  {Nagaoka}(1999)}]{GaspardP1999a}%
  \BibitemOpen
  \bibfield  {author} {\bibinfo {author} {\bibfnamefont {P.}~\bibnamefont
  {Gaspard}}\ and\ \bibinfo {author} {\bibfnamefont {M.}~\bibnamefont
  {Nagaoka}},\ }\href {https://doi.org/10.1063/1.479867} {\bibfield  {journal}
  {\bibinfo  {journal} {J. Chem. Phys.}\ }\textbf {\bibinfo {volume} {111}},\
  \bibinfo {pages} {5668} (\bibinfo {year} {1999})}\BibitemShut {NoStop}%
\bibitem [{\citenamefont {Farina}\ and\ \citenamefont
  {Giovannetti}(2019)}]{Farina2019}%
  \BibitemOpen
  \bibfield  {author} {\bibinfo {author} {\bibfnamefont {D.}~\bibnamefont
  {Farina}}\ and\ \bibinfo {author} {\bibfnamefont {V.}~\bibnamefont
  {Giovannetti}},\ }\href {https://doi.org/10.1103/PhysRevA.100.012107}
  {\bibfield  {journal} {\bibinfo  {journal} {Phys. Rev. A}\ }\textbf {\bibinfo
  {volume} {100}},\ \bibinfo {pages} {012107} (\bibinfo {year}
  {2019})}\BibitemShut {NoStop}%
\bibitem [{\citenamefont {Mozgunov}\ and\ \citenamefont
  {Lidar}(2020)}]{Mozgunov2020}%
  \BibitemOpen
  \bibfield  {author} {\bibinfo {author} {\bibfnamefont {E.}~\bibnamefont
  {Mozgunov}}\ and\ \bibinfo {author} {\bibfnamefont {D.}~\bibnamefont
  {Lidar}},\ }\href {https://doi.org/10.22331/q-2020-02-06-227} {\bibfield
  {journal} {\bibinfo  {journal} {Quantum}\ }\textbf {\bibinfo {volume} {4}},\
  \bibinfo {pages} {227} (\bibinfo {year} {2020})}\BibitemShut {NoStop}%
\bibitem [{\citenamefont {Davidovi\'c}(2020)}]{Davidovic2020}%
  \BibitemOpen
  \bibfield  {author} {\bibinfo {author} {\bibfnamefont {D.}~\bibnamefont
  {Davidovi\'c}},\ }\href {https://doi.org/10.22331/q-2020-09-21-326}
  {\bibfield  {journal} {\bibinfo  {journal} {Quantum}\ }\textbf {\bibinfo
  {volume} {4}},\ \bibinfo {pages} {326} (\bibinfo {year} {2020})}\BibitemShut
  {NoStop}%
\bibitem [{\citenamefont {Hartmann}\ and\ \citenamefont
  {Strunz}(2020)}]{Hartmann2020}%
  \BibitemOpen
  \bibfield  {author} {\bibinfo {author} {\bibfnamefont {R.}~\bibnamefont
  {Hartmann}}\ and\ \bibinfo {author} {\bibfnamefont {W.~T.}\ \bibnamefont
  {Strunz}},\ }\href {https://doi.org/10.1103/PhysRevA.101.012103} {\bibfield
  {journal} {\bibinfo  {journal} {Phys. Rev. A}\ }\textbf {\bibinfo {volume}
  {101}},\ \bibinfo {pages} {012103} (\bibinfo {year} {2020})}\BibitemShut
  {NoStop}%
\bibitem [{\citenamefont {Gorini}\ \emph {et~al.}(1976)\citenamefont {Gorini},
  \citenamefont {Kossakowski},\ and\ \citenamefont {Sudarshan}}]{Gorini1976}%
  \BibitemOpen
  \bibfield  {author} {\bibinfo {author} {\bibfnamefont {V.}~\bibnamefont
  {Gorini}}, \bibinfo {author} {\bibfnamefont {A.}~\bibnamefont
  {Kossakowski}},\ and\ \bibinfo {author} {\bibfnamefont {E.~C.~G.}\
  \bibnamefont {Sudarshan}},\ }\href {https://doi.org/10.1063/1.522979}
  {\bibfield  {journal} {\bibinfo  {journal} {J. Math. Phys.}\ }\textbf
  {\bibinfo {volume} {17}},\ \bibinfo {pages} {821} (\bibinfo {year}
  {1976})}\BibitemShut {NoStop}%
\bibitem [{\citenamefont {Lindblad}(1976)}]{Lindblad1976}%
  \BibitemOpen
  \bibfield  {author} {\bibinfo {author} {\bibfnamefont {G.}~\bibnamefont
  {Lindblad}},\ }\href {https://doi.org/10.1007/BF01608499} {\bibfield
  {journal} {\bibinfo  {journal} {Commun. Math. Phys.}\ }\textbf {\bibinfo
  {volume} {48}},\ \bibinfo {pages} {119} (\bibinfo {year} {1976})}\BibitemShut
  {NoStop}%
\bibitem [{\citenamefont {Manzano}(2020)}]{Manzano2020}%
  \BibitemOpen
  \bibfield  {author} {\bibinfo {author} {\bibfnamefont {D.}~\bibnamefont
  {Manzano}},\ }\href {https://doi.org/10.1063/1.5115323} {\bibfield  {journal}
  {\bibinfo  {journal} {AIP Adv.}\ }\textbf {\bibinfo {volume} {10}},\ \bibinfo
  {pages} {025106} (\bibinfo {year} {2020})}\BibitemShut {NoStop}%
\bibitem [{\citenamefont {Boltzmann}(1872)}]{Boltzmann1872}%
  \BibitemOpen
  \bibfield  {author} {\bibinfo {author} {\bibfnamefont {L.}~\bibnamefont
  {Boltzmann}},\ }\href@noop {} {\bibfield  {journal} {\bibinfo  {journal}
  {Wiener Berichte}\ }\textbf {\bibinfo {volume} {66}},\ \bibinfo {pages} {275}
  (\bibinfo {year} {1872})},\ \bibinfo {note} {in German}\BibitemShut {NoStop}%
\bibitem [{\citenamefont {Weinberg}\ and\ \citenamefont
  {Wigner}(1958)}]{Weinberg1958}%
  \BibitemOpen
  \bibfield  {author} {\bibinfo {author} {\bibfnamefont {A.~M.}\ \bibnamefont
  {Weinberg}}\ and\ \bibinfo {author} {\bibfnamefont {E.~P.}\ \bibnamefont
  {Wigner}},\ }\href {https://books.google.com/books?vid=ISBN9780226885179}
  {\emph {\bibinfo {title} {The Physical Theory of Neutron Chain Reactors}}},\
  University of Chicago Committee on Publications in the Physical Sciences\
  (\bibinfo  {publisher} {University of Chicago Press},\ \bibinfo {year}
  {1958})\BibitemShut {NoStop}%
\bibitem [{\citenamefont {Balescu}(1975)}]{Balescu1975}%
  \BibitemOpen
  \bibfield  {author} {\bibinfo {author} {\bibfnamefont {R.}~\bibnamefont
  {Balescu}},\ }\href {https://books.google.com/books?vid=ISBN9780471046004}
  {\emph {\bibinfo {title} {Equilibrium and Non-Equilibrium Statistical
  Mechanics}}},\ A Wiley interscience publication\ (\bibinfo  {publisher}
  {Wiley},\ \bibinfo {address} {New York},\ \bibinfo {year} {1975})\BibitemShut
  {NoStop}%
\bibitem [{\citenamefont {Huang}(1987)}]{Huang1987}%
  \BibitemOpen
  \bibfield  {author} {\bibinfo {author} {\bibfnamefont {K.}~\bibnamefont
  {Huang}},\ }\href {https://books.google.com/books?vid=ISBN9780471815181}
  {\emph {\bibinfo {title} {Statistical Mechanics}}},\ \bibinfo {edition}
  {2nd}\ ed.\ (\bibinfo  {publisher} {Wiley},\ \bibinfo {address} {New York},\
  \bibinfo {year} {1987})\BibitemShut {NoStop}%
\bibitem [{\citenamefont {Harris}(2004)}]{Harris2004}%
  \BibitemOpen
  \bibfield  {author} {\bibinfo {author} {\bibfnamefont {S.}~\bibnamefont
  {Harris}},\ }\href {https://books.google.com/books?vid=ISBN9780486438313}
  {\emph {\bibinfo {title} {An Introduction to the Theory of the Boltzmann
  Equation}}},\ Dover books on physics\ (\bibinfo  {publisher} {Dover
  Publications},\ \bibinfo {address} {Mineola},\ \bibinfo {year}
  {2004})\BibitemShut {NoStop}%
\bibitem [{\citenamefont {Nordheim}(1928)}]{Nordheim1928}%
  \BibitemOpen
  \bibfield  {author} {\bibinfo {author} {\bibfnamefont {L.~W.}\ \bibnamefont
  {Nordheim}},\ }\href {https://doi.org/10.1098/rspa.1928.0126} {\bibfield
  {journal} {\bibinfo  {journal} {Proc. R. Soc. A}\ }\textbf {\bibinfo {volume}
  {119}},\ \bibinfo {pages} {689} (\bibinfo {year} {1928})}\BibitemShut
  {NoStop}%
\bibitem [{\citenamefont {Uehling}\ and\ \citenamefont
  {Uhlenbeck}(1933)}]{Uehling1933}%
  \BibitemOpen
  \bibfield  {author} {\bibinfo {author} {\bibfnamefont {E.~A.}\ \bibnamefont
  {Uehling}}\ and\ \bibinfo {author} {\bibfnamefont {G.~E.}\ \bibnamefont
  {Uhlenbeck}},\ }\href {https://doi.org/10.1103/PhysRev.43.552} {\bibfield
  {journal} {\bibinfo  {journal} {Phys. Rev.}\ }\textbf {\bibinfo {volume}
  {43}},\ \bibinfo {pages} {552} (\bibinfo {year} {1933})}\BibitemShut
  {NoStop}%
\bibitem [{\citenamefont {Kadanoff}\ and\ \citenamefont
  {Baym}(1962)}]{Kadanoff1962}%
  \BibitemOpen
  \bibfield  {author} {\bibinfo {author} {\bibfnamefont {L.~P.}\ \bibnamefont
  {Kadanoff}}\ and\ \bibinfo {author} {\bibfnamefont {G.}~\bibnamefont
  {Baym}},\ }\href {https://doi.org/10.1201/9780429493218} {\emph {\bibinfo
  {title} {Quantum Statistical Mechanics: Green's Function Methods in
  Equilibrium and Nonequilibrium Problems}}}\ (\bibinfo  {publisher} {W. A.
  Benjamin},\ \bibinfo {address} {Boca Raton, Florida},\ \bibinfo {year}
  {1962})\BibitemShut {NoStop}%
\bibitem [{\citenamefont {Schlosshauer}(2019)}]{Schlosshauer2019}%
  \BibitemOpen
  \bibfield  {author} {\bibinfo {author} {\bibfnamefont {M.}~\bibnamefont
  {Schlosshauer}},\ }\href {https://doi.org/10.1016/j.physrep.2019.10.001}
  {\bibfield  {journal} {\bibinfo  {journal} {Phys. Rep.}\ }\textbf {\bibinfo
  {volume} {831}},\ \bibinfo {pages} {1} (\bibinfo {year} {2019})}\BibitemShut
  {NoStop}%
\bibitem [{\citenamefont {Hornberger}\ \emph {et~al.}(2003)\citenamefont
  {Hornberger}, \citenamefont {Uttenthaler}, \citenamefont {Brezger},
  \citenamefont {Hackerm\"{u}ller}, \citenamefont {Arndt},\ and\ \citenamefont
  {Zeilinger}}]{Hornberger2003a}%
  \BibitemOpen
  \bibfield  {author} {\bibinfo {author} {\bibfnamefont {K.}~\bibnamefont
  {Hornberger}}, \bibinfo {author} {\bibfnamefont {S.}~\bibnamefont
  {Uttenthaler}}, \bibinfo {author} {\bibfnamefont {B.}~\bibnamefont
  {Brezger}}, \bibinfo {author} {\bibfnamefont {L.}~\bibnamefont
  {Hackerm\"{u}ller}}, \bibinfo {author} {\bibfnamefont {M.}~\bibnamefont
  {Arndt}},\ and\ \bibinfo {author} {\bibfnamefont {A.}~\bibnamefont
  {Zeilinger}},\ }\href {https://doi.org/10.1103/PhysRevLett.90.160401}
  {\bibfield  {journal} {\bibinfo  {journal} {Phys. Rev. Lett.}\ }\textbf
  {\bibinfo {volume} {90}},\ \bibinfo {pages} {160401} (\bibinfo {year}
  {2003})}\BibitemShut {NoStop}%
\bibitem [{\citenamefont {Hornberger}\ and\ \citenamefont
  {Sipe}(2003)}]{Hornberger2003b}%
  \BibitemOpen
  \bibfield  {author} {\bibinfo {author} {\bibfnamefont {K.}~\bibnamefont
  {Hornberger}}\ and\ \bibinfo {author} {\bibfnamefont {J.~E.}\ \bibnamefont
  {Sipe}},\ }\href {https://doi.org/10.1103/PhysRevA.68.012105} {\bibfield
  {journal} {\bibinfo  {journal} {Phys. Rev. A}\ }\textbf {\bibinfo {volume}
  {68}},\ \bibinfo {pages} {012105} (\bibinfo {year} {2003})}\BibitemShut
  {NoStop}%
\bibitem [{\citenamefont {Hornberger}(2006)}]{Hornberger2006a}%
  \BibitemOpen
  \bibfield  {author} {\bibinfo {author} {\bibfnamefont {K.}~\bibnamefont
  {Hornberger}},\ }\href {https://doi.org/10.1103/PhysRevLett.97.060601}
  {\bibfield  {journal} {\bibinfo  {journal} {Phys. Rev. Lett.}\ }\textbf
  {\bibinfo {volume} {97}},\ \bibinfo {pages} {060601} (\bibinfo {year}
  {2006})}\BibitemShut {NoStop}%
\bibitem [{\citenamefont {Hornberger}\ and\ \citenamefont
  {Vacchini}(2008)}]{Hornberger2008}%
  \BibitemOpen
  \bibfield  {author} {\bibinfo {author} {\bibfnamefont {K.}~\bibnamefont
  {Hornberger}}\ and\ \bibinfo {author} {\bibfnamefont {B.}~\bibnamefont
  {Vacchini}},\ }\href {https://doi.org/10.1103/PhysRevA.77.022112} {\bibfield
  {journal} {\bibinfo  {journal} {Phys. Rev. A}\ }\textbf {\bibinfo {volume}
  {77}},\ \bibinfo {pages} {022112} (\bibinfo {year} {2008})}\BibitemShut
  {NoStop}%
\bibitem [{\citenamefont {Vacchini}\ and\ \citenamefont
  {Hornberger}(2009)}]{Vacchini2009}%
  \BibitemOpen
  \bibfield  {author} {\bibinfo {author} {\bibfnamefont {B.}~\bibnamefont
  {Vacchini}}\ and\ \bibinfo {author} {\bibfnamefont {K.}~\bibnamefont
  {Hornberger}},\ }\href {https://doi.org/10.1016/j.physrep.2009.06.001}
  {\bibfield  {journal} {\bibinfo  {journal} {Phys. Rep.}\ }\textbf {\bibinfo
  {volume} {478}},\ \bibinfo {pages} {71} (\bibinfo {year} {2009})}\BibitemShut
  {NoStop}%
\bibitem [{\citenamefont {Hornberger}(2009)}]{Hornberger2009}%
  \BibitemOpen
  \bibfield  {author} {\bibinfo {author} {\bibfnamefont {K.}~\bibnamefont
  {Hornberger}},\ }in\ \href {https://doi.org/10.1007/978-3-540-88169-8} {\emph
  {\bibinfo {booktitle} {Entanglement and Decoherence: Foundations and Modern
  Trends}}},\ \bibinfo {series} {Lecture Notes in Physics}, Vol.\ \bibinfo
  {volume} {768},\ \bibinfo {editor} {edited by\ \bibinfo {editor}
  {\bibfnamefont {A.}~\bibnamefont {Buchleitner}}, \bibinfo {editor}
  {\bibfnamefont {C.}~\bibnamefont {Viviescas}},\ and\ \bibinfo {editor}
  {\bibfnamefont {M.}~\bibnamefont {Tiersch}}}\ (\bibinfo  {publisher}
  {Springer},\ \bibinfo {address} {Berlin, Heidelberg},\ \bibinfo {year}
  {2009})\ Chap.~\bibinfo {chapter} {5}\BibitemShut {NoStop}%
\bibitem [{\citenamefont {Di\'osi}(1995)}]{Diosi1995}%
  \BibitemOpen
  \bibfield  {author} {\bibinfo {author} {\bibfnamefont {L.}~\bibnamefont
  {Di\'osi}},\ }\href {https://doi.org/10.1209/0295-5075/30/2/001} {\bibfield
  {journal} {\bibinfo  {journal} {Europhys. Lett.}\ }\textbf {\bibinfo {volume}
  {30}},\ \bibinfo {pages} {63} (\bibinfo {year} {1995})}\BibitemShut {NoStop}%
\bibitem [{\citenamefont {Di\'osi}(2009)}]{Diosi2009}%
  \BibitemOpen
  \bibfield  {author} {\bibinfo {author} {\bibfnamefont {L.}~\bibnamefont
  {Di\'osi}},\ }\href {https://doi.org/10.1103/PhysRevA.80.064104} {\bibfield
  {journal} {\bibinfo  {journal} {Phys. Rev. A}\ }\textbf {\bibinfo {volume}
  {80}},\ \bibinfo {pages} {064104} (\bibinfo {year} {2009})}\BibitemShut
  {NoStop}%
\bibitem [{\citenamefont {Di\'osi}(2022)}]{Diosi2022}%
  \BibitemOpen
  \bibfield  {author} {\bibinfo {author} {\bibfnamefont {L.}~\bibnamefont
  {Di\'osi}},\ }in\ \href {https://doi.org/10.1007/978-3-030-88781-0_10} {\emph
  {\bibinfo {booktitle} {From Quantum to Classical: Essays in Honour of
  H.-Dieter Zeh}}},\ \bibinfo {editor} {edited by\ \bibinfo {editor}
  {\bibfnamefont {C.}~\bibnamefont {Kiefer}}}\ (\bibinfo  {publisher}
  {Springer},\ \bibinfo {address} {Cham},\ \bibinfo {year} {2022})\ pp.\
  \bibinfo {pages} {217--224}\BibitemShut {NoStop}%
\bibitem [{\citenamefont {Halliwell}(2007)}]{Halliwell2007}%
  \BibitemOpen
  \bibfield  {author} {\bibinfo {author} {\bibfnamefont {J.~J.}\ \bibnamefont
  {Halliwell}},\ }\href {https://doi.org/10.1088/1751-8113/40/12/s11}
  {\bibfield  {journal} {\bibinfo  {journal} {J. Phys. A: Math. Theor.}\
  }\textbf {\bibinfo {volume} {40}},\ \bibinfo {pages} {3067} (\bibinfo {year}
  {2007})}\BibitemShut {NoStop}%
\bibitem [{\citenamefont {Kamleitner}(2010)}]{Kamleitner2010}%
  \BibitemOpen
  \bibfield  {author} {\bibinfo {author} {\bibfnamefont {I.}~\bibnamefont
  {Kamleitner}},\ }\emph {\bibinfo {title} {Open Quantum System Dynamics from a
  Measurement Perspective: Applications to Coherent Particle Transport and to
  Quantum Brownian Motion}},\ \href {https://arxiv.org/abs/1009.4349} {Ph.D.
  thesis},\ \bibinfo  {school} {Macquarie University} (\bibinfo {year}
  {2010}),\ \Eprint {https://arxiv.org/abs/1009.4349} {arXiv:1009.4349
  [quant-ph]} \BibitemShut {NoStop}%
\bibitem [{\citenamefont {Zeh}(1970)}]{Zeh1970}%
  \BibitemOpen
  \bibfield  {author} {\bibinfo {author} {\bibfnamefont {H.~D.}\ \bibnamefont
  {Zeh}},\ }\href {https://doi.org/10.1007/BF00708656} {\bibfield  {journal}
  {\bibinfo  {journal} {Found. Phys.}\ }\textbf {\bibinfo {volume} {1}},\
  \bibinfo {pages} {69} (\bibinfo {year} {1970})}\BibitemShut {NoStop}%
\bibitem [{\citenamefont {Zeh}(1973)}]{Zeh1973}%
  \BibitemOpen
  \bibfield  {author} {\bibinfo {author} {\bibfnamefont {H.~D.}\ \bibnamefont
  {Zeh}},\ }\href {https://doi.org/10.1007/BF00708603} {\bibfield  {journal}
  {\bibinfo  {journal} {Found. Phys.}\ }\textbf {\bibinfo {volume} {3}},\
  \bibinfo {pages} {109} (\bibinfo {year} {1973})}\BibitemShut {NoStop}%
\bibitem [{\citenamefont {Joos}\ and\ \citenamefont {Zeh}(1985)}]{Joos1985}%
  \BibitemOpen
  \bibfield  {author} {\bibinfo {author} {\bibfnamefont {E.}~\bibnamefont
  {Joos}}\ and\ \bibinfo {author} {\bibfnamefont {H.~D.}\ \bibnamefont {Zeh}},\
  }\href {https://doi.org/10.1007/BF01725541} {\bibfield  {journal} {\bibinfo
  {journal} {Z. Phys. B}\ }\textbf {\bibinfo {volume} {59}},\ \bibinfo {pages}
  {223} (\bibinfo {year} {1985})}\BibitemShut {NoStop}%
\bibitem [{\citenamefont {Zurek}(1991)}]{Zurek1991}%
  \BibitemOpen
  \bibfield  {author} {\bibinfo {author} {\bibfnamefont {W.~H.}\ \bibnamefont
  {Zurek}},\ }\href {https://doi.org/10.1063/1.881293} {\bibfield  {journal}
  {\bibinfo  {journal} {Phys. Today}\ }\textbf {\bibinfo {volume} {44}},\
  \bibinfo {pages} {36} (\bibinfo {year} {1991})},\ \bibinfo {note} {updated
  version
  \href{https://doi.org/10.48550/arXiv.quant-ph/0306072}{arXiv:quant-ph/0306072
  (2003)}}\BibitemShut {NoStop}%
\bibitem [{\citenamefont {Kiefer}\ and\ \citenamefont
  {Joos}(1999)}]{Kiefer1999}%
  \BibitemOpen
  \bibfield  {author} {\bibinfo {author} {\bibfnamefont {C.}~\bibnamefont
  {Kiefer}}\ and\ \bibinfo {author} {\bibfnamefont {E.}~\bibnamefont {Joos}},\
  }in\ \href {https://doi.org/10.1007/BFb0105342} {\emph {\bibinfo {booktitle}
  {Quantum Future From Volta and Como to the Present and Beyond}}},\ \bibinfo
  {series} {Lecture Notes in Physics}, Vol.\ \bibinfo {volume} {517},\ \bibinfo
  {editor} {edited by\ \bibinfo {editor} {\bibfnamefont {P.}~\bibnamefont
  {Blanchard}}\ and\ \bibinfo {editor} {\bibfnamefont {A.}~\bibnamefont
  {Jadczyk}}}\ (\bibinfo  {publisher} {Springer},\ \bibinfo {address} {Berlin,
  Heidelberg},\ \bibinfo {year} {1999})\ pp.\ \bibinfo {pages} {105--128},\
  \bibinfo {note} {proceedings of the 10th Born Symposium},\ \Eprint
  {https://arxiv.org/abs/quant-ph/9803052v1} {arXiv:quant-ph/9803052v1
  [quant-ph]} \BibitemShut {NoStop}%
\bibitem [{\citenamefont {Allahverdyan}\ \emph {et~al.}(2013)\citenamefont
  {Allahverdyan}, \citenamefont {Balian},\ and\ \citenamefont
  {Nieuwenhuizen}}]{Allahverdyan2013}%
  \BibitemOpen
  \bibfield  {author} {\bibinfo {author} {\bibfnamefont {A.~E.}\ \bibnamefont
  {Allahverdyan}}, \bibinfo {author} {\bibfnamefont {R.}~\bibnamefont
  {Balian}},\ and\ \bibinfo {author} {\bibfnamefont {T.~M.}\ \bibnamefont
  {Nieuwenhuizen}},\ }\href {https://doi.org/10.1016/j.physrep.2012.11.001}
  {\bibfield  {journal} {\bibinfo  {journal} {Phys. Rep.}\ }\textbf {\bibinfo
  {volume} {525}},\ \bibinfo {pages} {1} (\bibinfo {year} {2013})}\BibitemShut
  {NoStop}%
\bibitem [{\citenamefont {Arndt}\ \emph {et~al.}(1999)\citenamefont {Arndt},
  \citenamefont {Nairz}, \citenamefont {Voss-Andreae}, \citenamefont {Keller},
  \citenamefont {Van-Der-Zouw},\ and\ \citenamefont {Zeilinger}}]{Arndt1999}%
  \BibitemOpen
  \bibfield  {author} {\bibinfo {author} {\bibfnamefont {M.}~\bibnamefont
  {Arndt}}, \bibinfo {author} {\bibfnamefont {O.}~\bibnamefont {Nairz}},
  \bibinfo {author} {\bibfnamefont {J.}~\bibnamefont {Voss-Andreae}}, \bibinfo
  {author} {\bibfnamefont {C.}~\bibnamefont {Keller}}, \bibinfo {author}
  {\bibfnamefont {G.}~\bibnamefont {Van-Der-Zouw}},\ and\ \bibinfo {author}
  {\bibfnamefont {A.}~\bibnamefont {Zeilinger}},\ }\href
  {https://doi.org/10.1038/44348} {\bibfield  {journal} {\bibinfo  {journal}
  {Nature}\ }\textbf {\bibinfo {volume} {401}},\ \bibinfo {pages} {680}
  (\bibinfo {year} {1999})}\BibitemShut {NoStop}%
\bibitem [{\citenamefont {Arndt}\ and\ \citenamefont
  {Hornberger}(2009)}]{Arndt2009-arxiv}%
  \BibitemOpen
  \bibfield  {author} {\bibinfo {author} {\bibfnamefont {M.}~\bibnamefont
  {Arndt}}\ and\ \bibinfo {author} {\bibfnamefont {K.}~\bibnamefont
  {Hornberger}},\ }\Eprint {https://arxiv.org/abs/0903.1614} {arXiv:0903.1614
  [quant-ph]}  (\bibinfo {year} {2009}),\ \bibinfo {note} {published in B.
  Deveaud-Pledran, A. Quattropani, P. Schwendimann (Eds.), Quantum Coherence in
  Solid State Systems, International School of Physics ``Enrico Fermi'', Course
  CLXXI, Vol. 171, (IOS press, Amsterdam, 2009)}\BibitemShut {NoStop}%
\bibitem [{\citenamefont {Hackerm\"uller}\ \emph {et~al.}(2003)\citenamefont
  {Hackerm\"uller}, \citenamefont {Uttenthaler}, \citenamefont {Hornberger},
  \citenamefont {Reiger}, \citenamefont {Brezger}, \citenamefont {Zeilinger},\
  and\ \citenamefont {Arndt}}]{Hackermuller2003}%
  \BibitemOpen
  \bibfield  {author} {\bibinfo {author} {\bibfnamefont {L.}~\bibnamefont
  {Hackerm\"uller}}, \bibinfo {author} {\bibfnamefont {S.}~\bibnamefont
  {Uttenthaler}}, \bibinfo {author} {\bibfnamefont {K.}~\bibnamefont
  {Hornberger}}, \bibinfo {author} {\bibfnamefont {E.}~\bibnamefont {Reiger}},
  \bibinfo {author} {\bibfnamefont {B.}~\bibnamefont {Brezger}}, \bibinfo
  {author} {\bibfnamefont {A.}~\bibnamefont {Zeilinger}},\ and\ \bibinfo
  {author} {\bibfnamefont {M.}~\bibnamefont {Arndt}},\ }\href
  {https://doi.org/10.1103/PhysRevLett.91.090408} {\bibfield  {journal}
  {\bibinfo  {journal} {Phys. Rev. Lett.}\ }\textbf {\bibinfo {volume} {91}},\
  \bibinfo {pages} {090408} (\bibinfo {year} {2003})}\BibitemShut {NoStop}%
\bibitem [{\citenamefont {Hackermüller}\ \emph {et~al.}(2004)\citenamefont
  {Hackermüller}, \citenamefont {Hornberger}, \citenamefont {Brezger},
  \citenamefont {Zeilinger},\ and\ \citenamefont {Arndt}}]{Hackermuller2004}%
  \BibitemOpen
  \bibfield  {author} {\bibinfo {author} {\bibfnamefont {L.}~\bibnamefont
  {Hackermüller}}, \bibinfo {author} {\bibfnamefont {K.}~\bibnamefont
  {Hornberger}}, \bibinfo {author} {\bibfnamefont {B.}~\bibnamefont {Brezger}},
  \bibinfo {author} {\bibfnamefont {A.}~\bibnamefont {Zeilinger}},\ and\
  \bibinfo {author} {\bibfnamefont {M.}~\bibnamefont {Arndt}},\ }\href
  {https://doi.org/10.1038/nature02276} {\bibfield  {journal} {\bibinfo
  {journal} {Nature}\ }\textbf {\bibinfo {volume} {427}},\ \bibinfo {pages}
  {711} (\bibinfo {year} {2004})}\BibitemShut {NoStop}%
\bibitem [{\citenamefont {Stibor}\ \emph {et~al.}(2005)\citenamefont {Stibor},
  \citenamefont {Stefanov}, \citenamefont {Goldfarb}, \citenamefont {Reiger},\
  and\ \citenamefont {Arndt}}]{Stibor2005}%
  \BibitemOpen
  \bibfield  {author} {\bibinfo {author} {\bibfnamefont {A.}~\bibnamefont
  {Stibor}}, \bibinfo {author} {\bibfnamefont {A.}~\bibnamefont {Stefanov}},
  \bibinfo {author} {\bibfnamefont {F.}~\bibnamefont {Goldfarb}}, \bibinfo
  {author} {\bibfnamefont {E.}~\bibnamefont {Reiger}},\ and\ \bibinfo {author}
  {\bibfnamefont {M.}~\bibnamefont {Arndt}},\ }\href
  {https://doi.org/10.1088/1367-2630/7/1/224} {\bibfield  {journal} {\bibinfo
  {journal} {New J. Phys.}\ }\textbf {\bibinfo {volume} {7}},\ \bibinfo {pages}
  {224} (\bibinfo {year} {2005})}\BibitemShut {NoStop}%
\bibitem [{\citenamefont {Sonnentag}\ and\ \citenamefont
  {Hasselbach}(2007)}]{Sonnentag2007}%
  \BibitemOpen
  \bibfield  {author} {\bibinfo {author} {\bibfnamefont {P.}~\bibnamefont
  {Sonnentag}}\ and\ \bibinfo {author} {\bibfnamefont {F.}~\bibnamefont
  {Hasselbach}},\ }\href {https://doi.org/10.1103/PhysRevLett.98.200402}
  {\bibfield  {journal} {\bibinfo  {journal} {Phys. Rev. Lett.}\ }\textbf
  {\bibinfo {volume} {98}},\ \bibinfo {pages} {200402} (\bibinfo {year}
  {2007})}\BibitemShut {NoStop}%
\bibitem [{\citenamefont {Hasselbach}(2010)}]{Hasselbach2010}%
  \BibitemOpen
  \bibfield  {author} {\bibinfo {author} {\bibfnamefont {F.}~\bibnamefont
  {Hasselbach}},\ }\href {https://doi.org/10.1088/0034-4885/73/1/016101}
  {\bibfield  {journal} {\bibinfo  {journal} {Rep. Prog. Phys.}\ }\textbf
  {\bibinfo {volume} {73}},\ \bibinfo {pages} {016101} (\bibinfo {year}
  {2010})}\BibitemShut {NoStop}%
\bibitem [{\citenamefont {Nimmrichter}\ \emph {et~al.}(2011)\citenamefont
  {Nimmrichter}, \citenamefont {Haslinger}, \citenamefont {Hornberger},\ and\
  \citenamefont {Arndt}}]{Nimmrichter2011b}%
  \BibitemOpen
  \bibfield  {author} {\bibinfo {author} {\bibfnamefont {S.}~\bibnamefont
  {Nimmrichter}}, \bibinfo {author} {\bibfnamefont {P.}~\bibnamefont
  {Haslinger}}, \bibinfo {author} {\bibfnamefont {K.}~\bibnamefont
  {Hornberger}},\ and\ \bibinfo {author} {\bibfnamefont {M.}~\bibnamefont
  {Arndt}},\ }\href {https://doi.org/10.1088/1367-2630/13/7/075002} {\bibfield
  {journal} {\bibinfo  {journal} {New J. Phys.}\ }\textbf {\bibinfo {volume}
  {13}},\ \bibinfo {pages} {075002} (\bibinfo {year} {2011})}\BibitemShut
  {NoStop}%
\bibitem [{\citenamefont {Gerlich}\ \emph {et~al.}(2011)\citenamefont
  {Gerlich}, \citenamefont {Eibenberger}, \citenamefont {Tomandl},
  \citenamefont {Nimmrichter}, \citenamefont {Hornberger}, \citenamefont
  {Fagan}, \citenamefont {Tüxen}, \citenamefont {Mayor},\ and\ \citenamefont
  {Arndt}}]{Gerlich2011}%
  \BibitemOpen
  \bibfield  {author} {\bibinfo {author} {\bibfnamefont {S.}~\bibnamefont
  {Gerlich}}, \bibinfo {author} {\bibfnamefont {S.}~\bibnamefont
  {Eibenberger}}, \bibinfo {author} {\bibfnamefont {M.}~\bibnamefont
  {Tomandl}}, \bibinfo {author} {\bibfnamefont {S.}~\bibnamefont
  {Nimmrichter}}, \bibinfo {author} {\bibfnamefont {K.}~\bibnamefont
  {Hornberger}}, \bibinfo {author} {\bibfnamefont {P.~J.}\ \bibnamefont
  {Fagan}}, \bibinfo {author} {\bibfnamefont {J.}~\bibnamefont {Tüxen}},
  \bibinfo {author} {\bibfnamefont {M.}~\bibnamefont {Mayor}},\ and\ \bibinfo
  {author} {\bibfnamefont {M.}~\bibnamefont {Arndt}},\ }\href
  {https://doi.org/10.1038/ncomms1263} {\bibfield  {journal} {\bibinfo
  {journal} {Nat. Commun.}\ }\textbf {\bibinfo {volume} {2}},\ \bibinfo {pages}
  {263} (\bibinfo {year} {2011})}\BibitemShut {NoStop}%
\bibitem [{\citenamefont {Juffmann}\ \emph {et~al.}(2013)\citenamefont
  {Juffmann}, \citenamefont {Ulbricht},\ and\ \citenamefont
  {Arndt}}]{Juffmann2013}%
  \BibitemOpen
  \bibfield  {author} {\bibinfo {author} {\bibfnamefont {T.}~\bibnamefont
  {Juffmann}}, \bibinfo {author} {\bibfnamefont {H.}~\bibnamefont {Ulbricht}},\
  and\ \bibinfo {author} {\bibfnamefont {M.}~\bibnamefont {Arndt}},\ }\href
  {https://doi.org/10.1088/0034-4885/76/8/086402} {\bibfield  {journal}
  {\bibinfo  {journal} {Rep. Prog. Phys.}\ }\textbf {\bibinfo {volume} {76}},\
  \bibinfo {pages} {086402} (\bibinfo {year} {2013})}\BibitemShut {NoStop}%
\bibitem [{\citenamefont {Eibenberger}\ \emph {et~al.}(2013)\citenamefont
  {Eibenberger}, \citenamefont {Gerlich}, \citenamefont {Arndt}, \citenamefont
  {Mayor},\ and\ \citenamefont {T\"uxen}}]{Eibenberger2013}%
  \BibitemOpen
  \bibfield  {author} {\bibinfo {author} {\bibfnamefont {S.}~\bibnamefont
  {Eibenberger}}, \bibinfo {author} {\bibfnamefont {S.}~\bibnamefont
  {Gerlich}}, \bibinfo {author} {\bibfnamefont {M.}~\bibnamefont {Arndt}},
  \bibinfo {author} {\bibfnamefont {M.}~\bibnamefont {Mayor}},\ and\ \bibinfo
  {author} {\bibfnamefont {J.}~\bibnamefont {T\"uxen}},\ }\href
  {https://doi.org/10.1039/C3CP51500A} {\bibfield  {journal} {\bibinfo
  {journal} {Phys. Chem. Chem. Phys.}\ }\textbf {\bibinfo {volume} {15}},\
  \bibinfo {pages} {14696} (\bibinfo {year} {2013})}\BibitemShut {NoStop}%
\bibitem [{\citenamefont {Fein}\ \emph {et~al.}(2019)\citenamefont {Fein},
  \citenamefont {Geyer}, \citenamefont {Zwick}, \citenamefont {Kiałka},
  \citenamefont {Pedalino}, \citenamefont {Mayor}, \citenamefont {Gerlich},\
  and\ \citenamefont {Arndt}}]{Fein2019}%
  \BibitemOpen
  \bibfield  {author} {\bibinfo {author} {\bibfnamefont {Y.~Y.}\ \bibnamefont
  {Fein}}, \bibinfo {author} {\bibfnamefont {P.}~\bibnamefont {Geyer}},
  \bibinfo {author} {\bibfnamefont {P.}~\bibnamefont {Zwick}}, \bibinfo
  {author} {\bibfnamefont {F.}~\bibnamefont {Kiałka}}, \bibinfo {author}
  {\bibfnamefont {S.}~\bibnamefont {Pedalino}}, \bibinfo {author}
  {\bibfnamefont {M.}~\bibnamefont {Mayor}}, \bibinfo {author} {\bibfnamefont
  {S.}~\bibnamefont {Gerlich}},\ and\ \bibinfo {author} {\bibfnamefont
  {M.}~\bibnamefont {Arndt}},\ }\href
  {https://doi.org/10.1038/s41567-019-0663-9} {\bibfield  {journal} {\bibinfo
  {journal} {Nat. Phys.}\ }\textbf {\bibinfo {volume} {15}},\ \bibinfo {pages}
  {1242} (\bibinfo {year} {2019})}\BibitemShut {NoStop}%
\bibitem [{\citenamefont {Brand}\ \emph {et~al.}(2020)\citenamefont {Brand},
  \citenamefont {Kia\l{}ka}, \citenamefont {Troyer}, \citenamefont {Knobloch},
  \citenamefont {Simonovi\'c}, \citenamefont {Stickler}, \citenamefont
  {Hornberger},\ and\ \citenamefont {Arndt}}]{Brand2020}%
  \BibitemOpen
  \bibfield  {author} {\bibinfo {author} {\bibfnamefont {C.}~\bibnamefont
  {Brand}}, \bibinfo {author} {\bibfnamefont {F.}~\bibnamefont {Kia\l{}ka}},
  \bibinfo {author} {\bibfnamefont {S.}~\bibnamefont {Troyer}}, \bibinfo
  {author} {\bibfnamefont {C.}~\bibnamefont {Knobloch}}, \bibinfo {author}
  {\bibfnamefont {K.}~\bibnamefont {Simonovi\'c}}, \bibinfo {author}
  {\bibfnamefont {B.~A.}\ \bibnamefont {Stickler}}, \bibinfo {author}
  {\bibfnamefont {K.}~\bibnamefont {Hornberger}},\ and\ \bibinfo {author}
  {\bibfnamefont {M.}~\bibnamefont {Arndt}},\ }\href
  {https://doi.org/10.1103/PhysRevLett.125.033604} {\bibfield  {journal}
  {\bibinfo  {journal} {Phys. Rev. Lett.}\ }\textbf {\bibinfo {volume} {125}},\
  \bibinfo {pages} {033604} (\bibinfo {year} {2020})}\BibitemShut {NoStop}%
\bibitem [{\citenamefont {Schrinski}\ \emph {et~al.}(2020)\citenamefont
  {Schrinski}, \citenamefont {Nimmrichter},\ and\ \citenamefont
  {Hornberger}}]{Schrinski2020}%
  \BibitemOpen
  \bibfield  {author} {\bibinfo {author} {\bibfnamefont {B.}~\bibnamefont
  {Schrinski}}, \bibinfo {author} {\bibfnamefont {S.}~\bibnamefont
  {Nimmrichter}},\ and\ \bibinfo {author} {\bibfnamefont {K.}~\bibnamefont
  {Hornberger}},\ }\href {https://doi.org/10.1103/PhysRevResearch.2.033034}
  {\bibfield  {journal} {\bibinfo  {journal} {Phys. Rev. Res.}\ }\textbf
  {\bibinfo {volume} {2}},\ \bibinfo {pages} {033034} (\bibinfo {year}
  {2020})}\BibitemShut {NoStop}%
\bibitem [{\citenamefont {Holevo}(1993)}]{Holevo1993}%
  \BibitemOpen
  \bibfield  {author} {\bibinfo {author} {\bibfnamefont {A.~S.}\ \bibnamefont
  {Holevo}},\ }\href {https://doi.org/10.1016/0034-4877(93)90014-6} {\bibfield
  {journal} {\bibinfo  {journal} {Rep. Math. Phys.}\ }\textbf {\bibinfo
  {volume} {32}},\ \bibinfo {pages} {211} (\bibinfo {year} {1993})}\BibitemShut
  {NoStop}%
\bibitem [{\citenamefont {Vacchini}(2001)}]{Vacchini2001}%
  \BibitemOpen
  \bibfield  {author} {\bibinfo {author} {\bibfnamefont {B.}~\bibnamefont
  {Vacchini}},\ }\href {https://doi.org/10.1063/1.1386409} {\bibfield
  {journal} {\bibinfo  {journal} {J. Math. Phys.}\ }\textbf {\bibinfo {volume}
  {42}},\ \bibinfo {pages} {4291} (\bibinfo {year} {2001})}\BibitemShut
  {NoStop}%
\bibitem [{\citenamefont {Vacchini}(2005)}]{Vacchini2005a}%
  \BibitemOpen
  \bibfield  {author} {\bibinfo {author} {\bibfnamefont {B.}~\bibnamefont
  {Vacchini}},\ }\href {https://doi.org/10.1007/s10773-005-7077-4} {\bibfield
  {journal} {\bibinfo  {journal} {Int. J. Theor. Phys.}\ }\textbf {\bibinfo
  {volume} {44}},\ \bibinfo {pages} {1011} (\bibinfo {year}
  {2005})}\BibitemShut {NoStop}%
\bibitem [{\citenamefont {Segr\`e}(1977)}]{Segre1977}%
  \BibitemOpen
  \bibfield  {author} {\bibinfo {author} {\bibfnamefont {E.~G.}\ \bibnamefont
  {Segr\`e}},\ }\href {https://books.google.com/books?vid=ISBN9780805386011}
  {\emph {\bibinfo {title} {Nuclei and Particles: An Introduction to Nuclear
  and Subnuclear Physics}}},\ \bibinfo {edition} {2nd}\ ed.\ (\bibinfo
  {publisher} {W. A. Benjamin},\ \bibinfo {address} {Reading, Massachusetts},\
  \bibinfo {year} {1977})\BibitemShut {NoStop}%
\bibitem [{\citenamefont {Sigmund}(2006)}]{Sigmund2006}%
  \BibitemOpen
  \bibfield  {author} {\bibinfo {author} {\bibfnamefont {P.}~\bibnamefont
  {Sigmund}},\ }\href {https://doi.org/10.1007/3-540-31718-X} {\emph {\bibinfo
  {title} {Particle Penetration and Radiation Effects: General Aspects and
  Stopping of Swift Point Charges}}},\ \bibinfo {edition} {1st}\ ed.,\ \bibinfo
  {series} {Springer Series in Solid-State Sciences}, Vol.\ \bibinfo {volume}
  {151}\ (\bibinfo  {publisher} {Springer},\ \bibinfo {address} {Berlin,
  Heidelberg},\ \bibinfo {year} {2006})\BibitemShut {NoStop}%
\bibitem [{\citenamefont {Sigmund}(2014)}]{Sigmund2014}%
  \BibitemOpen
  \bibfield  {author} {\bibinfo {author} {\bibfnamefont {P.}~\bibnamefont
  {Sigmund}},\ }\href {https://doi.org/10.1007/978-3-319-05564-0} {\emph
  {\bibinfo {title} {Particle Penetration and Radiation Effects Volume 2:
  Penetration of Atomic and Molecular Ions}}},\ \bibinfo {edition} {1st}\ ed.,\
  \bibinfo {series} {Springer Series in Solid-State Sciences}, Vol.\ \bibinfo
  {volume} {179}\ (\bibinfo  {publisher} {Springer},\ \bibinfo {address}
  {Switzerland},\ \bibinfo {year} {2014})\BibitemShut {NoStop}%
\bibitem [{\citenamefont {Akkermans}\ and\ \citenamefont
  {Montambaux}(2007)}]{Akkermans2007}%
  \BibitemOpen
  \bibfield  {author} {\bibinfo {author} {\bibfnamefont {E.}~\bibnamefont
  {Akkermans}}\ and\ \bibinfo {author} {\bibfnamefont {G.}~\bibnamefont
  {Montambaux}},\ }\href {https://doi.org/10.1017/CBO9780511618833} {\emph
  {\bibinfo {title} {Mesoscopic Physics of Electrons and Photons}}},\ \bibinfo
  {edition} {1st}\ ed.\ (\bibinfo  {publisher} {Cambridge University Press},\
  \bibinfo {year} {2007})\BibitemShut {NoStop}%
\bibitem [{\citenamefont {Sheng}(2006)}]{ShengP2006}%
  \BibitemOpen
  \bibfield  {author} {\bibinfo {author} {\bibfnamefont {P.}~\bibnamefont
  {Sheng}},\ }\href {https://doi.org/10.1007/3-540-29156-3} {\emph {\bibinfo
  {title} {Introduction to Wave Scattering, Localization and Mesoscopic
  Phenomena}}},\ \bibinfo {series} {Springer Series in Materials Science},
  Vol.~\bibinfo {volume} {88}\ (\bibinfo  {publisher} {Springer},\ \bibinfo
  {address} {Berlin},\ \bibinfo {year} {2006})\BibitemShut {NoStop}%
\bibitem [{\citenamefont {Gaspard}\ and\ \citenamefont
  {Sparenberg}(2022{\natexlab{a}})}]{GaspardD2022a}%
  \BibitemOpen
  \bibfield  {author} {\bibinfo {author} {\bibfnamefont {D.}~\bibnamefont
  {Gaspard}}\ and\ \bibinfo {author} {\bibfnamefont {J.-M.}\ \bibnamefont
  {Sparenberg}},\ }\href {https://doi.org/10.1103/PhysRevA.105.042204}
  {\bibfield  {journal} {\bibinfo  {journal} {Phys. Rev. A}\ }\textbf {\bibinfo
  {volume} {105}},\ \bibinfo {pages} {042204} (\bibinfo {year}
  {2022}{\natexlab{a}})},\ \Eprint {https://arxiv.org/abs/2111.03136}
  {arXiv:2111.03136 [quant-ph]} \BibitemShut {NoStop}%
\bibitem [{\citenamefont {Gaspard}\ and\ \citenamefont
  {Sparenberg}(2022{\natexlab{b}})}]{GaspardD2022b}%
  \BibitemOpen
  \bibfield  {author} {\bibinfo {author} {\bibfnamefont {D.}~\bibnamefont
  {Gaspard}}\ and\ \bibinfo {author} {\bibfnamefont {J.-M.}\ \bibnamefont
  {Sparenberg}},\ }\href {https://doi.org/10.1103/PhysRevA.105.042205}
  {\bibfield  {journal} {\bibinfo  {journal} {Phys. Rev. A}\ }\textbf {\bibinfo
  {volume} {105}},\ \bibinfo {pages} {042205} (\bibinfo {year}
  {2022}{\natexlab{b}})},\ \Eprint {https://arxiv.org/abs/2111.04410}
  {arXiv:2111.04410 [quant-ph]} \BibitemShut {NoStop}%
\bibitem [{\citenamefont {Cohen-Tannoudji}\ \emph {et~al.}(2020)\citenamefont
  {Cohen-Tannoudji}, \citenamefont {Diu},\ and\ \citenamefont
  {Lalo\"e}}]{Cohen2020-vol3}%
  \BibitemOpen
  \bibfield  {author} {\bibinfo {author} {\bibfnamefont {C.}~\bibnamefont
  {Cohen-Tannoudji}}, \bibinfo {author} {\bibfnamefont {B.}~\bibnamefont
  {Diu}},\ and\ \bibinfo {author} {\bibfnamefont {F.}~\bibnamefont {Lalo\"e}},\
  }\href {https://books.google.com/books?vid=ISBN9783527345557} {\emph
  {\bibinfo {title} {Quantum Mechanics, Volume 3: Fermions, Bosons, Photons,
  Correlations, and Entanglement}}},\ \bibinfo {edition} {1st}\ ed.\ (\bibinfo
  {publisher} {Wiley},\ \bibinfo {address} {Weinheim},\ \bibinfo {year}
  {2020})\BibitemShut {NoStop}%
\bibitem [{\citenamefont {Barnett}\ \emph {et~al.}(2000)\citenamefont
  {Barnett}, \citenamefont {Franke-Arnold}, \citenamefont {Arnold},\ and\
  \citenamefont {Baxter}}]{Barnett2000}%
  \BibitemOpen
  \bibfield  {author} {\bibinfo {author} {\bibfnamefont {S.~M.}\ \bibnamefont
  {Barnett}}, \bibinfo {author} {\bibfnamefont {S.}~\bibnamefont
  {Franke-Arnold}}, \bibinfo {author} {\bibfnamefont {A.~S.}\ \bibnamefont
  {Arnold}},\ and\ \bibinfo {author} {\bibfnamefont {C.}~\bibnamefont
  {Baxter}},\ }\href {https://doi.org/10.1088/0953-4075/33/19/325} {\bibfield
  {journal} {\bibinfo  {journal} {J. Phys. B: At. Mol. Opt. Phys.}\ }\textbf
  {\bibinfo {volume} {33}},\ \bibinfo {pages} {4177} (\bibinfo {year}
  {2000})}\BibitemShut {NoStop}%
\bibitem [{\citenamefont {Born}\ and\ \citenamefont {Wolf}(2019)}]{Born2019}%
  \BibitemOpen
  \bibfield  {author} {\bibinfo {author} {\bibfnamefont {M.}~\bibnamefont
  {Born}}\ and\ \bibinfo {author} {\bibfnamefont {E.}~\bibnamefont {Wolf}},\
  }\href {https://www.cambridge.org/9781108477437} {\emph {\bibinfo {title}
  {Principles of Optics: 60th Anniversary Edition}}},\ \bibinfo {edition}
  {7th}\ ed.\ (\bibinfo  {publisher} {Cambridge University Press},\ \bibinfo
  {year} {2019})\BibitemShut {NoStop}%
\bibitem [{\citenamefont {Dirac}(1927)}]{Dirac1927}%
  \BibitemOpen
  \bibfield  {author} {\bibinfo {author} {\bibfnamefont {P.~A.~M.}\
  \bibnamefont {Dirac}},\ }\href {https://doi.org/10.1098/rspa.1927.0039}
  {\bibfield  {journal} {\bibinfo  {journal} {Proc. R. Soc. A}\ }\textbf
  {\bibinfo {volume} {114}},\ \bibinfo {pages} {243} (\bibinfo {year}
  {1927})}\BibitemShut {NoStop}%
\bibitem [{\citenamefont {Fermi}(1950)}]{Fermi1950}%
  \BibitemOpen
  \bibfield  {author} {\bibinfo {author} {\bibfnamefont {E.}~\bibnamefont
  {Fermi}},\ }\href {https://books.google.com/books?vid=ISBN9780226243658}
  {\emph {\bibinfo {title} {Nuclear Physics: A Course Given by Enrico Fermi at
  the University of Chicago}}}\ (\bibinfo  {publisher} {University of Chicago
  Press},\ \bibinfo {year} {1950})\ \bibinfo {note} {notes compiled by Jay
  Orear, A. H. Rosenfeld, and R. A. Schluter}\BibitemShut {NoStop}%
\bibitem [{\citenamefont {Visser}(2009)}]{Visser2009}%
  \BibitemOpen
  \bibfield  {author} {\bibinfo {author} {\bibfnamefont {T.~D.}\ \bibnamefont
  {Visser}},\ }\href {https://doi.org/10.1119/1.3096649} {\bibfield  {journal}
  {\bibinfo  {journal} {Am. J. Phys.}\ }\textbf {\bibinfo {volume} {77}},\
  \bibinfo {pages} {487} (\bibinfo {year} {2009})}\BibitemShut {NoStop}%
\bibitem [{\citenamefont {Landau}\ and\ \citenamefont
  {Lifshitz}(1967)}]{Landau1967}%
  \BibitemOpen
  \bibfield  {author} {\bibinfo {author} {\bibfnamefont {L.~D.}\ \bibnamefont
  {Landau}}\ and\ \bibinfo {author} {\bibfnamefont {E.~M.}\ \bibnamefont
  {Lifshitz}},\ }\href@noop {} {\emph {\bibinfo {title} {Quantum Mechanics:
  Non-Relativistic Theory}}},\ \bibinfo {edition} {2nd}\ ed.,\ \bibinfo
  {series} {Course of Theoretical Physics}, Vol.~\bibinfo {volume} {3}\
  (\bibinfo  {publisher} {Mir Editions},\ \bibinfo {address} {Moscow},\
  \bibinfo {year} {1967})\ \bibinfo {note} {translated from the Russian by
  Edouard Gloukhian}\BibitemShut {NoStop}%
\bibitem [{\citenamefont {Sakurai}\ and\ \citenamefont
  {Napolitano}(2020)}]{Sakurai2020}%
  \BibitemOpen
  \bibfield  {author} {\bibinfo {author} {\bibfnamefont {J.~J.}\ \bibnamefont
  {Sakurai}}\ and\ \bibinfo {author} {\bibfnamefont {J.}~\bibnamefont
  {Napolitano}},\ }\href {https://doi.org/10.1017/9781108587280} {\emph
  {\bibinfo {title} {Modern Quantum Mechanics}}},\ \bibinfo {edition} {3rd}\
  ed.\ (\bibinfo  {publisher} {Cambridge University Press},\ \bibinfo {year}
  {2020})\BibitemShut {NoStop}%
\bibitem [{\citenamefont {Joachain}(1979)}]{Joachain1979}%
  \BibitemOpen
  \bibfield  {author} {\bibinfo {author} {\bibfnamefont {C.~J.}\ \bibnamefont
  {Joachain}},\ }\href {https://books.google.com/books?vid=ISBN9780444852359}
  {\emph {\bibinfo {title} {Quantum Collision Theory}}},\ \bibinfo {edition}
  {2nd}\ ed.\ (\bibinfo  {publisher} {North-Holland},\ \bibinfo {address}
  {Amsterdam},\ \bibinfo {year} {1979})\BibitemShut {NoStop}%
\bibitem [{\citenamefont {Newton}(1982)}]{Newton1982}%
  \BibitemOpen
  \bibfield  {author} {\bibinfo {author} {\bibfnamefont {R.~G.}\ \bibnamefont
  {Newton}},\ }\href {https://books.google.com/books?vid=ISBN9780486425351}
  {\emph {\bibinfo {title} {Scattering Theory of Waves and Particles}}},\
  \bibinfo {edition} {2nd}\ ed.,\ Dover Books on Physics\ (\bibinfo
  {publisher} {Dover},\ \bibinfo {address} {Mineola},\ \bibinfo {year}
  {1982})\BibitemShut {NoStop}%
\bibitem [{\citenamefont {Taylor}(2006)}]{Taylor2006}%
  \BibitemOpen
  \bibfield  {author} {\bibinfo {author} {\bibfnamefont {J.~R.}\ \bibnamefont
  {Taylor}},\ }\href {https://books.google.com/books?vid=ISBN9780486450131}
  {\emph {\bibinfo {title} {Scattering Theory: The Quantum Theory of
  Nonrelativistic Collisions}}},\ Dover Books on Engineering\ (\bibinfo
  {publisher} {Dover Publications},\ \bibinfo {address} {Mineola},\ \bibinfo
  {year} {2006})\BibitemShut {NoStop}%
\bibitem [{\citenamefont {Strunz}\ and\ \citenamefont {Yu}(2004)}]{Strunz2004}%
  \BibitemOpen
  \bibfield  {author} {\bibinfo {author} {\bibfnamefont {W.~T.}\ \bibnamefont
  {Strunz}}\ and\ \bibinfo {author} {\bibfnamefont {T.}~\bibnamefont {Yu}},\
  }\href {https://doi.org/10.1103/PhysRevA.69.052115} {\bibfield  {journal}
  {\bibinfo  {journal} {Phys. Rev. A}\ }\textbf {\bibinfo {volume} {69}},\
  \bibinfo {pages} {052115} (\bibinfo {year} {2004})}\BibitemShut {NoStop}%
\bibitem [{\citenamefont {Vladimirov}(1971)}]{Vladimirov1971}%
  \BibitemOpen
  \bibfield  {author} {\bibinfo {author} {\bibfnamefont {V.~S.}\ \bibnamefont
  {Vladimirov}},\ }\href {https://books.google.com/books?vid=ISBN9780824717131}
  {\emph {\bibinfo {title} {Equations of Mathematical Physics}}},\ edited by\
  \bibinfo {editor} {\bibfnamefont {A.}~\bibnamefont {Jeffrey}},\ \bibinfo
  {series} {Monographs and textbooks in pure and applied mathematics},
  Vol.~\bibinfo {volume} {3}\ (\bibinfo  {publisher} {Marcel Dekker},\ \bibinfo
  {address} {New York},\ \bibinfo {year} {1971})\ \bibinfo {note} {translated
  from Russian by Audrey Littlewood}\BibitemShut {NoStop}%
\bibitem [{\citenamefont {Adler}(2006)}]{Adler2006}%
  \BibitemOpen
  \bibfield  {author} {\bibinfo {author} {\bibfnamefont {S.~L.}\ \bibnamefont
  {Adler}},\ }\href {https://doi.org/10.1088/0305-4470/39/45/015} {\bibfield
  {journal} {\bibinfo  {journal} {J. Phys. A: Math. Gen.}\ }\textbf {\bibinfo
  {volume} {39}},\ \bibinfo {pages} {14067} (\bibinfo {year}
  {2006})}\BibitemShut {NoStop}%
\bibitem [{\citenamefont {Wigner}(1932)}]{Wigner1932}%
  \BibitemOpen
  \bibfield  {author} {\bibinfo {author} {\bibfnamefont {E.~P.}\ \bibnamefont
  {Wigner}},\ }\href {https://doi.org/10.1103/PhysRev.40.749} {\bibfield
  {journal} {\bibinfo  {journal} {Phys. Rev.}\ }\textbf {\bibinfo {volume}
  {40}},\ \bibinfo {pages} {749} (\bibinfo {year} {1932})}\BibitemShut
  {NoStop}%
\bibitem [{\citenamefont {Moyal}(1949)}]{Moyal1949a}%
  \BibitemOpen
  \bibfield  {author} {\bibinfo {author} {\bibfnamefont {J.~E.}\ \bibnamefont
  {Moyal}},\ }\href {https://doi.org/10.1017/S0305004100000487} {\bibfield
  {journal} {\bibinfo  {journal} {Math. Proc. Camb. Philos. Soc.}\ }\textbf
  {\bibinfo {volume} {45}},\ \bibinfo {pages} {99} (\bibinfo {year}
  {1949})}\BibitemShut {NoStop}%
\bibitem [{\citenamefont {Basdevant}\ and\ \citenamefont
  {Dalibard}(2002)}]{Basdevant2002}%
  \BibitemOpen
  \bibfield  {author} {\bibinfo {author} {\bibfnamefont {J.-L.}\ \bibnamefont
  {Basdevant}}\ and\ \bibinfo {author} {\bibfnamefont {J.}~\bibnamefont
  {Dalibard}},\ }\href {https://doi.org/10.1007/3-540-28805-8} {\emph {\bibinfo
  {title} {Quantum Mechanics}}},\ Advanced Texts in Physics\ (\bibinfo
  {publisher} {Springer},\ \bibinfo {address} {Berlin, Heidelberg},\ \bibinfo
  {year} {2002})\BibitemShut {NoStop}%
\end{thebibliography}
\end{document}